\documentclass[paper]{agujournal2019}
\usepackage{url} 
\usepackage{lineno}
\usepackage[inline]{trackchanges} 
\usepackage{soul}
\nolinenumbers

\graphicspath{{./figures/}}

\usepackage{graphicx}	
\usepackage{amsmath}	
\usepackage{amssymb}	

\usepackage{booktabs}
\usepackage{physics}
\usepackage[flushleft]{threeparttable}

\newcommand\aap{A\&A}                
\newcommand\apjl{ApJ}                
\newcommand\apjs{ApJS}               

\newcommand{\specialcell}[2][c]{%
  \begin{tabular}[#1]{@{}c@{}}#2\end{tabular}}

\journalname{Journal of Geophysical Research: Planets}

\begin{document}

\title{Stability and Evolution of Fallen Particles Around\\ the Surface of Asteroid (101955) Bennu}

%
%



\authors{A. Amarante\affil{1,2,3}, O.C. Winter\affil{2}\,and R. Sfair\affil{2}}

\affiliation{1}{Grupo de Matem\'atica Aplicada e Processamento de Sinais, State University of Mato Grosso do Sul, Cassil\^andia, MS, Brazil}
\affiliation{2}{Grupo de Din\^amica Orbital e Planetologia, S\~ao Paulo State University, Guaratinguet\'a, SP, Brazil}
\affiliation{3}{Grupo de Simula\c c\~ao Num\'erica Computacional, IFSP, Cubat\~ao, SP, Brazil}





\correspondingauthor{A. Amarante}{andre.amarante@unesp.br}




\begin{keypoints}
\item The dynamics of particles orbiting near the equator of Bennu are dominated by eight equilibrium points irregularly located around the asteroid.
\item Particles smaller than a centimeter are quickly removed from the system by the solar radiation pressure.
\item Particles larger than a few centimeters, initially orbiting in a spherical cloud around Bennu, preferentially fall onto the surface near high-altitude regions at low equatorial latitudes and close to the north pole.
\end{keypoints}


%
%

%
%


\begin{abstract}
In this work, we study the dynamics of particles around Bennu.
The goal is to understand the stability, evolution, and final outcome of the simulated particles around the asteroid.
According to the results, the particle sizes can be divided into two main groups depending on their behavior. Particles smaller than a centimeter are quickly removed from the system by solar radiation pressure, while the dynamics of particles larger than a few centimeters is dominated by the gravitational field of Bennu.
Because of its shape and spin period, Bennu has eight equilibrium points around it. The structure of the phase space near its equatorial surface is directly connected to these equilibrium points.
Therefore, we performed numerical simulations to obtain information about the orbital evolution near the equilibrium points.
The results show that most of the particles larger than a few centimeters fall in the equatorial region close to the Kingfisher area or close to the region diametrically opposite to it. In contrast, almost none of these particles fall in the equatorial region close to the Osprey area.
In addition, we also performed computational experiments considering a spherical cloud of particles initially orbiting Bennu. Most of the particles in prograde orbits fall on the surface within our integration period, which was limited to 1.14 years.
The particles preferentially fall near high-altitude regions at low equatorial latitudes and close to the north pole.
The mid-latitudes are those more depleted of falls, as in the Nightingale and Sandpiper areas.

\end{abstract}

\section*{Plain Language Summary}
In general, asteroids are small bodies with a very irregular shape that is not spherical.
Bennu is an asteroid being explored by the spacecraft mission OSIRIS-REx.
This asteroid has a size smaller than five hundred meters (1/3 mile), and its shape is a bouldery spheroid with an equatorial ridge.
Bennu's gravity is very weak and complicated because of its shape, and it rotates around itself about every four hours.
The evolution of the trajectories of small boulders around Bennu is a puzzling problem that is explored in this work.
Performing computational simulations, we identified equilibrium positions and the preferred paths.
Particles smaller than a centimeter are quickly removed from the system,
while most particles of a few centimeters or larger collide with the surface of Bennu.
The majority of the impacts are in the equatorial region where the locations near valleys of the surface did not have a significant number of falls, while the number of falls is high near peaks.
The polar regions also show a considerable number of impacts, and there is a location in Bennu's north pole with a large number of falls.
Lastly, this work presents animated movies showing the simulations.

%
%

%


%
%
%
%

\section{Introduction}
\label{sec:intro}
In 2018, the OSIRIS-REx mission \cite{Lauretta2017} started the rendezvous with its target asteroid, (101955) Bennu. Many key parameters were estimated during its preliminary survey phase. An overview of the results was presented in \citeA{Lauretta2019}. Some important parameters such as the shape, mass, density, and spin state were derived in \citeA{Barnouin2019}, \citeA{Scheeres2019}, and \citeA{Hergenrother2019}.

The irregular shape of Bennu has been represented in detail by a polyhedron and is covered with boulders along its surface \cite{Lauretta2019,Walsh2019}. The current work is a study on the orbital dynamics near the surface of Bennu taking into account the solar radiation pressure (SRP) effects. We focused on the dynamics of particles with negligible mass under the gravitational effect because of Bennu's irregular shape, considering a uniform density. \citeA{McMahon2020} focused on the dynamics of ejected particles observed by OSIRIS-REx, while the present work attempts to understand the general dynamics of small particles around Bennu, giving new understanding of dust particle behaviors around this asteroid.

Because there are eight equilibrium points in this region \cite{Winter2017,Scheeres2019,Scheeres2019b}, the structure of the phase space of such a region might be strongly dependent on those equilibrium points of the system. Therefore, it is important to know what occurs to the particles that are in the vicinity of Bennu's equilibrium points. What is the influence of these equilibrium points on the dynamics of particles near Bennu's surface? For example, are the particles distributed in a torus around Bennu containing these points, or in local clouds surrounding the eight equilibrium points, or even in a spherical cloud around the surface of Bennu stable? These particles eventually could fall back to Bennu's surface. Thus, where are the preferred locations for reaccumulation, how long do the particles survive before colliding with Bennu's surface, and how does the initial inclination of the particles changes the particle distributions over the surface of asteroid Bennu?

Thus, to answer these questions, we performed sets of numerical simulations exploring the orbital evolution of particles in terms of the location of temporarily stable regions and the distribution of impacts of particles orbiting Bennu.

This paper is structured as follows. The next section presents the mathematical models adopted in the numerical simulations. We used a mascon approach to compute the geopotential around the surface of asteroid Bennu. The SRP is also considered for the adopted model. The results associated with the equilibrium points under SRP and their stabilities are given in Section \ref{sec:equi}. The numerical simulations that were performed are discussed and presented in Section \ref{sec:sim}. We used three different types of initial conditions: particles were placed in local disks surrounding the eight equilibrium points of Bennu, or synchronous orbits where a particle is stationary in the asteroid-fixed frame. A torus filled with particles from the Bennu center of mass and containing equilibrium points is also investigated. Section \ref{sec:fall} shows the results of the density of falls along the surface for particles orbiting in an initially spherical cloud around Bennu. Finally, the last section presents our final comments about the present manuscript.

\section{Mathematical Model}
\label{sec:math}
There are currently three known models for the shape of asteroid Bennu. The OSIRIS-REx team planned the spacecraft's mission using a first radar-based shape model ($1,348$ vertices, $4,038$ edges, and $2,692$ faces) that was developed in 2013 based on radar and optical photometric observations made in 1999 and 2005 from the Arecibo Observatory and the Goldstone Tracking Station \cite{Nolan2013}. Optical observations are conducted at the Catalina Station 1.5-m telescope and the Chuguev Observation Station 0.7-m telescope. This model resolution is approximately 25\,m between vertices. Although this original model did not predict the surface topography of Bennu in detail, at least one large rock formation was predicted in the southern hemisphere. The preliminary shape model ($25,350$ vertices, $73,728$ edges, and $49,152$ faces) of asteroid Bennu was created from a compilation of images taken by OSIRIS-REx's PolyCam camera during the spacecraft's approach toward Bennu during November 2018. This shape model shows features on Bennu as small as 6\,m. After studying asteroid Bennu at close range for three months, the OSIRIS-REx mission team has developed a new 3-D shape model with a 75-cm resolution. This more precise model shows features smaller than 1\,m on the surface of Bennu (the v20 model of \cite{Barnouin2019}). The 75-cm shape model of asteroid Bennu was adopted in the current work.

We obtained the polyhedral data for Bennu from the following website: \url{https://www.asteroidmission.org/updated-bennu-shape-model-3d-files/}. For our purposes, we are only interested in the geometric vertices (v) of the 75-cm shape model. Then, we extracted the 3-D Bennu polyhedral model using $99,846$ vertices and $294,912$ edges combined into $196,608$ triangular faces. We used the GNUPlot program \cite{Williams2011} to build the triangular mesh of the model. The 3-D Bennu polyhedral shape model is also available at Data Publisher for Earth $\&$ Environmental Science \cite{Amarante2020data1}. The polyhedral model of Bennu was also constructed using the same volume and overall dimensions computed from the best-fitting model by Barnouin et al. \citeA{Barnouin2019}. It has a volume of $61,588,235$\,m$^3$ with an equivalent approximated spherical radius of $245$\,m.

The mathematical model considered in this work considers two main significant perturbations for minor bodies: the gravity that comes from the geopotential generated by the irregular Bennu polyhedral shape model and the SRP. This work examines the dynamics of particles near the surface of Bennu and also examines the shadowing. Previous studies \cite{McMahon2020} found that all of these perturbations are crucial in producing the correct evolutionary behavior for particles around asteroid Bennu. These studies are described in some detail in the following sections.

\subsection{Geopotential}
\label{sec:math:geo}
A commonly used approach for modeling the gravitational force potential field of an irregular minor body is to fill its volume with point masses (mascons) on an evenly spaced cubic grid in a suitable way that best reproduces the mass distribution of the body shape \cite{Geissler1996}. Mascon models are particularly suited to minor bodies because of their ability to model irregular shapes and density distributions at arbitrary resolution \cite{Scheeres1998,Park2010}. Each mascon has its own mass and volume, and the sum of them must be the total mass and the entire volume of the body \cite{Werner1997b,Scheeres1998}. In general, the body volume is filled by point masses and the force exerted on a particle in orbit around the body is given by the vector sum of the forces generated by each mass concentration. Intuitively, even with a large number of mascons, there would still be empty spaces, i.e., the best solution would be an infinite number of small masses, but this situation does not exist, and a larger number of small masses slows the computational processing speed. The number of mascons is inversely proportional to the integration speed. Therefore, the number of mass concentrations to be used depends on the desired accuracy and purpose. A great advantage of this method is the simple conceptual approach, and depending on the number of point masses used, the accuracy can be considered quite satisfactory for the study of irregular minor objects, as shown in later sections. Even when near the body, this method can be used, unlike the method of spherical harmonics \cite{MacMillan1936,Kaula1966}. This procedure using the mascon model has been extensively tested in previous work to study the dynamical environment and surface characteristics of an asteroid \cite{Amarante2020,Winter2020,Moura2020}. Then, we use the mascons technique to compute Bennu's gravitational force potential numerically as follows:
\begin{eqnarray}
U(x,y,z) & = & \sum_{i=1}^{N}\frac{\mu_i}{|\textbf{r}-\textbf{r}_i|},
\label{eq:sim_2}
\end{eqnarray}
\noindent where $x$, $y$, and $z$ represent the coordinates of a particle in the Bennu-fixed frame measured from the Bennu barycenter, with the unit vectors defined along the minimum, intermediate, and maximum moments of inertia. $\mu_i=Gm_i$ are the gravitational parameters of each mascon with gravitational constant (CODATA - \url{http://physics.nist.gov/constants}) $G=6.67408 \times 10^{-20}$\, km$^3$\, kg$^{-1}$\, s$^{-2}$. $N$ indicates the number of mascon cells confined in the Bennu volume, $\textbf{r}$ shows the particle radius vector from the Bennu centroid, $\textbf{r}_i$ indicates the mascon distance relative to the center of mass of Bennu, $|\textbf{r}-\textbf{r}_i|=r_i=\sqrt{(x-x_i)^2+(y-y_i)^2+(z-z_i)^2}$ represents the distance of the orbiting particle in relation to each mascon, whose mass is given by $m_i=M/N$, and $M$ is the polyhedron mass given in Table \ref{tab:math_0}. In this approach, we considered $12$-m spaced mascon cubic cells following the irregular shape model of Bennu, resulting in a model composed of $N = 35,640$ point masses with an equal mass of $m_i = 2.05632 \times 10^{6}$ kg each.

We compute the mascon gravity attraction vector as follows:
\begin{eqnarray}
\nabla U(x,y,z) & = & -\sum_{i=1}^{N}\frac{\mu_i}{|\textbf{r}-\textbf{r}_i|^3}(\textbf{r}-\textbf{r}_i),
\label{eq:sim_3}
\end{eqnarray}
\noindent where $\nabla$ represents the Hamiltonian operator.

In this method, the gravity gradient tensor can be computed as shown in Eq. (\ref{eq:sim_4}):
\begin{eqnarray}
\nabla\nabla U(x_1,x_2,x_3) & = & {{\partial U(x_1,x_2,x_3)}\over{\partial x_m}{\partial x_n}} = \sum_{i=1}^{N}\frac{\mu_i}{r_i^3}\Bigg[\frac{3(x_m-x_{m_i})(x_n-x_{n_i})}{r_i^2}-\delta_{m,n}\Bigg]
\qquad (m,n = 1,2,3),
\label{eq:sim_4}
\end{eqnarray}
\noindent where $(x_1,x_2,x_3)\equiv(x,y,z)$ and $\delta_{m,n}$ is the Kronecker delta.

These assumptions lead us to compute Bennu's geopotential \cite{Scheeres2016}, as shown in Eq. (\ref{eq:sim_1b}):
\begin{eqnarray}
V(x,y,z)=-{1\over 2}\omega^2(x^2+y^2)-U(x,y,z)
\label{eq:sim_1b}
\end{eqnarray}
\noindent where $\omega$ is the spin rate of asteroid Bennu.
\begin{table}
\centering
  \caption{The SRP parameters used in the mathematical model.}
 \label{tab:math_1}
 \scalebox{1.0}
{
 \begin{tabular}{cccc}
  \toprule
  Parameter & Value & Units & Comments \\
  \hline
  $Q_{pr}$ & 1 & -- & \text{efficiency factor} \\
  $S_0$ & $1.36\times 10^3$ & \text{kg/s$^3$} & \text{solar constant} \\
  $R_0$ & $1.495978707\times 10^8$ & \text{km} & astronomical unit distance \\
  $c$ & $2.99792197442722\times 10^5$ & \text{km/s} & speed of light \\
  $\rho$ & 1.190 & \text{g/cm$^{3}$} & bulk density \\
  $r_p$ & 10$^{-4}$--10 & \text{cm} & particles spherical radius \\
  \hline
 \end{tabular}}
 \end{table}

\subsection{Solar Radiation Pressure}
\label{sec:math:srp}
The stability and orbital evolution of particles around minor bodies are also affected by the radial SRP force because of the Sun. The equation of motion of a particle with geometric cross section $A$ and mass $m$ moving under the influence of the SRP acceleration can be written as \cite{Burns1979,Mignard1984}:
\begin{eqnarray}
\textbf{a}_{SRP} & = & -\frac{Q_{pr}S_0R_0^2}{c|\textbf{r}_{s}-\textbf{r}|^3}\frac{A}{m}W(\theta ,r_\theta)(\textbf{r}_{s}-\textbf{r}),
\label{eq:math_1}
\end{eqnarray}
\noindent where $\textbf{r}_{s}$ is the position vector of the Sun with respect to Bennu, $|\textbf{r}_{s}-\textbf{r}|$ is the distance from the particle's surface element to the Sun, $S_0$ is the solar constant or radiation flux density at astronomical unit distance $R_0$, $c$ is the speed of light, and $Q_{pr}$ is the dimensionless efficiency factor for radiation pressure, which depends on the properties (e.g., density, shape, size) of the particle. In the mathematical model, it is assumed to be equal to unity to represent the value of an ideal material. The values of the adopted SRP parameters are given in Table \ref{tab:math_1}.
\begin{table}
\centering
\begin{threeparttable}
  \caption{Orbital and physical properties of asteroid Bennu.}
 \label{tab:math_0}
 \begin{tabular}{ccc}
  \toprule
   Parameter & Value & Units \\
  \hline
  Semimajor axis\tnote{\textit{a}} & $1.68507040768\times 10^8$ & km \\
  Eccentricity\tnote{\textit{a}} & 0.20375 & -- \\
  Inclination\tnote{\textit{a}} & 6.0349 & degrees \\
  Arg. of perihelion\tnote{\textit{a}} & 66.2231 & degrees \\
  Long. of asc. node\tnote{\textit{a}} & 182.0609 & degrees \\
  Mean anomaly\tnote{\textit{a}} & 101.7039 & degrees \\
  Obliquity\tnote{\textit{b}} & 177.6 & degrees \\
  Rotation period\tnote{\textit{b}} & 4.296057 & hours \\
  Mass\tnote{\textit{b}} & $7.329\times 10^{10}$  & kg \\
  \hline
 \end{tabular}
\footnotesize
    \begin{tablenotes}
        \item[\textit{a}] \cite{Chesley2014}.
        \item[\textit{b}] \cite{Lauretta2019}.
    \end{tablenotes}
\end{threeparttable}
 \end{table}


Because the distributed particles are located in Bennu's neighborhood, we need to consider a shadowing model to compute SRP acceleration. The shadowing function $W(\theta ,r_\theta)$ from Eq. (\ref{eq:math_1}) is assumed to be cylindrical to simplify the computational time effort of the numerical integration. The cylindrical shadow has a radius value of 290\,m, which is close to the polyhedral mesh of Bennu's surface. Its shadowing function angle $\theta$ is computed as follows:
\begin{eqnarray}
\zeta & = & \cos(\theta),
\label{eq:math_2}
\end{eqnarray}
\noindent where $\zeta=\frac{\textbf{r}\cdot\textbf{r}_s}{|\textbf{r}||\textbf{r}_s|}$ and $0<\theta<\pi$.

The shadowing radius $r_\theta$ can be computed from Eq. (\ref{eq:math_2}) as:
\begin{eqnarray}
r_\theta & = & |\textbf{r}|\sin(\theta).
\label{eq:math_3}
\end{eqnarray}
If $\theta>\pi/2$ and $r_\theta > 290$\,m, then $W(\theta ,r_\theta)=1$. Otherwise, $W(\theta ,r_\theta)=0$.

Note that the negative sign in Eq. (\ref{eq:math_1}) makes the SRP acceleration act in the anti-Sun direction. In addition, we assume that the Sun describes a Keplerian orbit around Bennu, in the asteroid-fixed frame, whose orbital elements are given by Table \ref{tab:math_0}.

Finally, particles in the mathematical model are modeled as spheres such that the area-to-mass ratio varies, as shown in Eq. (\ref{eq:math_4}):
\begin{eqnarray}
\frac{A}{m} & = & {3\over 4}\frac{1}{\rho r_p},
\label{eq:math_4}
\end{eqnarray}
\noindent where $\rho$ is the particle density, with the assumption that it is the same as Bennu's bulk density for all particles, and $r_p$ is the particle spherical radius, from fine dust grains up to small particles (Table \ref{tab:math_1}). We also provide at the end of subsection \ref{sec:to} a discussion of particle radius limits in terms of the density of meteorite analogs \cite{Hamilton2019,Lauretta2019b}.



\subsection{Analytical Approach}
\label{sec:sim:srp}
\begin{figure}
  \centering
  \includegraphics[width=7.0cm]{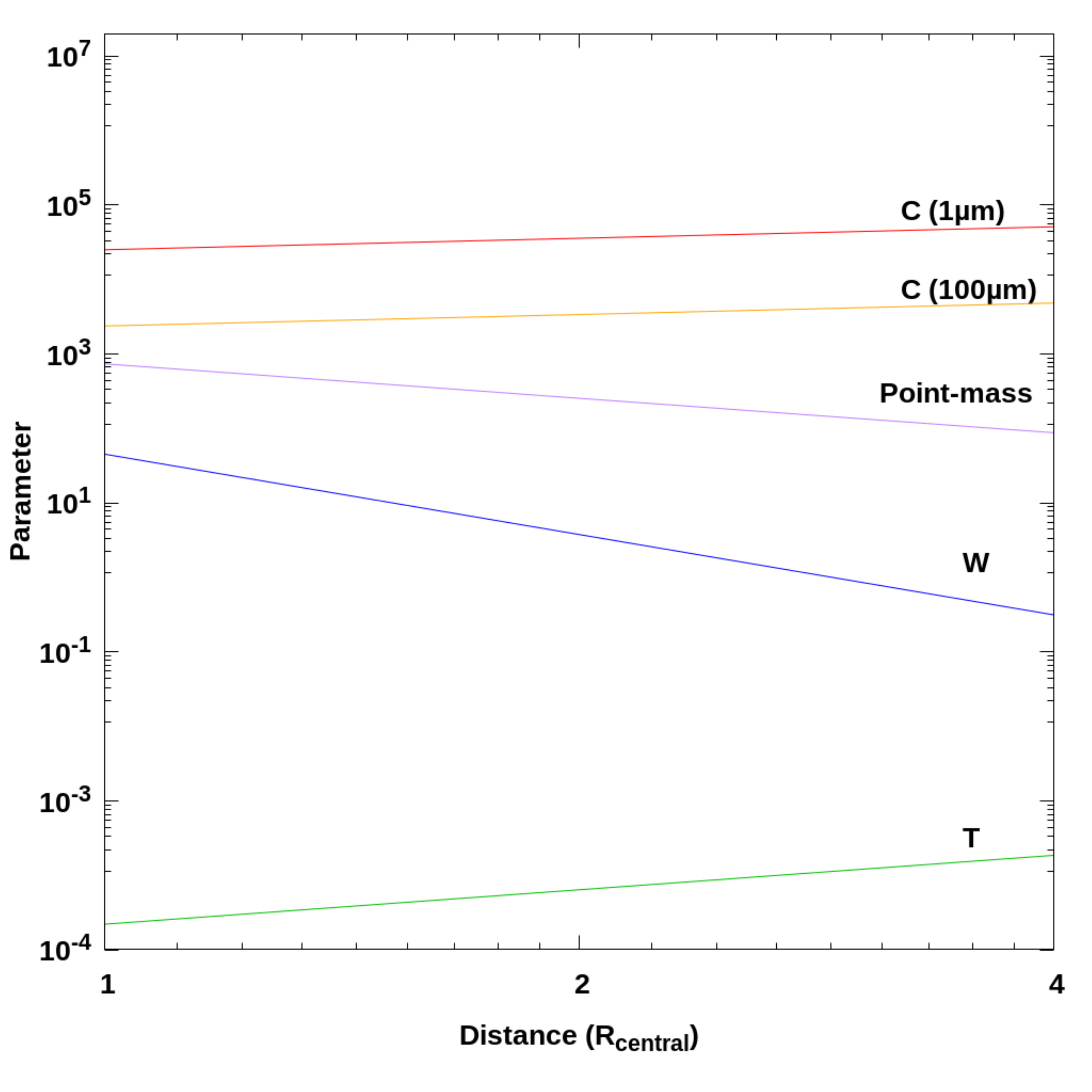}
  \caption{Dimensionless parameters to compare the strengths of the solar tide ($A$), the radiation pressure ($C$), and the
oblateness ($W$) as a function of the distance from Bennu in units of its volumetric radius. Derivation of these parameters
can be found in \cite{hamilton96}.}
  \label{fig:sim_0}
\end{figure}
\citeA{hamilton96} presented dimensionless parameters that allow the strengths of some perturbations to be compared as a function of the distance of the central body. Following their notation, Figure \ref{fig:sim_0} shows the parameters calculated for Bennu's environment in the range from 1 to 4 volumetric radii from the asteroid. The computed parameters correspond to the oblateness ($W$), solar tide ($T$), and SRP ($C$) for two different sizes of particles.

Assuming that the asteroid follows a circular heliocentric orbit, the solar tide parameter is defined as
\begin{align}
    T = \frac{3}{4} \frac{n_\odot}{n},
\end{align}
\noindent where $n$ is the mean motion of the particle and $n_\odot$ is the mean motion of the asteroid around the Sun.
The parameter $W$ depends on the radius of the asteroid ($R$) and its second zonal coefficient $J_2$:
\begin{align}
    W = \frac{3}{2} J_2 \left(\frac{R}{a}\right)^2\frac{n}{n_\odot},
\end{align}
\noindent where $J2=-C_{20}=0.0581$.

To compute the SRP parameter for a given particle with radius $r_p$ at a distance $r$ from the asteroid, the expression is

\begin{align}
    C = \frac{9}{8}  \frac{n}{n_\odot} Q_{pr} \frac{F_\odot r^2}{Gmc \rho r_p},
\end{align}

\noindent where $F_\odot$ is the solar flux at the asteroid heliocentric distance.

These variables were determined by assuming a circular orbit for Bennu, although no qualitative difference was found when considering the orbital eccentricity of the asteroid. For completeness, a scaled strength for the point-mass attraction of the asteroid is also presented.

As is expected, the forces because of Bennu's gravity ($W$ and point-mass attraction) decrease according to the distance, while solar tide ($T$) increased for external particles. The radiation pressure parameter is mostly constant in the interval, and it is stronger for smaller grains.

The comparison shows that the solar tide can be safely ignored in the region close to the surface because it is at least two orders of magnitude weaker than the other parameters. In contrast, even for grains as large as 10$^{-2}$\,cm, the SRP easily surpasses all the forces, indicating that any dust eventually present in the vicinity of the asteroid experiences a strong perturbation and is quickly removed.


\section{Equilibrium Points under Solar Radiation Pressure}
\label{sec:equi}
The SRP perturbation force might significantly change the location and stability of the equilibrium points. Given this condition, it is important to consider the SRP perturbation in the equilibria approach for asteroid Bennu. Considering the geopotential from the irregular polyhedral shape of asteroid Bennu (Eq. (\ref{eq:sim_1b})) and the SRP perturbation (Eq. (\ref{eq:math_1})), the locations of the instantaneous equilibrium points can be found when:
 \begin{equation}
   \nabla  V(x,y,z) -\textbf{a}_{SRP} = \textbf{0}.
 \label{eq:equi_1}
 \end{equation}
The locations of the instantaneous equilibrium points are computed for a given relative position of the orientation of Bennu and the Sun. This relative position changes in time because of the rotational and orbital motions of Bennu. In the cases where the effect of SRP does not significantly affect the system, the instantaneous equilibrium points do not change their locations in the Bennu-fixed frame. Otherwise, in the cases where the SRP significantly affects the dynamics of the system, the locations of the instantaneous equilibrium points change with time, generating a kind of equilibrium curve, known as dynamical equivalent \cite{Xin2015}.
\begin{table}
\centering
  \caption{Location of equilibrium points about asteroid Bennu, and their values of longitude $\lambda$, radial barycentric distance $r_{eq}$, the geopotential, $V(x,y,z)$, the orbital velocity  at the inertial frame $v_{eq}$, and the equilibrium lifetime. Equilibria were computed through the mascons technique with Minor-Equilibria-NR \cite{minor-equilibria-nr} package} using an accuracy of $10^{-5}$, and assuming a constant density and a uniform rotation period for Bennu.
 \label{tab:equi_mas}
 \scalebox{1.0}
{
 \begin{tabular}{ccccccccc}
  \toprule
  Point & X (m) & Y (m) & Z (m) & $\lambda$ (deg) & $r_{eq}$ (m) & \specialcell{$V(x,y,z)$\\ \small(cm$^2$/s$^{2}$)} & \specialcell{$v_{eq}$\\ \small(cm/s)} & \specialcell{equilibrium\\ lifetime (h)} \\
  \hline
  E$_1$ & 311.806 & 83.0970 & -0.622273 & 14.9226 & 322.689 & -242.533 & 13.1094 & 5 \\
  E$_2$ & 80.9594 & 303.489 & 1.79614 & 75.0635 & 314.107 & -239.197 & 12.7654 & 15 \\
  E$_3$ & -192.882 & 257.253 & -0.208082 & 126.862 & 321.532 & -241.766 & 13.0513 & 8 \\
  E$_4$ & -272.788 & 163.934 & -2.31600 & 148.996 & 318.265 & -241.237 & 12.9187 & 10 \\
  E$_5$ & -296.945 & -127.131 & -2.57372 & 203.177 & 323.025 & -242.594 & 13.1183 & 8 \\
  E$_6$ & -2.79430 & -314.841 & -1.533460 & 269.491 & 314.858 & -239.558 & 12.7969 & 17 \\
  E$_7$ & 187.591 & -257.634 & -1.56659 & 306.059 & 318.697 & -240.720 & 12.9411 & 8 \\
  E$_8$ & 269.591 & -164.149 & -0.862951 & 328.663 & 315.635 & -240.350 & 12.8280 & 10 \\
  E$_9$ & 0.887886 & -0.311608 & 0.0217195 & 340.661 & 0.941230 & -298.443 & - & - \\
  \hline
 \end{tabular}}
 \end{table}

\begin{figure}
  \centering
  \fbox{\includegraphics[width=7.0cm]{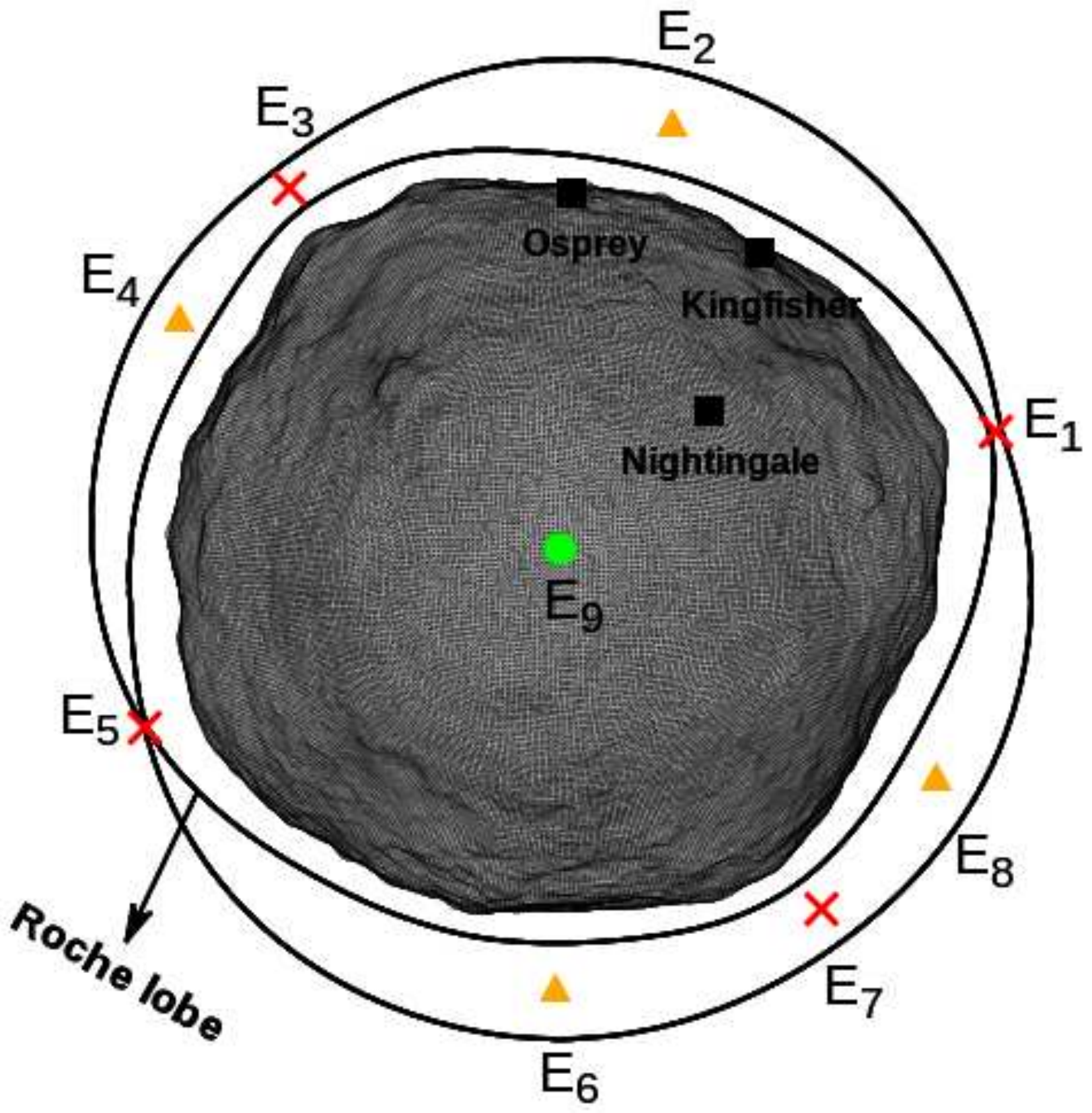}}
  \caption{Location of the nine equilibrium points, stationary points in the Bennu-fixed frame, projected in the equatorial plane, for a constant density and a uniform rotation period. Red X-dots are topologically classified as saddle--center--center points (hyperbolically unstable), orange triangular-dots as sink--source--center points (complexly unstable), and green circle-dot as center--center--center point (linearly stable). The rotational Roche lobe is represented by the outer region of the contour black line that envelops the Bennu shape and the black square boxes are the locations of OSIRIS-REx's candidate sample sites.}
  \label{fig:equi_1}
\end{figure}

\subsection{Equilibria Location}
\label{sec:equi:eqloc}
The locations of our computed equilibrium points for particle sizes that are not significantly affected by SRP, i.e., stationary points in the Bennu-fixed frame, are shown in Table \ref{tab:equi_mas}.
They were computed through the mascons technique (Eq. \ref{eq:sim_3}) with the Minor-Equilibria-NR package \cite{Amarante2020} where the SRP was implemented \cite{minor-equilibria-nr}.
The number of solutions of Eq. (\ref{eq:equi_1}) depends on the shape and spin period of the body.
It is already known that Bennu has eight equilibrium points around it \cite{Scheeres2016}. Bennu also has a single internal equilibrium point that has a slight offset of 1\,m from its centroid. Table \ref{tab:equi_mas} shows that all of Bennu's nine equilibrium points are slightly out-of-plane. In other words, they are not in the equatorial plane because of its asymmetrical shape in the latitudinal direction. Note that only the equilibrium points E$_2$ and E$_9$ are located slightly above the equatorial plane of Bennu.

Fig. \ref{fig:equi_1} shows the arrangement of all Bennu's equilibrium points in its projection plane $xOy$. The black line is the zero-velocity curve corresponding to the equilibrium point E$_5$, which is the point that has the minimum geopotential value ($-242.594$\,cm$^2$/s$^{2}$ from Table \ref{tab:equi_mas}). The eight external equilibrium points of asteroid Bennu are bounded by this curve, this region we call the inner region of this curve. Meanwhile, the outer region of this curve is composed of two regions connected by equilibrium point E$_5$, in which one of them encompasses the Bennu shape. This is the region that delimits Bennu's rotational Roche lobe. Besides, we indicate the locations of the OSIRIS-REx's candidate sample sites \cite{Lauretta2019}.


\begin{figure}
  \centering
  \includegraphics[width=6.68cm]{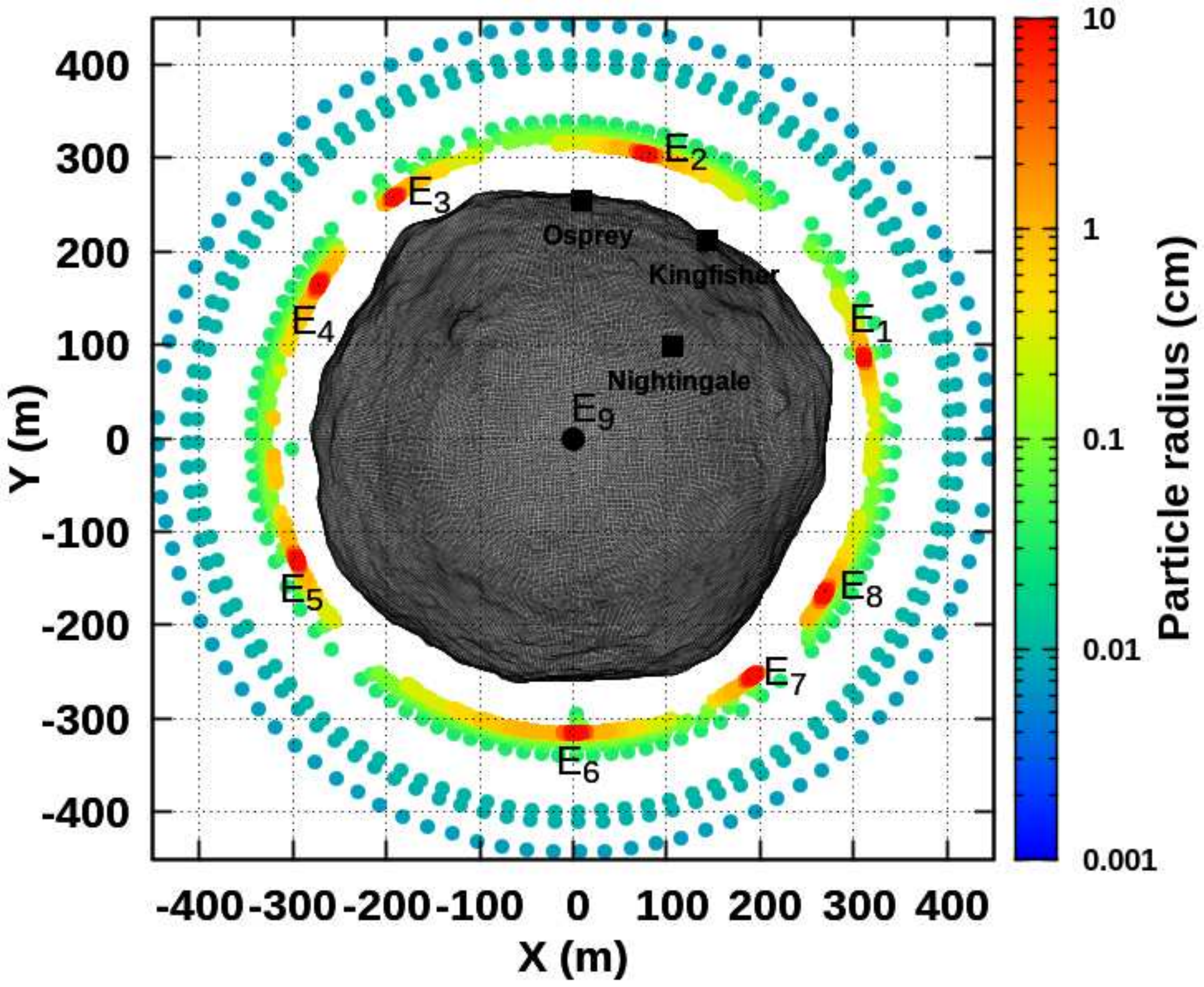}
  \includegraphics[width=6.68cm]{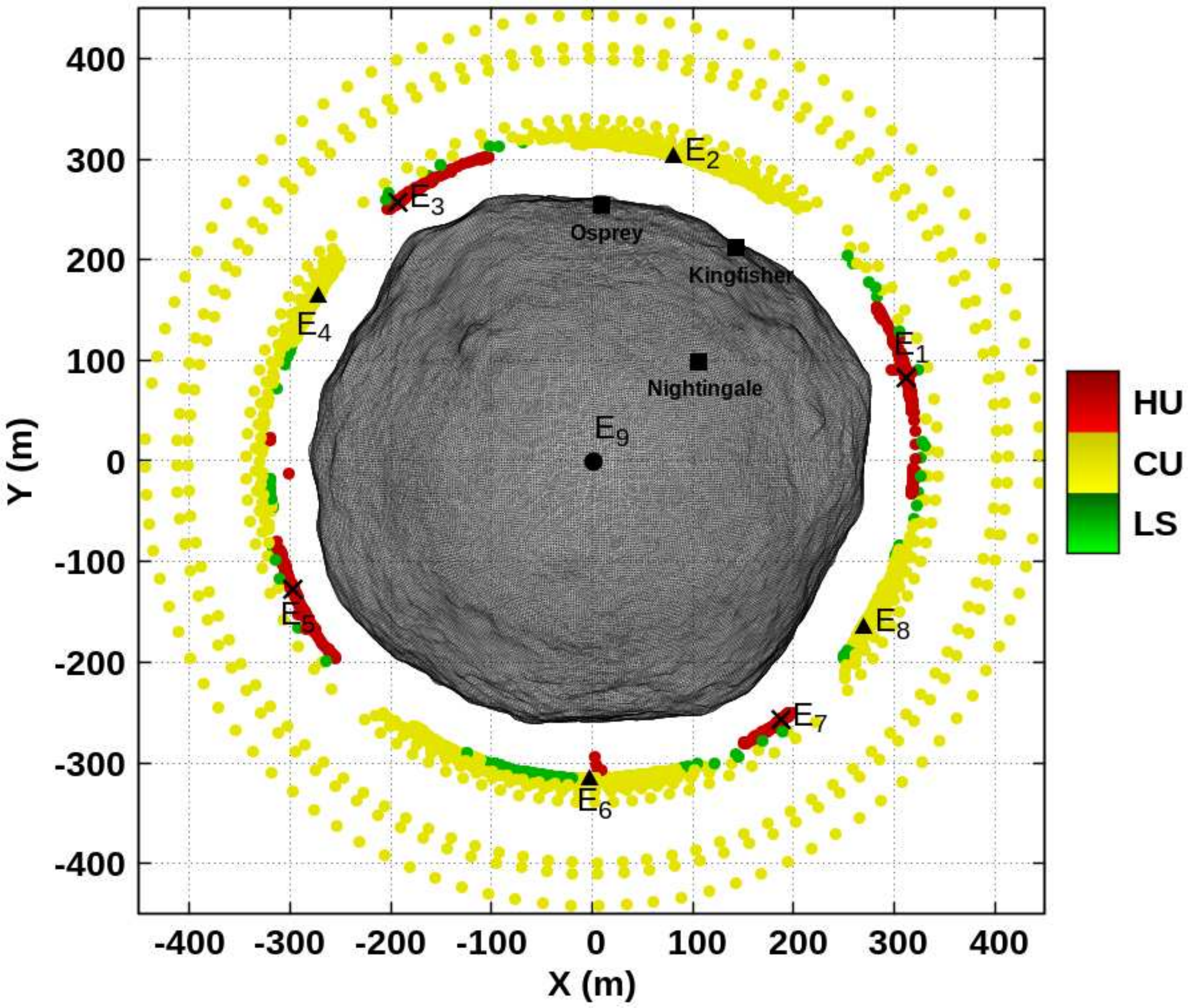}
  \caption{Locations of the instantaneous equilibrium points and their topological structures about Bennu, under radial SRP perturbation and considering the shadowing for particle sizes in a range of 10$^{-3}$--10\,cm. The asteroid's rotation angle increment is taken to be $3.6^\circ$. (left side) Equilibrium point locations around Bennu varying the asteroid's rotation angle and considering that the relative movement Sun-Bennu completes a loop around Bennu. The color box gives the particle sizes. (right side) Topological stabilities of the equilibrium points about Bennu. The red dots are hyperbolically unstable points (HU), the yellow dots are complexly unstable points (CU) and the green ones are linearly stable points (LS).}
  \label{fig:equi_2}
\end{figure}
Considering the case of particle sizes that are affected by SRP, we have chosen a range of 10$^{-3}$--10\,cm for the size of particles. We gradually varied the particle size using the upper and lower values, as shown in Figure \ref{fig:equi_2}. The colored points represent the instantaneous equilibrium points.
The SRP is considered while implementing the shadowing function $W(\theta ,r_\theta)$ for the radial SRP acceleration (Eq. (\ref{eq:math_1})). We search for instantaneous equilibrium points in a bounding box of $-550 \leq x \leq 550$, $-550 \leq y \leq 550$, and $-500 \leq z \leq 500$\,m, centered at Bennu's barycenter. The plot shows the instantaneous equilibrium points around Bennu keeping the Sun mean anomaly fixed at zero and varying the rotation angle of the asteroid every $3.6^\circ$. Initially, with the Sun in a given position around Bennu, we obtain the eight equilibrium points for the large particles. As the particle size is decreased, the equilibrium points move away from Bennu's surface to balance the radiation pressure force that points toward the surface of Bennu on the sunlit side, meaning that the SRP driven acceleration is always inward. In contrast, because the smaller particles in the unlit side are not affected by SRP, then the locations of their equilibrium points are the same as the large particles. Consequently, this situation generates an asymmetry in the locations of the equilibrium points of the particles around Bennu. However, because the relative movement Sun--Bennu changes over time, this equilibria location asymmetry completes a loop around Bennu, making the location of the equilibrium points symmetrical. Figure \ref{fig:equi_2} was made with the composition of all these frames, showing this symmetry.

The left side of Fig. \ref{fig:equi_2} shows the instantaneous equilibrium point location for particle sizes from 10$^{-3}$ up to 10\,cm. The color code gives the size of particles. For smaller particles, the instantaneous equilibrium points move far away from Bennu's surface. Otherwise, if the particle size is increased, then the instantaneous equilibrium points get closer to the asteroid.

That is expected. As particle size decreases, the radial SRP acceleration becomes stronger ($\propto 1/r_p$); and depending on the adopted Sun direction, the balance between the gravitational, centripetal, and SRP accelerations can be reestablished (Eq. (\ref{eq:equi_1})) to configure a dynamical equivalent \cite{Xin2015}.

Fig. \ref{fig:equi_2} shows that for particles that have sizes smaller than $\sim\,2\times10^{-2}$\,cm, the points are farther away from Bennu's surface (blue dots). Meanwhile, for particles with sizes $\sim\,\geq2\times10^{-2}$\, cm, they surround Bennu's surface (yellow dots) and the actual equilibrium points (red dots) (Table \ref{tab:equi_mas}). For particles with a size of nearly 1 cm (orange dots), the SRP acceleration still influences the instantaneous equilibrium point locations. Nevertheless, for particles of a few centimeters or larger (red dots), the particles are no longer significantly affected by the SRP force, reaching the actual equilibrium points.

\subsection{Equilibria Stability}
\label{sec:equi:stab}
We also examined the topological stability of the actual equilibrium points (Fig. \ref{fig:equi_1}) and instantaneous equilibrium points (right side of Fig. \ref{fig:equi_2}) through their linear stability \cite{Jiang2014}. The unnormalized eigenvalues for the actual equilibrium points and their corresponding topological stabilities are given in Table \ref{tab:app_1} (\ref{sec:eigen}). According to the eigenvalues, all outer equilibrium points are unstable. However, they have different topological stability. The odd indices of the equilibrium points have a saddle--center--center topological structure (HU), while the even indices are associated with a sink--source--center stability (CU). In addition, the inner equilibrium point E$_9$ has a center--center--center topological structure that is LS. It is important to mention that the topological stability and existence of an equilibrium point are sensitive to the density and shape of the central body \cite{Scheeres2016}. For the 75-cm shape model (v20) of asteroid Bennu adopted in this work, all outer equilibria are unstable. On the other hand, \citeA{Scheeres2019} using the v14 model \cite{Barnouin2019} of asteroid Bennu found that the equilibrium point E$_6$ is LS, and \citeA{Barnouin2019} show that there are differences between the models v14 and v20. However, \citeA{Scheeres2019} also comment about the sensibility of the stability of the particular external equilibrium point E$_6$. Therefore, we attribute that the difference between the models v14 and v20 is responsible for the difference in the stability of this point. Using the value of the geopotential to list the equilibrium points from lowest to highest, we get E$_9$, E$_5$, E$_1$, E$_3$, E$_4$, E$_7$, E$_8$, E$_6$ as well as E$_2$. Thus, one can conclude that E$_2$ is the most unstable equilibrium point, and E$_5$ is the least unstable external equilibrium point.

The right side of Fig. \ref{fig:equi_2} shows that the topological stabilities of the equilibrium points also change when the radial SRP acceleration is considered for different particle sizes. Particles that have sizes of about $2\times10^{-2}$\,cm are more sensitive to change their linear stability from CU (yellow) to HU (red), or in a couple of cases to LS (green), than particles smaller than $2\times10^{-2}$\,cm. These particles mostly have topological structures as CU (yellow).

\section{Numerical Simulations}
\label{sec:sim}
The geophysical connection and relevance between the rotational Roche lobe (Fig. \ref{fig:equi_1}) and the surface properties of asteroid Bennu was shown by OSIRIS-REx's mission team \cite{Scheeres2019}. Asteroid Bennu was seen to have a distinct transition in its surface slope distribution. This can directly affect the boulder distribution across Bennu's surface. Thus, the understanding of the orbital evolution of particles close to the surface of asteroid Bennu and how the equilibria stabilities affect them are key information to provide clues on the surface evolution of Bennu. Because of the centrifugal force, particles that depart from the surface may enter into a synchronous orbit and eventually settle into the regions within the Roche lobe \cite{Scheeres2019}.

Therefore, we performed three main types of numerical simulations to study the dynamical evolution of fallen particles around the surface of asteroid Bennu and how the topological stabilities of the equilibrium points can influence it. They are categorized as:


(I) torus cloud

(II) local disk

(III) spherical cloud

Types (I) and (II) are made to obtain information about the stability and evolution of samples of particles near the equilibrium points and close to the equatorial region of Bennu. They are presented and discussed in subsections \ref{sec:to} and \ref{sec:lod}. In Section \ref{sec:fall}, we use integrations of type (III) to study the density of particles that fall across the entire surface of Bennu. All integrations have an evolution time of about $1.14$\,yr (10,000\,h). Most of the particles in our integrations collide or escape from the system in less than 1000\,h, so this integration time is sufficient to evolve all samples of particles in our three types of simulations. We developed the \textit{Minor-Mercury package} \cite{minor-mercury}: a modified version of the original Mercury package \cite{Chambers1999} to handle an irregular-shaped minor body. The integration model details with our collisional and escape criteria are presented in \ref{sec:minor-mercury}. We give details about each numerical experiment and how they are simulated in the following sections.

\subsection{The Torus Cloud type}
\label{sec:to}
The initial conditions of type (I) integrations are built through a uniform torus cloud of particles distributed involving the eight external equilibrium points of asteroid Bennu. The torus cloud width lies between 280\,m and 360\,m from Bennu's center of mass, and the particles' orbits are inclined up to $0.45^\circ$. The lower limit of 280\,m is chosen to fill Bennu's Roche lobe of particles and the upper one (360\,m) is to take into account the dominant region by the equilibrium points. The inclination values are chosen to be near Bennu's equatorial area. The torus cloud was filled with 10,000 particles uniformly distributed. All particles were initially placed in a circular Keplerian orbit around Bennu \cite{hnm-ring}. The longitude of the ascending node and the mean anomaly angles were taken randomly from $0^\circ$ up to $360^\circ$. An animated movie showing the time evolution of the Keplerian initial conditions of the remaining particles is available online at Supporting Information file (Movie S1: Torus Kep IC). The SRP was also considered in this type of numerical experiment. We adopted particle sizes from 10$^{-4}$ up to 10\,cm distributed into five intervals: $10^{-4}$--$10^{-3}$, $10^{-3}$--$10^{-2}$, $10^{-2}$--$10^{-1}$, $10^{-1}$--1, and 1--10\,cm. Each interval was filled uniformly with 10,000 particles.

\subsection*{\textbf{Bigger Particles}}
\label{sec:to:big}
\begin{figure}
  \centering
  \includegraphics[width=6.68cm]{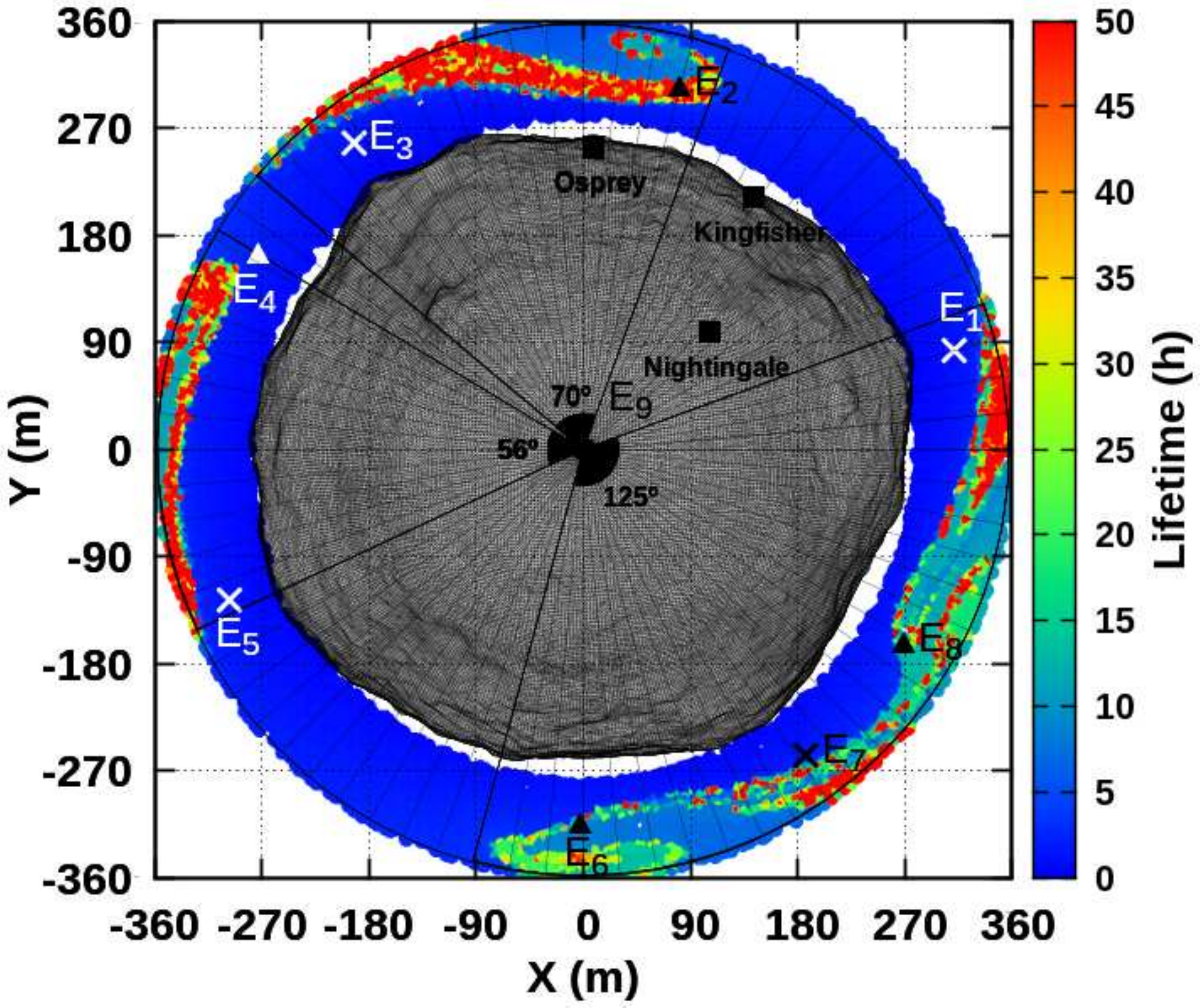}
  \includegraphics[width=6.68cm]{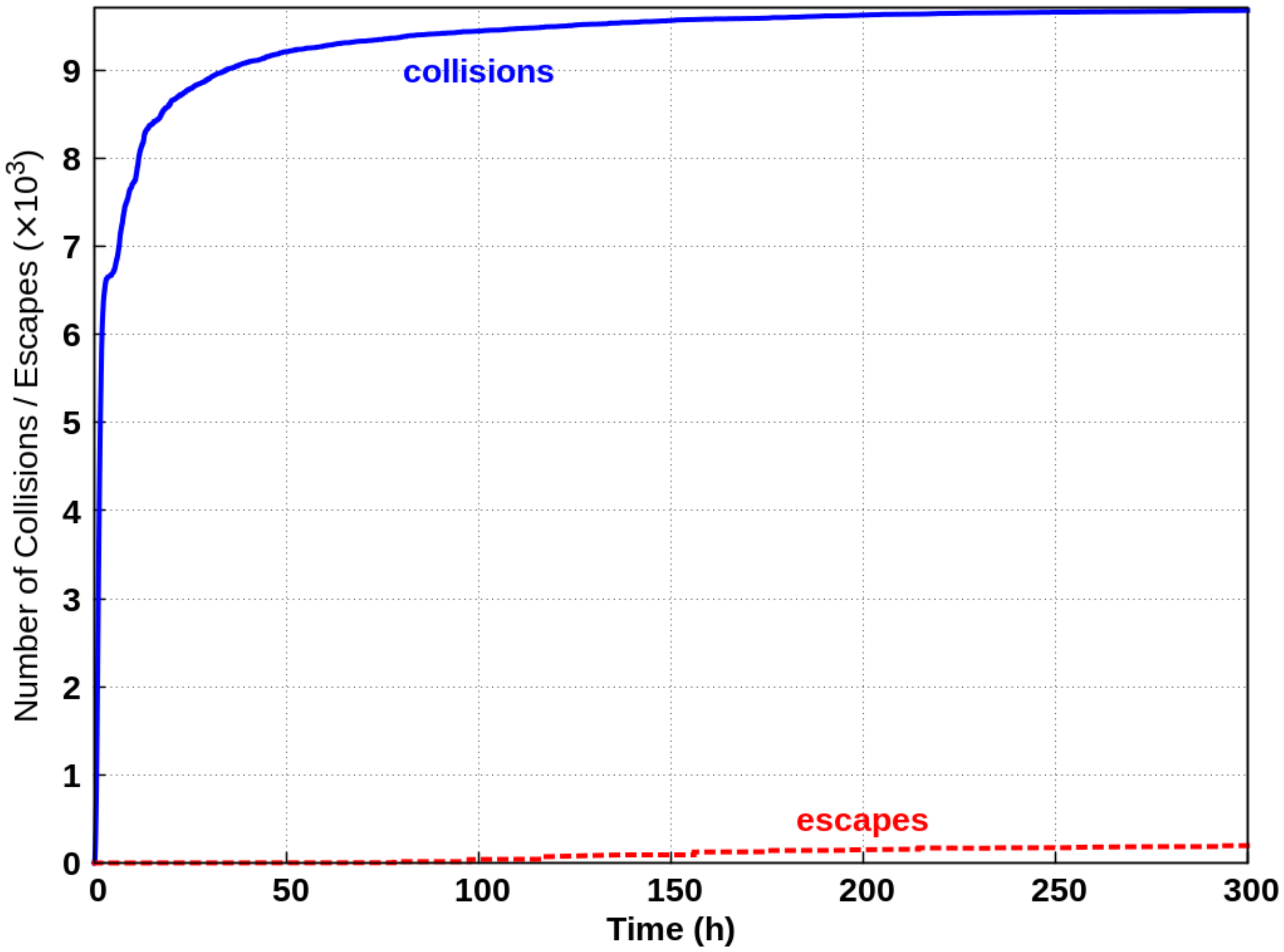}
  \includegraphics[width=6.68cm]{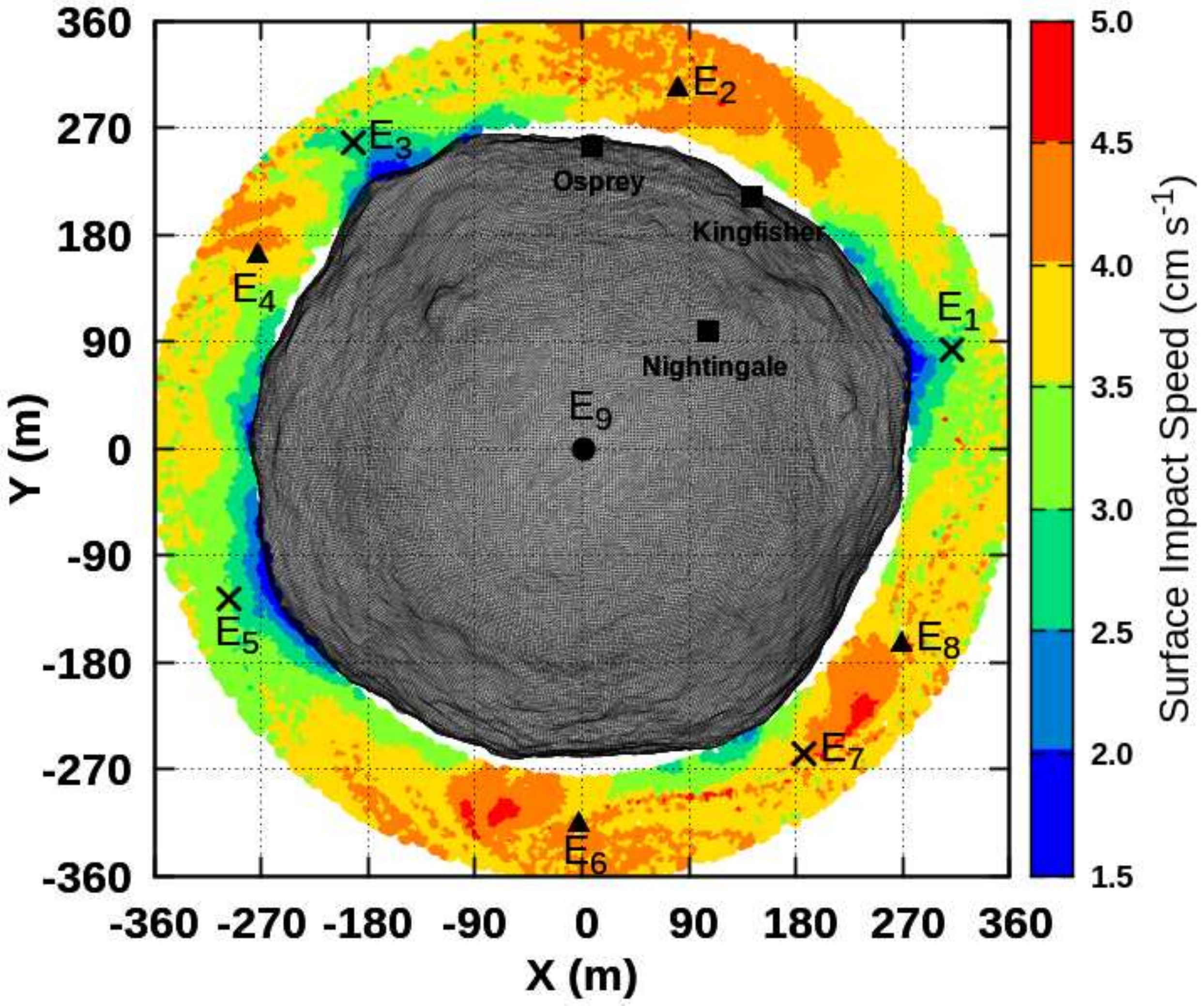}
  \includegraphics[width=6.68cm]{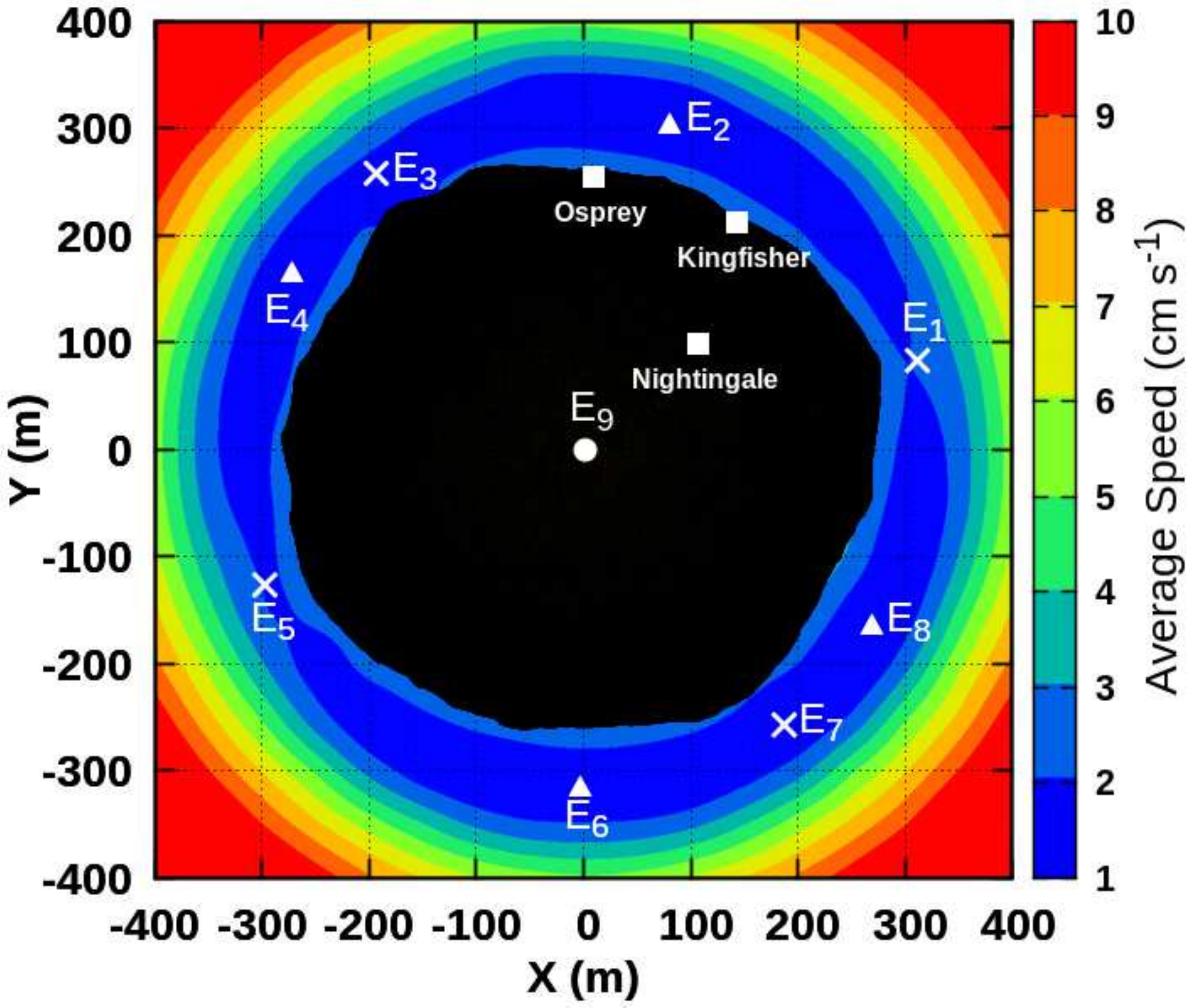}
  \caption{Results from initial conditions of the type (I) integration for big particles (larger than a few centimeters). (top-left side) Lifetime map of the initial conditions. The color box gives the particle's lifetime (h). (top-right side) Number of collisions (blue line) / escapes (dashed red line) over the time. (bottom-left side) Surface impact speed, in the Bennu-fixed frame. The color box represents the surface impact speed, in cm/s. (bottom-right side) Particles' average speed (cm/s), in the Bennu-fixed frame, over the entire integration.}
  \label{fig:to_1}
\end{figure}
On the top-left side of Figure \ref{fig:to_1}, we show the lifetime map of a sample of the Keplerian initial conditions for the simulation type (I). These are the initial conditions for particles that survived for the given amount of time. The color code gives the particles' lifetimes. We consider bigger particle sizes that are not significantly affected by SRP (larger than a few centimeters). The size of the torus cloud decreases over time (see animated Movie S1: Torus Kep IC Supporting Information). The dark-blue region shows that most of the particles are quickly removed from the system in less than tens of hours ($\leq 50$\,h). However, there are three distinguishable longitudinal regions around external equilibrium points, where the particles survive for longer times ($> 50$\,h). The first region is the smaller one, and it is located around equilibrium points E$_4$--E$_5$ and has an angular size of $\sim 56^\circ$. The second region is near the equilibrium points E$_2$--E$_3$ with an angular aperture of $\sim 70^\circ$. The third area is the bigger one and connects equilibrium points E$_6$--E$_7$--E$_8$--E$_1$ with an angular range of $\sim 125^\circ$. The bigger particles in the longitudinal equatorial area near the Kingfisher sample collection site (E$_1$--E$_2$) are quickly removed. This same behavior can be observed in the diametrically opposite longitudinal region between equilibrium points E$_5$--E$_6$. Otherwise, in the longitudinal equatorial region close to the Osprey sample collection site, the particles remain for long times before colliding with the surface of asteroid Bennu. The results show longitudinal symmetric stability and a tadpole stability region around equilibrium point E$_2$. After $\sim\,55$\,h, the size of the tadpole stability region around the equilibrium point E$_2$ had a significant decrease (see animated Movie S1: Torus Kep IC Supporting Information file).
The top-right side of Fig. \ref{fig:to_1} shows the time evolution of the number of collisions/escapes for the Keplerian torus cloud. Most of the bigger particles are removed from the system because of collisions with the surface of the asteroid. They cover 97.1\% of the removed particles. Half of the torus cloud of particles collided with Bennu's surface in less than $\sim 1.54$\,h. Only 2.9\% of the removed particles escape from the system, and they start to escape only after $58.5$\,h.

The bottom-left side of Fig. \ref{fig:to_1} shows the surface impact speed (cm/s), in the Bennu-fixed frame, for the Keplerian torus cloud. The particles fall back to the surface of asteroid Bennu with impact speeds $\leq5$\,cm/s. This suggests that particles with speeds $>5$\,cm/s can escape from the Bennu neighborhood, agreeing with the results of the episodes of particle ejection from the surface of Bennu that were found in \citeA{Lauretta2019b} and simulated by \citeA{McMahon2020}. Most particles have surface impact speeds from 3 to 4.5\,cm/s (from green to brown). The Kingfisher and Osprey sample sites are reached by particles with speeds in a range of 4--4.5\,cm/s. In addition, the bottom-right side of Fig. \ref{fig:to_1} plots the particles' average speed in the Bennu-fixed frame over the entire integration. The average speed of the bigger particles in the proximity of Bennu's surface is less than 10\,cm/s. Besides, most particles have average speeds from 2 to 3\,cm/s before colliding with the surface of asteroid Bennu.

\subsection*{\textbf{Particle Flux}}
\label{sec:to:vol}
\begin{figure}
  \centering
  \includegraphics[width=6.68cm]{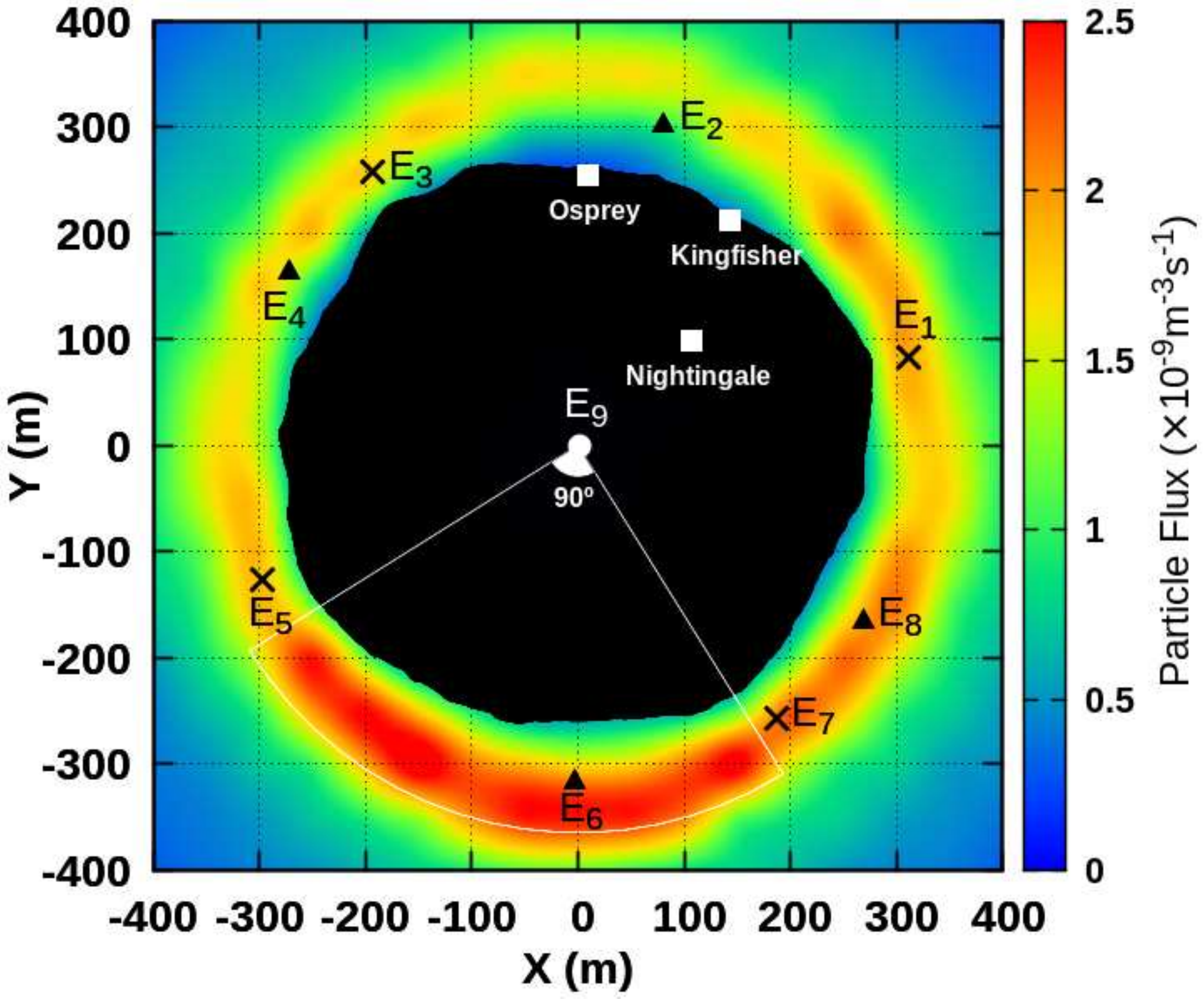}
  \includegraphics[width=6.68cm]{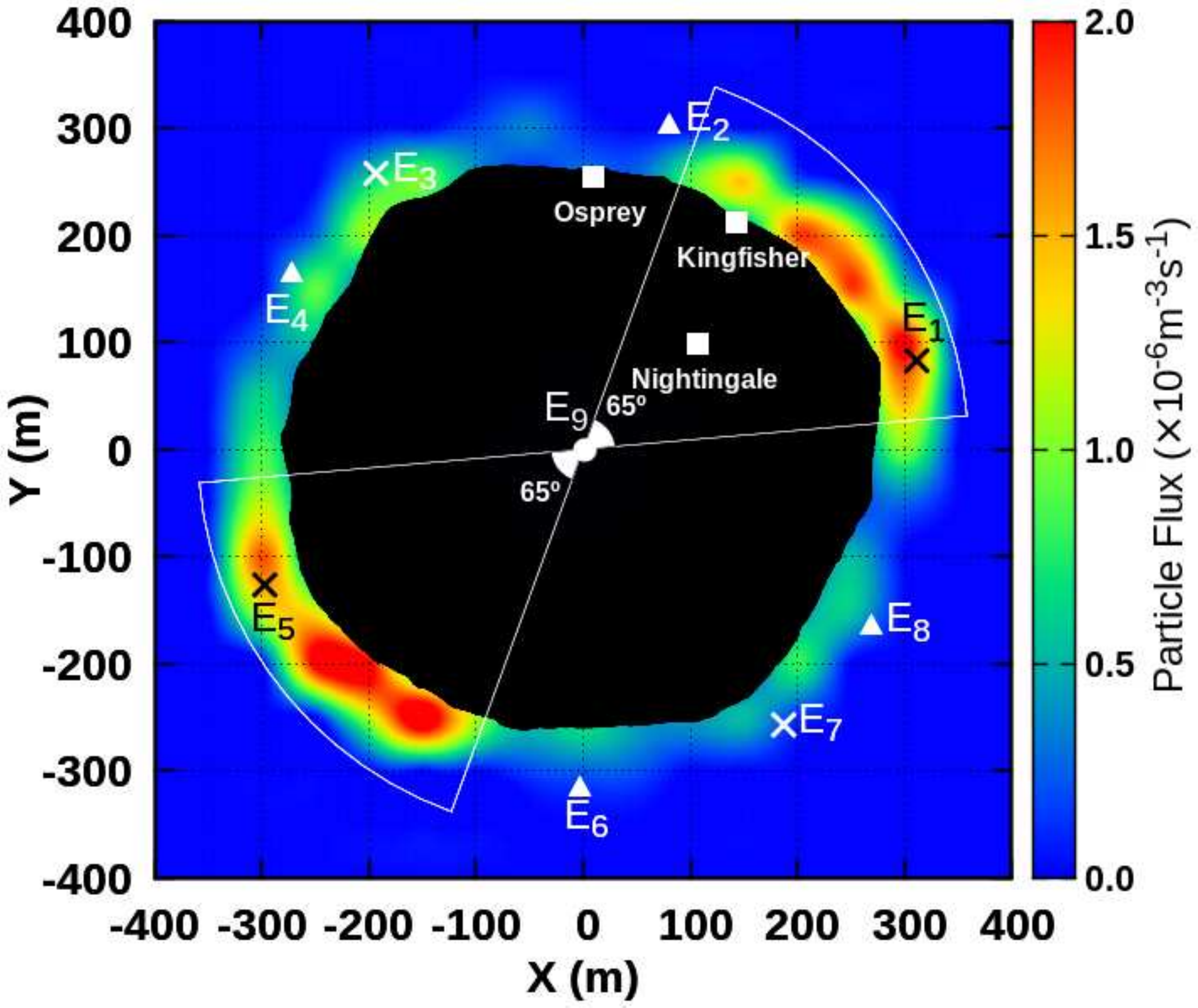}
  \caption{(left side) Mapping of the particle flux around Bennu averaged over the entire integration of type (I). The color code denotes the particle flux, in m$^{-3}$\,s$^{-1}$. (right side) Mapping of the particle flux around Bennu averaged over the last five time steps (30 min) before the particles impact on Bennu's surface.}
  \label{fig:to_2}
\end{figure}

The particle flux is defined as the rate of the number of particles through a unit volume averaged over a period of time. The mapping of the particle flux around Bennu for simulation type (I) is also explored in Figure \ref{fig:to_2}. The color boxes denote the particle flux. The difference between the left and right sides of Fig. \ref{fig:to_2} is that on the left side, we show the particle flux over the entire integration. This can be thought of as the preferred region of the remaining particles during the simulation. It shows that this preferred region (See animated Movie S2: Torus Kep available online in the Supporting Information) is around equilibrium point E$_6$ and has a longitudinal range of $\sim 90^\circ$. Meanwhile, the right side of Fig. \ref{fig:to_2} indicates the preferred site locations just before the particles fall on the surface of Bennu. The figure shows that most of the particles fall in the longitudinal regions that connect equilibrium points E$_1$--E$_2$ and E$_5$--E$_6$, with higher particle flux values around equilibrium points E$_1$ and E$_5$. The figure also shows that these preferred site locations have an approximated angular range of $65^\circ$. This is in accordance with the particle's lifetime map of Fig. \ref{fig:to_1}. This feature suggests that most of the bigger particles that are initially placed in a Keplerian orbit inside the Roche lobe involving the equilibrium points will fall back across its surface near the Kingfisher sample region or near the region diametrically opposite to it. In contrast, almost none of them will fall back near the Osprey area.
\begin{figure}
  \centering
  \includegraphics[width=8.0cm]{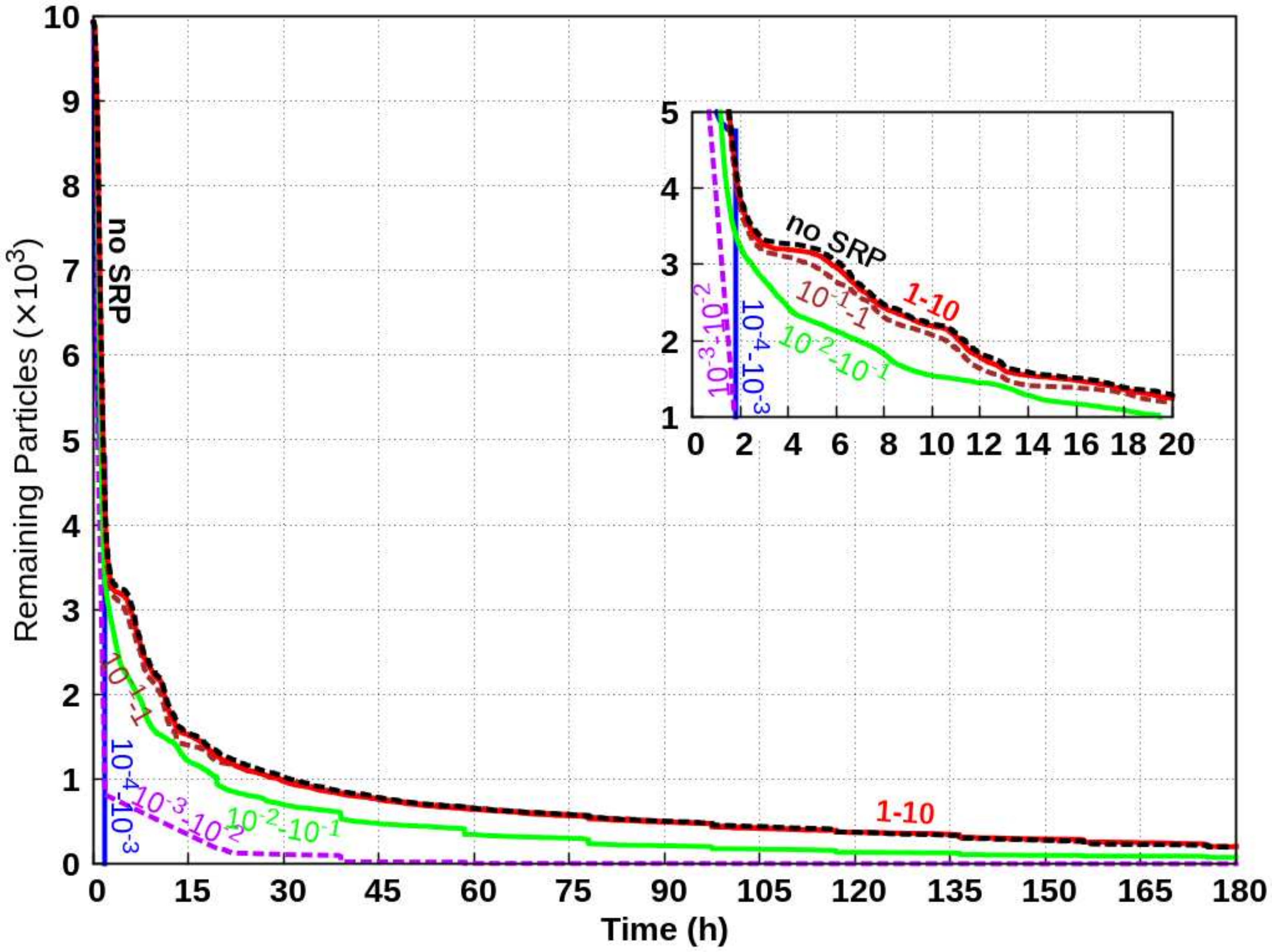}
  \caption{Number of remaining particles ($\times 10^3$) over time for all particle sizes of the integrations of type (I). The color of the line indicates the size range of particles:
  Blue, 10$^{-4}$--10$^{-3}$\,cm;
  Dark-magenta, 10$^{-3}$--10$^{-2}$\,cm;
  Green, 10$^{-2}$--10$^{-1}$\,cm;
  Brown, 10$^{-1}$--1\,cm;
  Red, 1--10\,cm.
  The black line shows the case without considering the SRP.}
  \label{fig:to_3}
\end{figure}

An animated Movie S2: Torus Kep is available online in the Supporting Information. It shows the particle flux behavior over time, for integrations of type (I). Note in this animation that most of the particles fall in the region near the Kingfisher sample site and in the region diametrically opposite to it. Note also that most of the particles librate around the equilibrium point E$_6$ in accordance with the particle flux of Fig. \ref{fig:to_2}.

\subsection*{\textbf{Smaller particles}}
\label{sec:to:sml}
The simulations of the torus cloud type (I) consider 10,000 particles for each of the five intervals of particle sizes and for a case without considering the SRP. Along with the simulations, the particles were removed when they collided with the surface of Bennu or were ejected from the system.
Figure \ref{fig:to_3} shows the number of the remaining particles over the integration time for all particle sizes adopted in the simulations of the torus cloud, type (I). The results show that more than half of the particles, independently of the size, are removed from the system in just a few hours.
More than $90\%$ of all particles with sizes $<10^{-2}$\,cm are removed very quickly. Most of them escaped from Bennu's neighborhood because of the strong effect of the SRP.
In contrast, the temporal evolution of the remaining biggest particles, 1--10 cm, is very close to the case of the simulation without considering the SRP. These results for the torus simulation type (I) are in agreement with those for the equilibrium point locations from 10$^{-3}$-cm up to 10-cm particle sizes, suggesting that particles larger than a few centimeters are not significantly affected by the SRP.


It is also interesting to note that if we assume a constant particle density of $\rho$ = 2 g/cm$^3$, which is consistent with meteorite analogs \cite{Hamilton2019, Lauretta2019b}, keeping the same area-to-mass ratio $\frac{A}{m}$ as Eq. \ref{eq:math_4}, then, our adopted particle size limits would change by a factor of $\sim3/5$. In this case, the particle sizes range goes from $\sim 0.6\times 10^{-4}$ up to $\sim 6$\,cm, and smaller particles with radius $<0.6$\,cm are preferentially removed from Bennu's neighborhood. \citeA{McMahon2020} using a different approach for simulations of ejected particles observed by the OSIRIS-REx space probe found that small particles $<1$\,cm radius are preferentially removed from the system.

An animated Movie S3: Torus Kep SRP is available online in the Supporting Information. It shows the time evolution of the torus cloud for particle sizes from 10$^{-4}$ up to 10\,cm for simulations of type (I).

\subsection{The Local Disk type}
\label{sec:lod}
In this section, we study the stability and evolution in the neighborhoods of the outer equilibrium points of asteroid Bennu. We considered the local disk samples for simulation type (II). Our goal is to analyze how the evolution of the particles with bigger sizes (neglecting the SRP) can be affected by the equilibria stabilities. In the local disk type (II), we randomly distributed eight samples of particles around the outer equilibrium points, one around each point. Each sample was evenly filled by 1,000 particles. To find the upper and lower radii limits of the samples, we computed the minimum distance from Bennu's surface to each equilibrium point. Then, we built a local circular and planar disk with a radius equal to this distance. The center of each local disk is located at its corresponding equilibrium point. All particles were initially distributed at the inertial frame, and they also had an initial orbital velocity value proportional to the orbital velocity of its corresponding equilibrium point. The direction of the particle velocity vector is perpendicular to the particle radius vector. The orbital velocity of each equilibrium point is presented in Table \ref{tab:equi_mas} and it was computed using:
\begin{eqnarray}
v_{eq} & = & \omega r_{eq},
\label{eq:lod_1}
\end{eqnarray}
\noindent where $r_{eq}$ is the equilibrium point distance from Bennu's center of mass (Table \ref{tab:equi_mas}).
Thus, the proportional initial orbital velocity for each particle was computed adopting:
\begin{eqnarray}
v_{lo} & = & v_{eq}\sqrt{\frac{r_{eq}}{r}},
\label{eq:lod_2}
\end{eqnarray}
\noindent where $r$ is the particle radius distance from Bennu's center of mass.
\begin{figure}
  \centering
  \includegraphics[width=6.68cm]{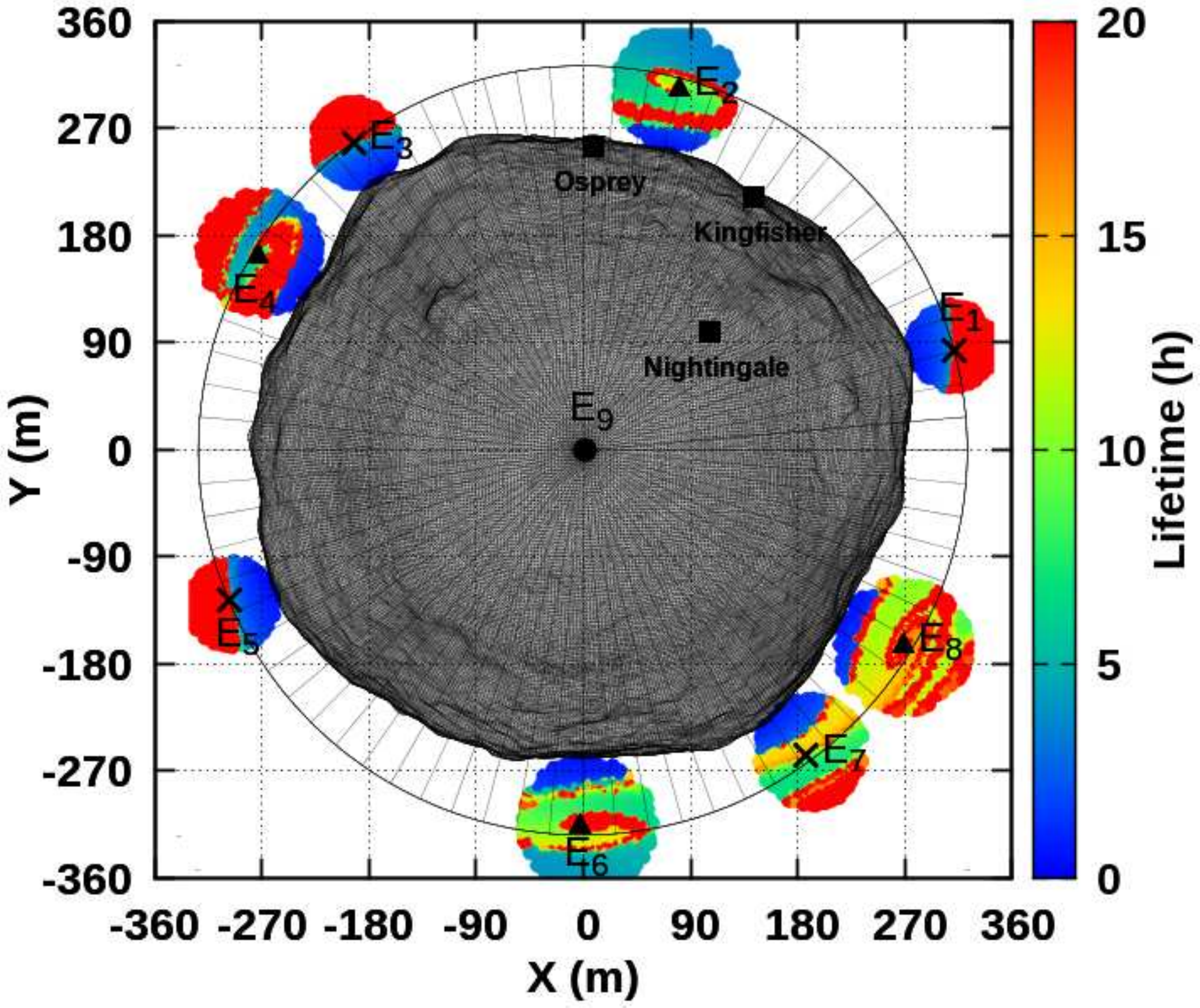}
  \includegraphics[width=6.68cm]{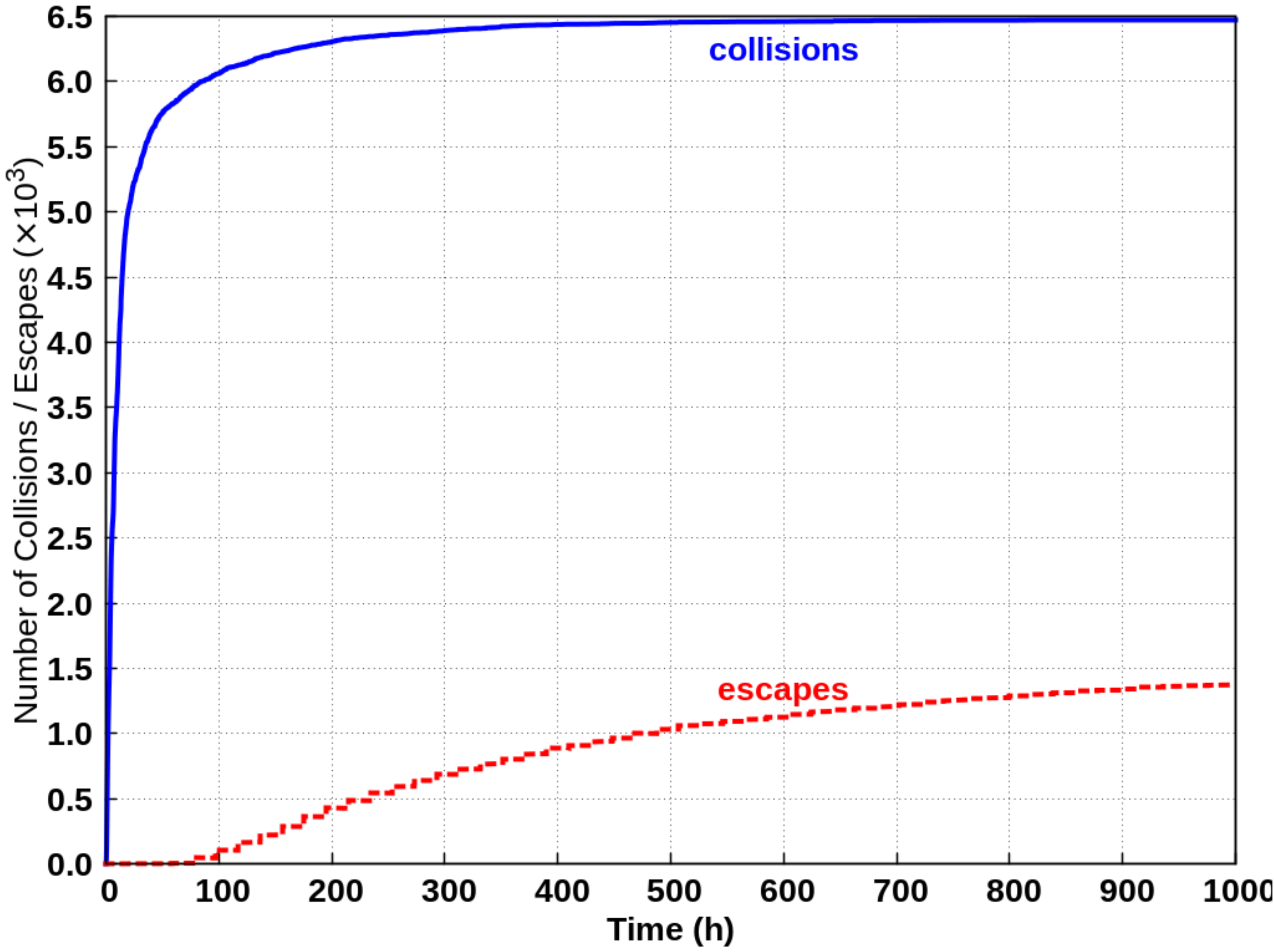}
  \includegraphics[width=6.68cm]{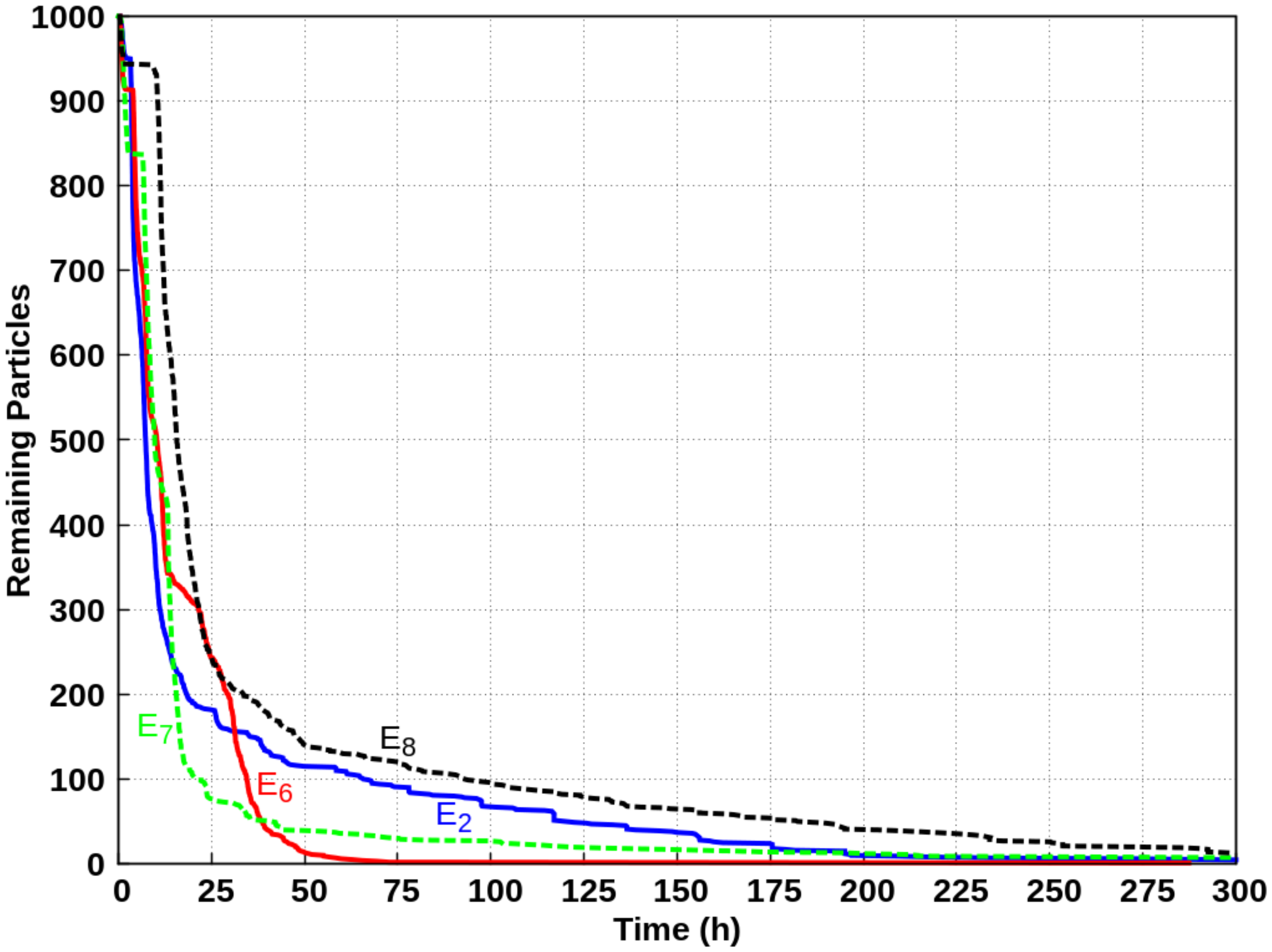}
  \includegraphics[width=6.68cm]{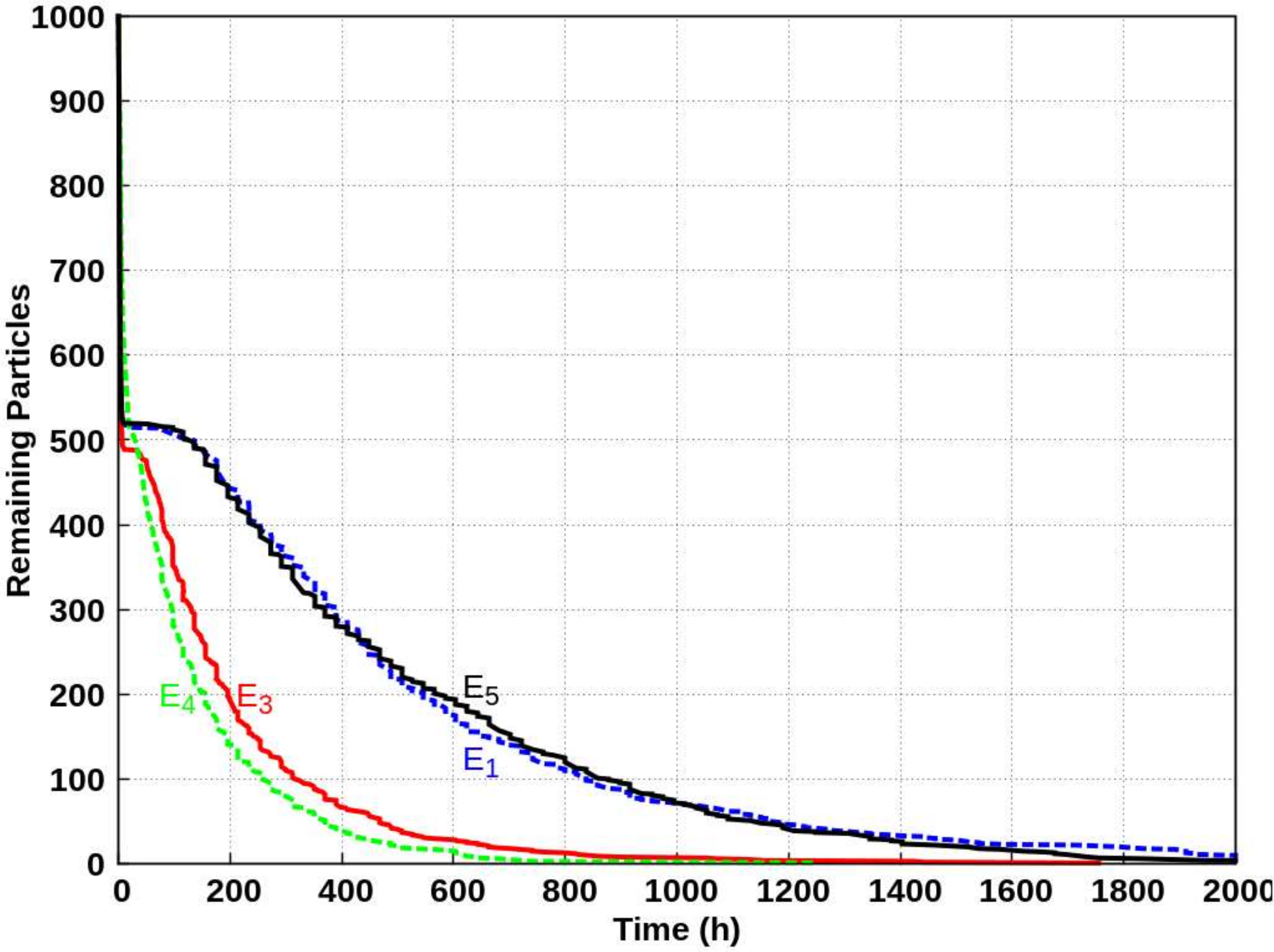}
  \caption{(top-left side) Lifetime map of the initial conditions for the integrations of the type (II). The particles are initially distributed uniformly in a filled local disk that considers the eight external equilibrium points. The color box gives the particle's lifetime (h). (top-right side) Total number of collisions (blue line)/escapes (dashed red line) over the time for the integrations type (II). (bottom-left side) Number of the remaining particles over time for the local disks around equilibrium points E$_2$ (blue), E$_6$ (red), E$_7$ (dashed green), and E$_8$ (dashed black). (bottom-right side) Number of remaining particles over time for the local disks around equilibrium points E$_1$ (dashed blue), E$_3$ (red), E$_4$ (dashed green), and E$_5$ (black).}
  \label{fig:lod_1}
\end{figure}

The top-left side of Figure \ref{fig:lod_1} shows the lifetime map of the initial condition samples around the eight outer equilibrium points for simulation type (II). Again, these are the initial conditions of particles that survived for a given amount of time. The color box denotes the lifetime of the particles. The SRP was not considered in this type of integration. The black circular line represents approximately the 1:1 resonance radius. Most of the particles that are initially situated close to the surface of Bennu are quickly removed in just a couple of hours (See animated Movie S4: Local Disk IC available online in the Supporting Information). The dark-blue dots of the local disks E$_1$, E$_3$, and E$_5$ initially situated inside the 1:1 resonance radius are particles more susceptible to be removed from Bennu's neighborhood than those in the other local disks. Meanwhile, for these three local disks plus the one around E$_4$, the particles located initially outside the 1:1 resonance radius (red dots) survive for longer times ($>20$\,h) than those from the other local disks.

The top-right side of Fig. \ref{fig:lod_1} shows the time evolution of the number of collisions/escapes for the eight local disk samples from simulation type (II). Most of the particles were removed because of collisions with Bennu's surface. They cover 80.9\% of the removed particles, while the remaining 19.1\% escaped from the system. Comparing with the type torus cloud integrations, there are more escapes in simulation type (II). In addition, they remain for a longer time. For example, after 300\,h, the collisions were almost done. However, the local disks need a considerable number of escapes to occur.

The lifetime of the particles from different local disks can be divided into two groups. The bottom-left side of Fig. \ref{fig:lod_1} gives the number of remaining particles over time for the local disks E$_2$ (blue), E$_6$ (red), E$_7$ (dashed green), and E$_8$ (dashed black). It shows that more than half of the local disks collided with the surface of asteroid Bennu in a few tens of hours ($<20$\,h). In most of the initial hours, the number of the remaining particles that depart from local disk E$_2$ decreases the fastest until 14.8\,h, remaining with 23.6\% of the particles. After this time, local disk E$_7$ starts to shrink more quickly than local disk E$_2$ up to 39.3\,h, remaining with only 5.1\% of the particles. Finally, after 39.3\,h the remaining particles from local disk E$_6$ are removed more quickly from the system. Besides, particles that depart from local disk E$_8$ are the slowest to be removed from the proximity of Bennu, except between 22.1 and 27.7\,h ($\sim$ 29.6--22.0\% of the particles), when particles from local disk E$_6$ are also being removed more slowly from Bennu's vicinity. This group of local disks had their particles removed from the system faster ($<300$\,h) than the group of local disks E$_1$, E$_3$, E$_4$, and E$_5$ (bottom-right side of Fig. \ref{fig:lod_1}). For these local disks, few particles remain in the system for long-time integrations. For example, 0.8\% of the particles initially situated in the local disks E$_3$/E$_4$ stay in the simulation for more than 1,000\,h, while 1.2\% of the particles initially located in the local disks E$_1$/E$_5$ can survive up to 2,000\,h. The bottom-right side of Fig. \ref{fig:lod_1} shows that almost half of the particles that depart from local disks E$_1$, E$_3$, E$_4$, and E$_5$ are pruned from the system in a few tens of hours ($<29.6$\,h).

An animated movie (See animated Movie S5: Local Disk available online in the Supporting Information) is available online. It shows the time evolution of the local disks for simulations of type (II). Note in this animation that the particles have a preference to reach the Kingfisher and Osprey sample sites and the region diametrically opposite to them.

\subsection*{The Local type}
\label{sec:lo}
\begin{figure}
  \centering
  \includegraphics[width=6.68cm]{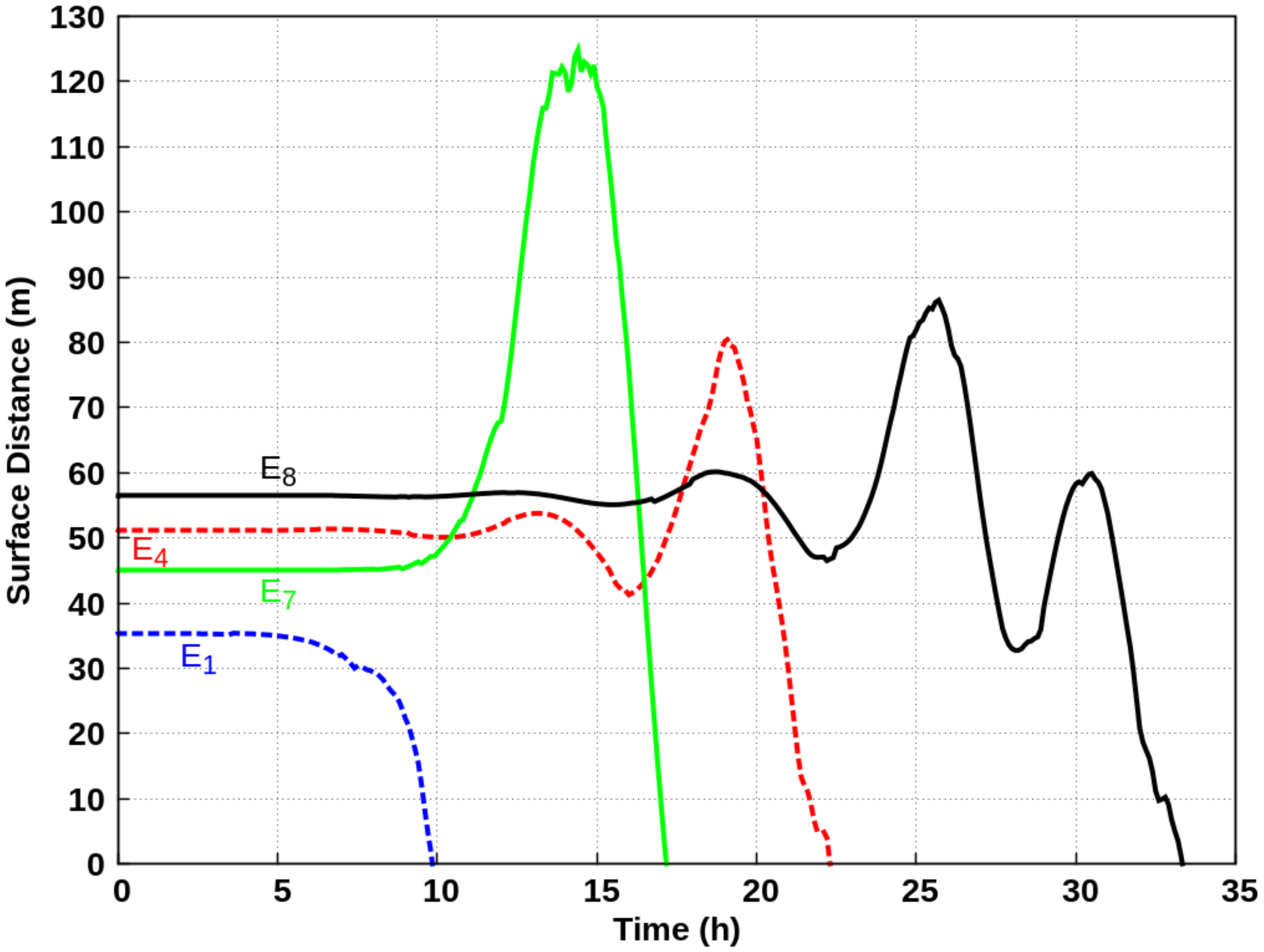}
  \includegraphics[width=6.68cm]{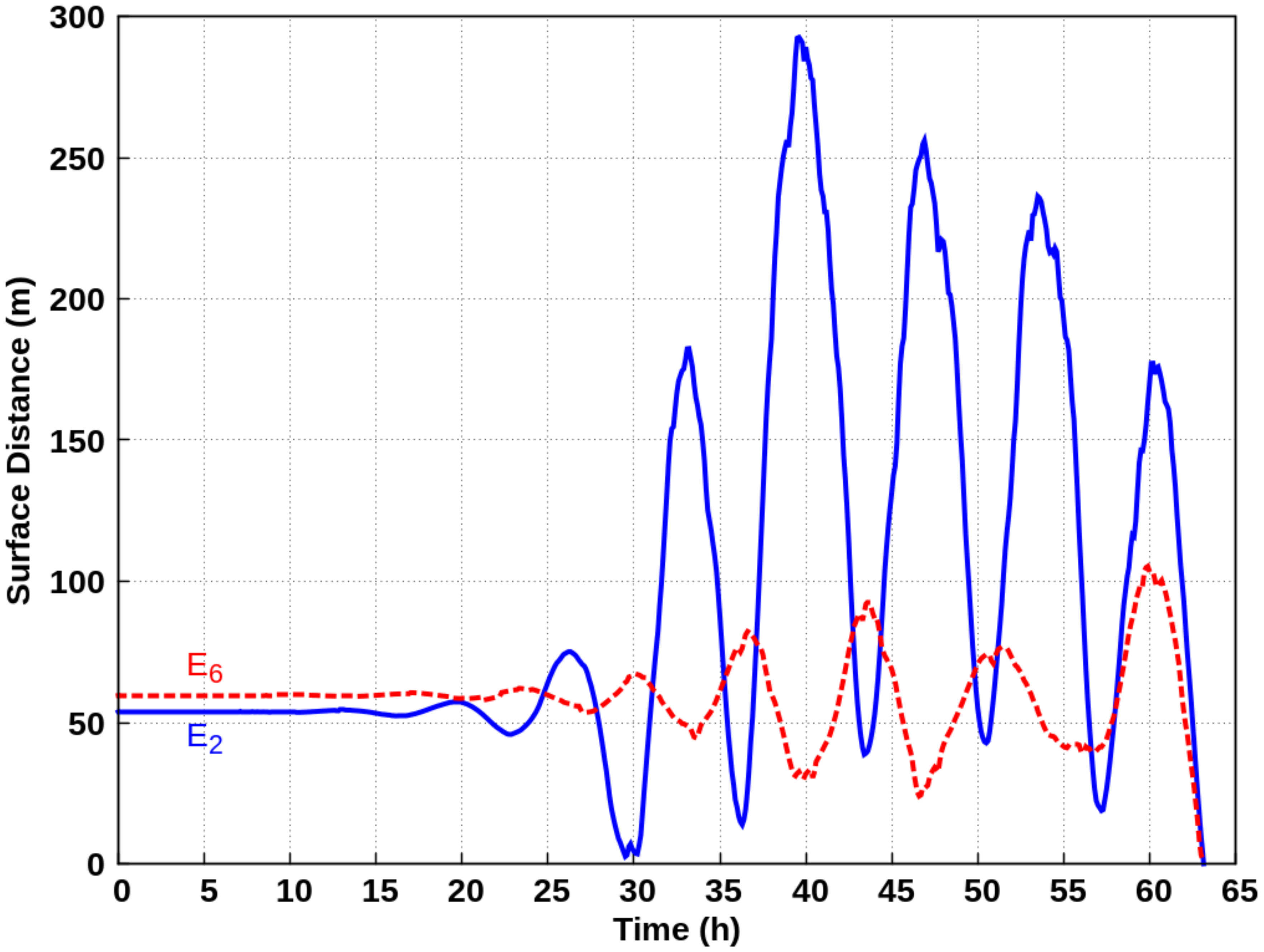}
  \includegraphics[width=7.0cm]{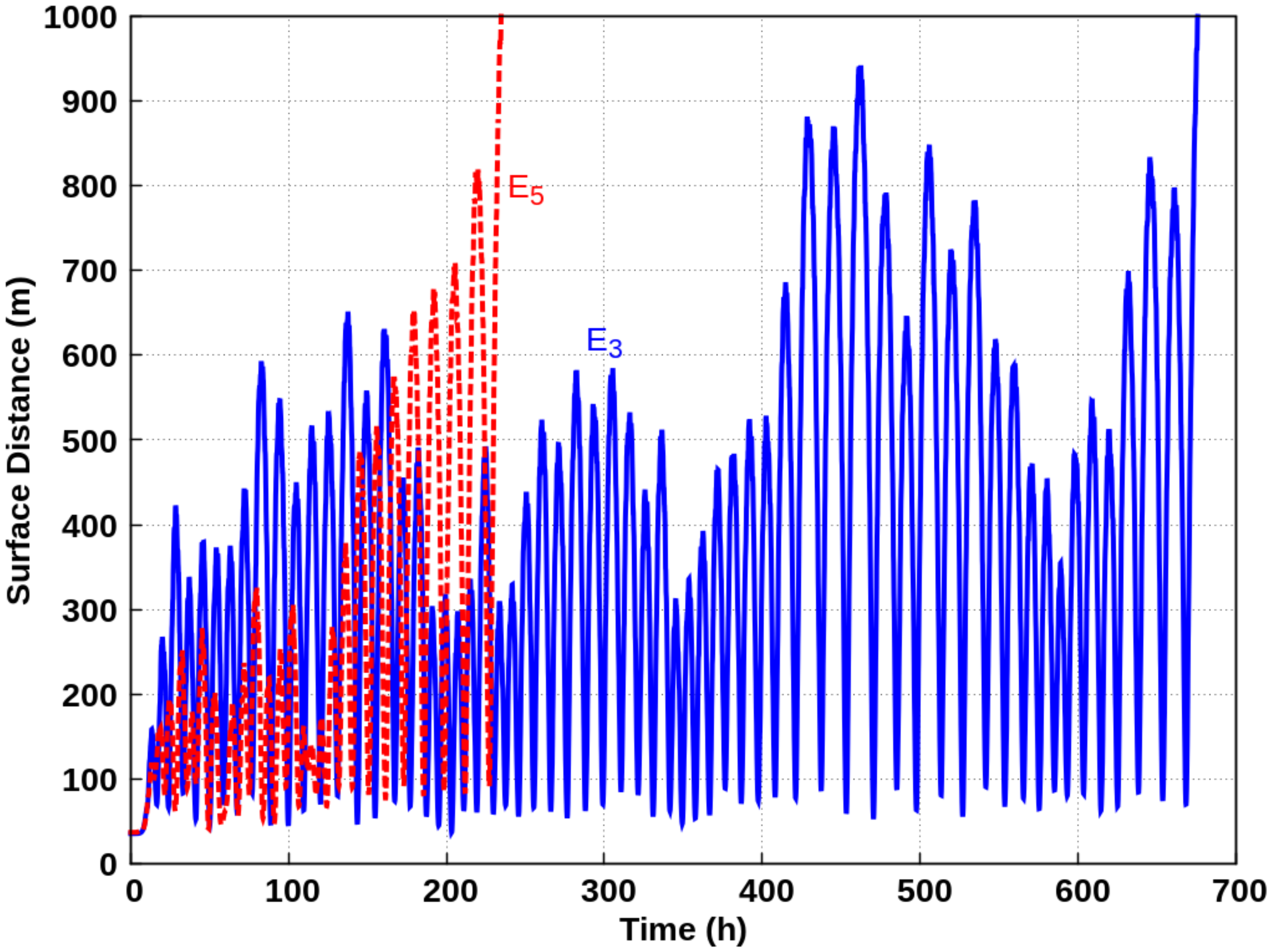}
  \caption{Surface distance (m) over time (h) for particles initially placed in the location of each external equilibrium point (local types) and computed through the mascons approach. The particles are initially stationary in the Bennu-fixed frame. (top-left side) Surface distance over time for the local types of points E$_1$ (dashed blue), E$_4$ (dashed red), E$_7$ (green), and E$_8$ (black). (top-right side) Surface distance over time for the local types of points E$_2$ (blue) and E$_6$ (dashed red). (bottom) Surface distance over time for the local types of points E$_3$ (blue) and E$_5$ (dashed red).}
  \label{fig:lo_1}
\end{figure}
In addition, we analyzed the stability of the eight outer equilibrium points around Bennu, placing a single particle in the location of each computed equilibrium point (Table \ref{tab:equi_mas}). All particles are initially stationary in the Bennu-fixed frame. Then, we propagate each orbit while computing the time that particles collide with Bennu's surface or escape from its proximity. The times when the particles leave their equilibrium point neighborhoods are also computed. We defined this time as the equilibrium lifetime (Table \ref{tab:equi_mas}). To make such measurements, we analyzed the surface distance of the particle. It was computed by finding the difference of the particle distance from the center of Bennu and the point of intersection of the particle's vector radius with the surface of Bennu at each timestep.

The results are presented in Fig. \ref{fig:lo_1}, and an animated Movie S6: Local is available online in the Supporting Information.
The top-left side of Fig. \ref{fig:lo_1} shows that a particle located at E$_1$ (dashed blue line) is the first to collide with the surface of Bennu, at $\sim 10$\,h. This particle starts to leave the vicinity of E$_1$ at about 5\,h (equilibrium lifetime). After 17\,h, the particle from the equilibrium point E$_7$ (green line) impacts the surface of Bennu at a local site close to equilibrium point E$_5$. Note that this is the region of the highest particle flux around Bennu the last five time steps before the particles impact on Bennu's surface, as shown in the right side of Fig. \ref{fig:to_2}. This particle has an equilibrium lifetime of approximately 8\,h. After 22\,h, a particle that departs from equilibrium point E$_4$ (dashed red line) collides with the surface of asteroid Bennu in a local site close to equilibrium point E$_4$. This particle starts to move away from its equilibrium point in $\sim 10$\,h. Finally, a particle from equilibrium point E$_8$ (black line) impacts Bennu's surface after 33\,h, and it has an equilibrium lifetime of $\sim 10$\,h. The previous equilibrium lifetime analyses are made for the set of particles that have the shortest lifetime and that intersect with the Bennu surface very quickly (less than 35\,h). Note that there are two pairs of equilibrium points, E$_1$/E$_7$ and E$_4$/E$_8$, that have topological linear stabilities that are HU and CU, respectively (Fig. \ref{fig:equi_1}).

The top-left side of Fig. \ref{fig:lo_1} shows that the particles from equilibrium points E$_2$ (blue line) and E$_6$ (dashed red line) impact the surface of the asteroid almost at the same time, after 63\,h. These particles also have equilibrium lifetimes close to each other, $\sim 15$\,h for the particles that depart from equilibrium point E$_2$ and $\sim 17$\,h for the particles initially located at equilibrium point E$_6$. Note also that equilibrium points E$_2$ and E$_6$ both have HU and CU topological structures, respectively.

The bottom side of Fig. \ref{fig:lo_1} shows the time evolution of the surface distance for particles initially placed at equilibrium points E$_3$ (blue line) and E$_5$ (dashed red line). The orbit of a particle from equilibrium point E$_5$ escapes from Bennu's neighborhood after 235\,h, and the orbit of the particle from the equilibrium point E$_2$ evolves around Bennu until escaping from the system after 676\,h. In addition, these particles move away from their equilibrium points around 8\,h.

\section{Falls}
\label{sec:fall}
The last type of initial conditions in our numerical simulations is the spherical one (III). Because we are interested in particles larger than a few centimeters, the SRP was not considered in these simulations.

\subsection{The Spherical Cloud type}
\label{sec:sph}
\begin{figure}
  \centering
  \includegraphics[width=\textwidth]{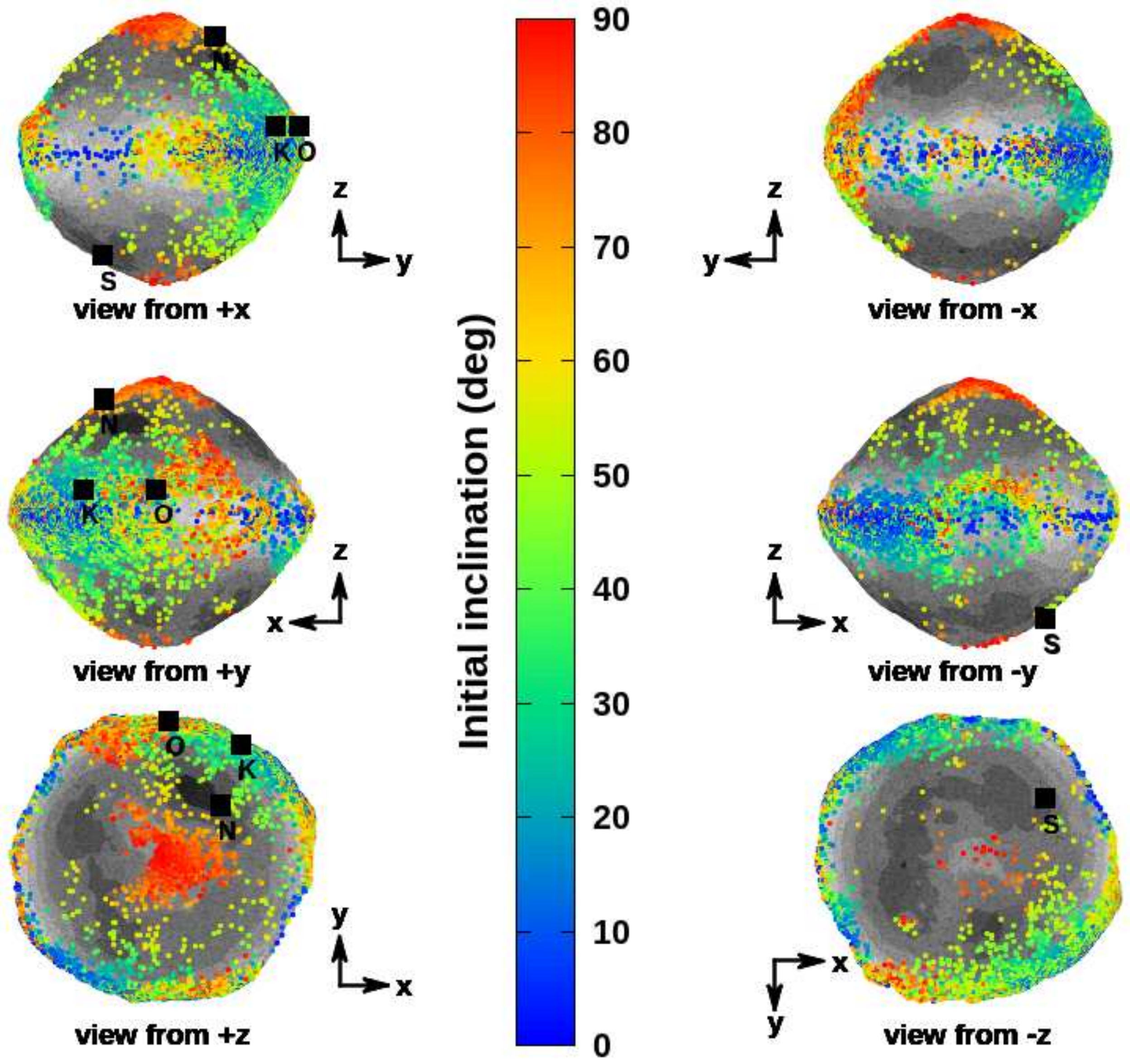}
  \caption{Results from initial conditions of type (III) integration for big particles (larger than a few centimeters). Mapping of the fall density across the Bennu surface. The color dots represent prograde impact sites from an initially spherical cloud of particles with initial inclination (degrees) in the inertial frame given by the color box. The gray color scale measures the surface distance (altitude) of asteroid Bennu, in m.}
  \label{fig:fall_1}
\end{figure}
\begin{figure}
  \centering
  \includegraphics[width=\textwidth]{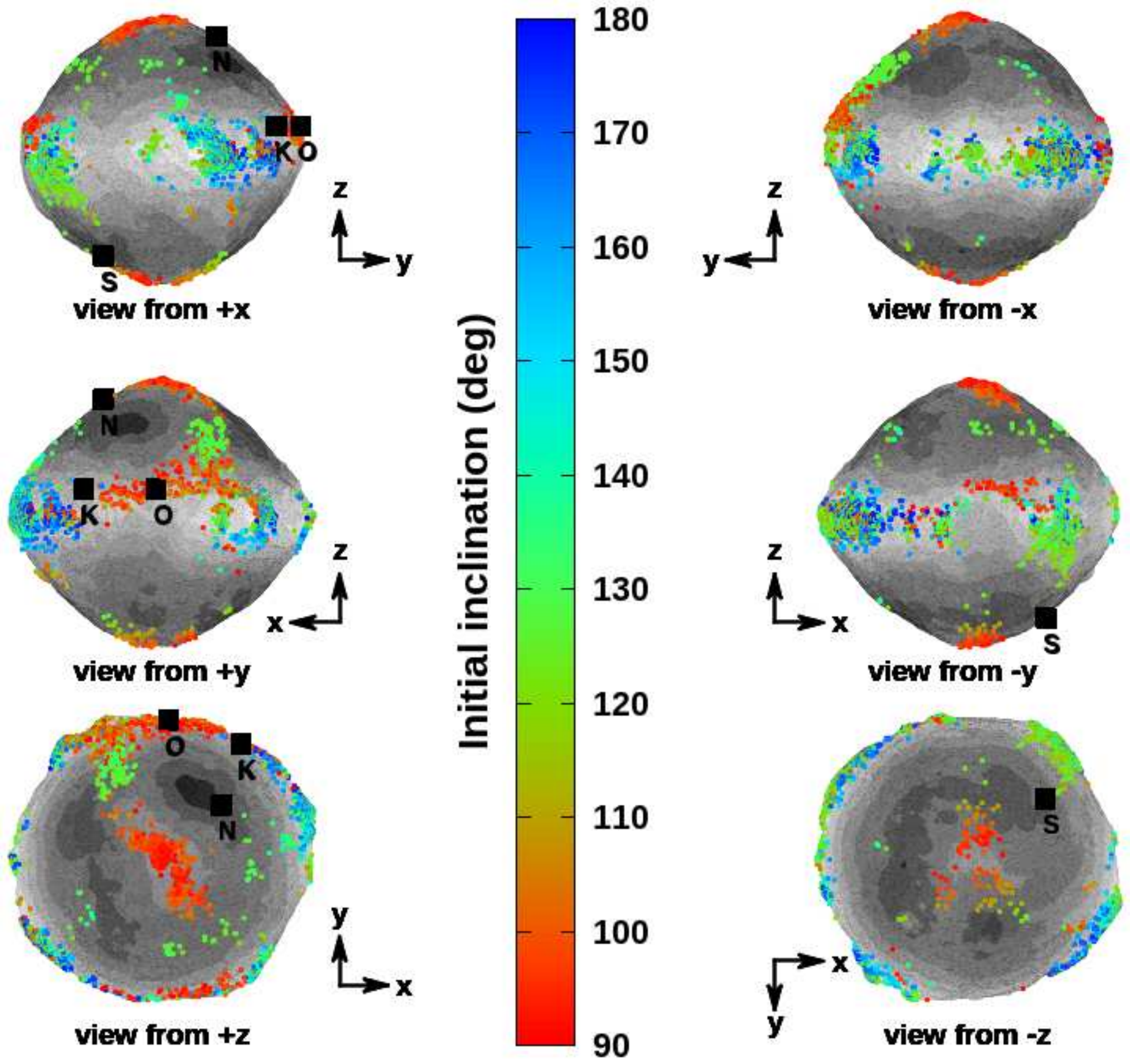}
  \caption{Results from initial conditions of type (III) integration for big particles (larger than a few centimeters). Mapping of the fall density across the Bennu surface. The color dots denote retrograde impact sites from an initially spherical cloud of particles with initial inclination (degrees) in the inertial frame given by the color box. The gray color scale measures the surface distance (altitude) of asteroid Bennu, in m.}
  \label{fig:fall_2}
\end{figure}

This section presents the results of a numerical experiment consisting of an initially spherical cloud of particles that are orbiting around Bennu and end by falling on its surface. The particles are initially placed evenly around Bennu in a spherical cloud sample with the center at Bennu's barycenter. The initial semimajor axis of the orbit of each particle varies randomly between 290\,m and 490\,m, two times the distance of the equivalent spherical radius of asteroid Bennu. The sample is filled with 20,000 particles that have osculating orbits around Bennu with initially null eccentricity and random asteroidal angles. The initial orbital inclination is in the interval of [$0^\circ$,$180^\circ$]. We considered two samples, one of prograde (inclination $< 90^\circ$) orbits and other of retrograde (inclination $> 90^\circ$) orbits. The study of the falling density of bigger particles is made through areal number density maps.

Figures \ref{fig:fall_1} and \ref{fig:fall_2} show the fall locations of the larger particles on the surface of asteroid Bennu for prograde and retrograde orbits, respectively. At first, we note that most of the particles fall near the equatorial regions and that a considerable number also falls on the polar regions. The mid-latitudes are those more depleted of falls. Note that the falls in the equatorial area are not uniformly distributed. There are accumulations and empty regions. A more careful comparison of these locations with the altitudes indicates that falls occur preferentially onto and near the high altitudes.

The particles were divided into two groups with each containing 20,000 bigger particles: those that were initially on prograde trajectories (Fig. \ref{fig:fall_1}) and those that were initially on retrograde trajectories (Fig. \ref{fig:fall_2}). Apart from the highest altitude peaks, where falls of particles from both groups occurred, we have that in the regions near those peaks, and the prograde and retrograde trajectories fall spread in complementary locations. This is an effect seen before in other work \cite{Winter2017, Moura2020, Winter2020}. The additional gravitational attraction because of a high peak tends to force a fall soon after the peak. Thus, for prograde trajectories, the fall occurs on one side after the peak, while for retrograde ones, it occurs on the other side. In summary, a general path is that the falls usually occurred in high-altitude locations. This can also be explained by the power-gravity behavior, as shown in the following paragraphs.

There are no significant differences between the falls in the prograde and retrograde orbits, except in the time of the falls. For the total integration time of $1.14$\,yr, $\sim 50$\% of the initial number of prograde particles fall over the surface of asteroid Bennu. Meanwhile, for retrograde orbits, $\sim 30$\% of the initial number of particles hit Bennu's surface. Therefore, a retrograde orbit is more stable against perturbing effects of the nonuniform gravitational field of an irregular-shaped minor body. This same behavior for retrograde orbits can be shown for other perturbation effects, such as oblateness \cite{Scheeres1994} and solar tides \cite{Hamilton1991}. For irregular-shaped minor bodies such as asteroid Bennu, this feature can be explained by the length of time that a particle spends in the proximity of a bulge of the surface of Bennu. By mutual gravitational perturbation, if the particle is orbiting Bennu in the same direction as its rotation (prograde), then the particle will be close to the bulge for a long time and have more time to have its trajectory disturbed. In contrast, if the particle is orbiting Bennu in the opposite direction to its rotation period (retrograde), then the particle spends much less time near the bulge, so the gravitational perturbation is minimized.

Figures \ref{fig:fall_1} and \ref{fig:fall_2} also show that the particle distributions over the surface of asteroid Bennu are highly dependent on the initial inclination of the particles. For example, particles initially with low inclinations in the inertial frame are more susceptible to hit the equatorial area of the surface of Bennu. On the other hand, fallen particles with high initial inclinations have impact sites in the polar areas of Bennu's surface. The middle-latitude regions of the surface of asteroid Bennu are filled mostly with particles with moderate initial inclinations (between $40^\circ$ and $60^\circ$). $\sim 60$\% of the prograde fallen particles have impact sites between latitudes -10$^\circ$ and 10$^\circ$ with average inclinations of $\sim 40^\circ$. For the retrograde sample, we have a higher value of $\sim 75$\% with average inclinations of $\sim 140^\circ$. There is also a significant number of impact sites in the north polar area of Bennu ($\sim 8$\%) than in its south polar region ($\sim 0.4$\%) with average inclinations of $\sim 86^\circ$. This same asymmetric polar behavior can be evinced for retrograde orbits. The Kingfisher and Osprey latitude regions are those more filled with falls than Nightingale and Sandpiper areas. Additionally, most of the fallen particles account between 0$^\circ$ and 120$^\circ$ longitudes.
\begin{figure}
  \centering
  \hspace{0.1cm}
  \includegraphics[width=9.7cm]{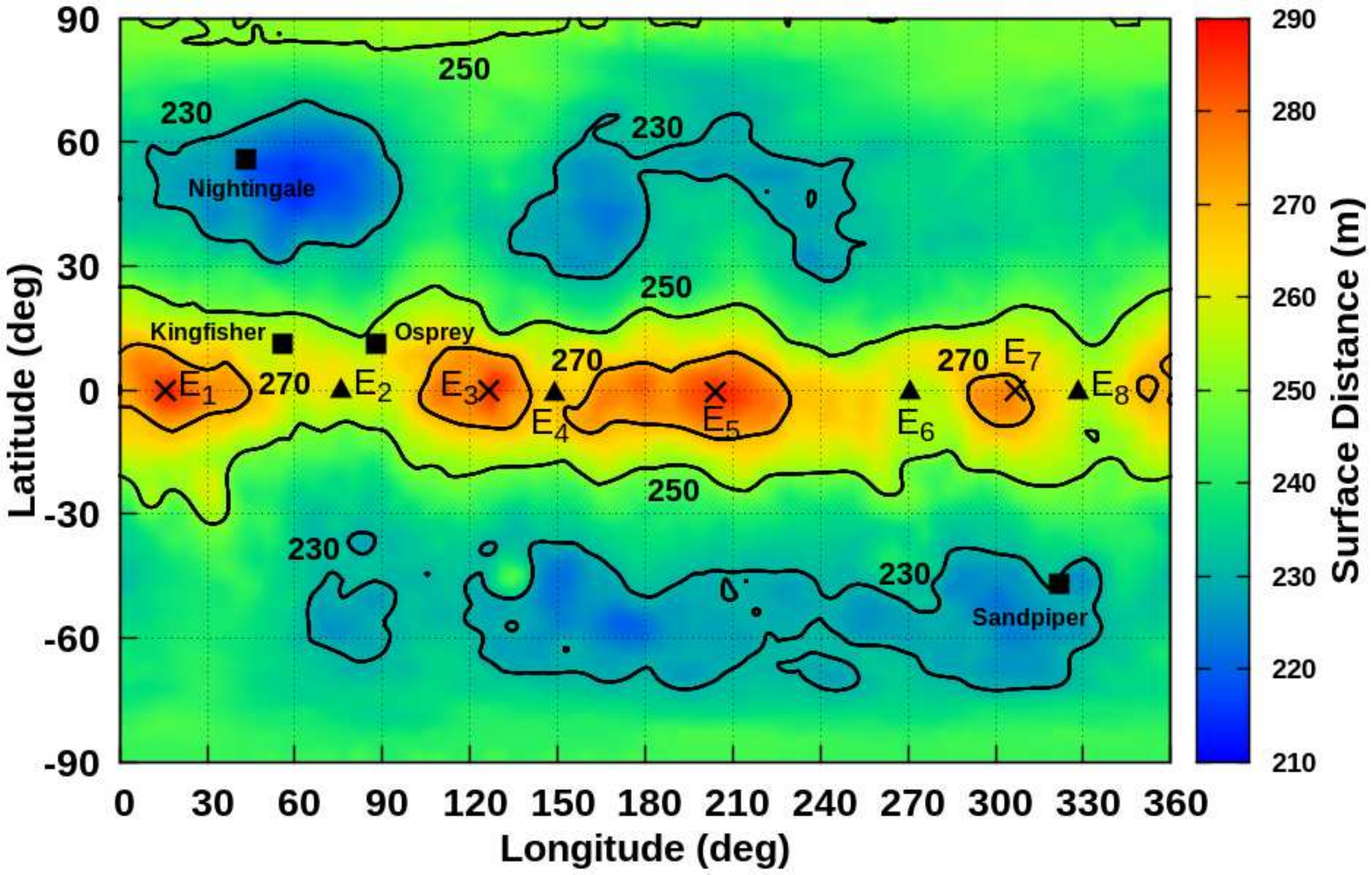}
  \includegraphics[width=9.5cm]{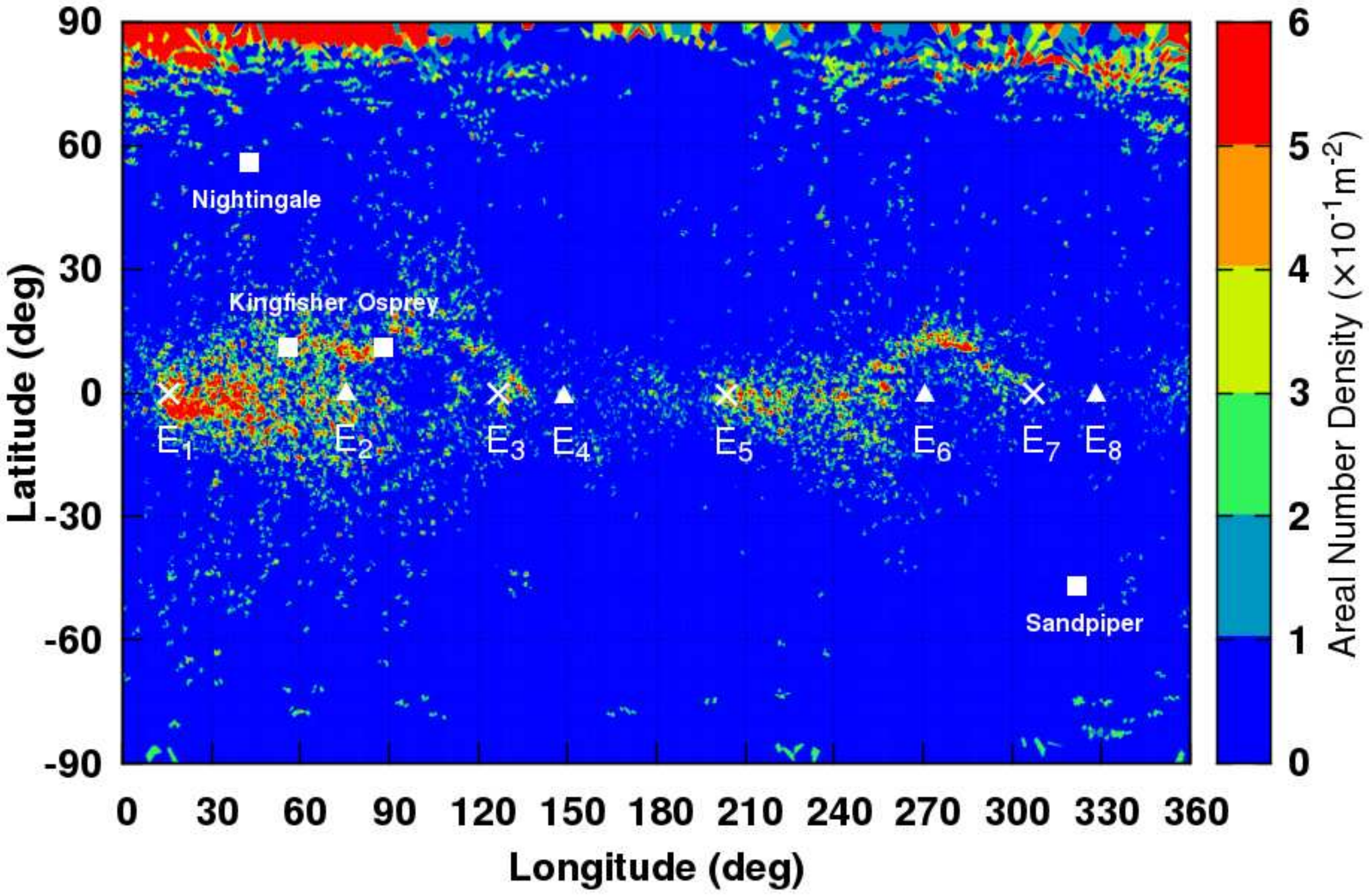}
  \caption{Results from initial conditions of type (III) integration for big particles (larger than a few centimeters). Equilibrium point projections over the Bennu surface are represented by X-cross and triangle marks and the squared-boxes are the locations of OSIRIS-REx's candidate sample sites. (top side) Level curve contours of Bennu's topography. The color code indicates the depth to Bennu's centroid in meters (altitude). The black contour lines denote levels every $20$\,m. (bottom side) Areal number density map ($\times 10^{-1}/m^2$) across the surface of Bennu for prograde orbits. The areal number density of falls is calculated by counting the number of impact sites within each facet and then normalized to the triangular face area.}
  \label{fig:fall_3}
\end{figure}

To gain a better visualization of these results, the top and bottom of Figure \ref{fig:fall_3} show a map of the topographic altitudes of Bennu and a map of the areal number density, respectively. For the sake of comparison of the locations, both plots are given in terms of latitude and longitude. In addition, square boxes in these figures indicate the final four potential sample sites of the OSIRIS-REx mission \cite{Lauretta2019}. The bottom side of Fig. \ref{fig:fall_3} shows the areal number density map for the particles in prograde orbits only. The abundance of falls for each location on the surface of Bennu is based on a 75-cm shape model of its surface, where the areal number density ($\times 10^{-1}/m^2$) is computed by counting the number of impacts into a facet and then is normalized to the corresponding triangular face area. This plot also indicates the projection of the location of the equilibrium points. Note that all equilibrium points that are over valleys of the surface (even-numbered) did not have a significant number of falls, while for equilibrium points on and near peaks (odd-numbered), the areal number density is higher.
\begin{figure}
  \centering
  \includegraphics[width=6.68cm]{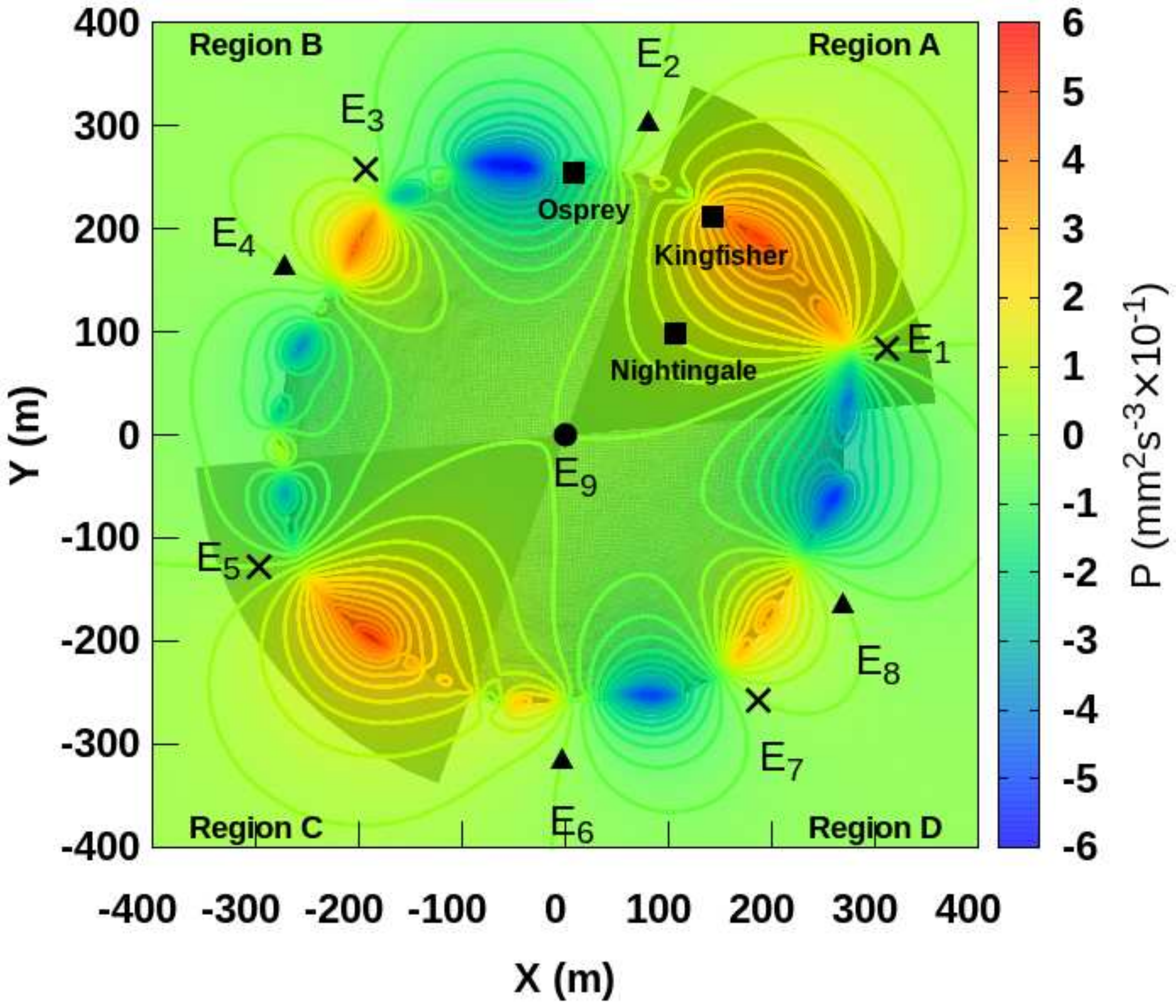}
  \includegraphics[width=6.68cm]{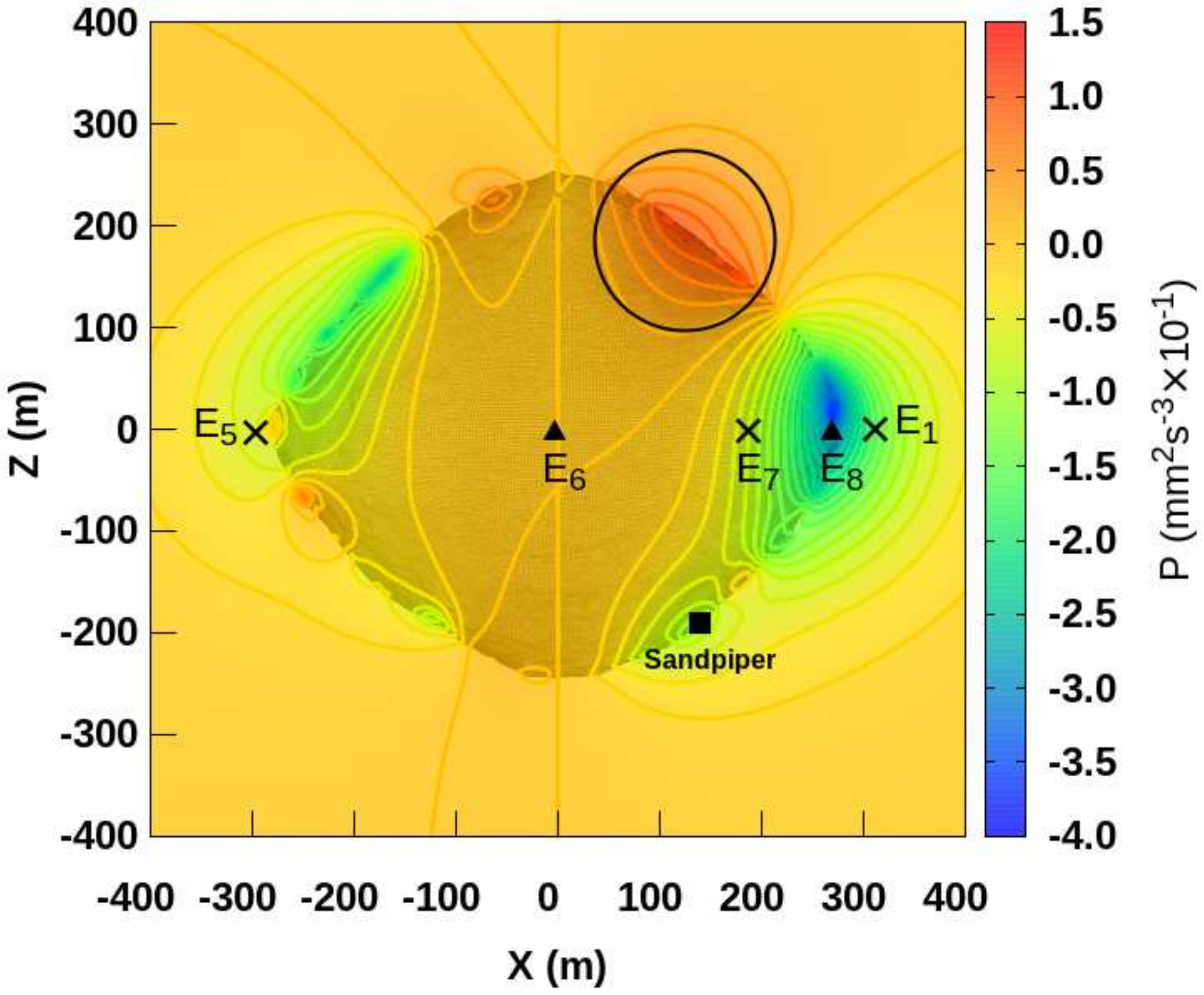}
  \caption{Gravity-Power map for asteroid Bennu. (Left side) The gravity-power map $P$ in the equatorial plane. (Right side) The gravity-power curves in the projection plane $xOz$. The color panels indicate the values of $P$, in mm$^2$\,s$^{-3}\times 10^{-1}$. Shadowed areas sketch the shape of asteroid Bennu.}
  \label{fig:fall_4}
\end{figure}

Combining the spin rate vector $\pmb{\Omega}$ and the gravity attraction (Eq. \ref{eq:sim_3}) of Bennu, the formula for Bennu's gravity power can be obtained \cite{Amarante2020}:
\begin{align}
P(x,y,z) & = (\pmb{\Omega}\times\textbf{r})\cdot(-\nabla U).
\label{eq:fall_1}
\end{align}
Eq. \eqref{eq:fall_1} is only particle position-dependent (\textbf{r}). Bennu's potential (Eq. \ref{eq:sim_2}) is fully determined by the geometry of the gravitational field, which is useful in measuring the enhancing and receding orbital energy of a particle. The left side of Fig. \ref{fig:fall_4} illustrates the gravity-power field of asteroid Bennu that divides the equatorial plane $xOy$ into four regions: `A' (between equilibrium points E$_1$ and E$_2$), `B' (between equilibrium points E$_2$ and E$_5$), `C' (between equilibrium points E$_5$ and E$_6$), and `D' (between equilibrium points E$_6$ and E$_1$), respectively. We also show the locations of all nine equilibrium points, and the shadowed area sketches the shape of Bennu. In regions A and C, the gravity-power equation is defined positively. These areas also have higher values of gravity power close to the surface of Bennu. However, in regions B and D, the gravity-power field is increased or decreased symmetrically depending on the equilibrium point region. For example, between equilibrium point pairs E$_2$-E$_3$ and E$_6$-E$_7$, the gravity power is defined negatively. These regions also have lower measures of the gravity power around Bennu. Meanwhile, between equilibrium point pairs E$_3$-E$_4$ and E$_7$-E$_8$, we have $P>0$. In addition, between equilibrium point pairs E$_4$-E$_5$ and E$_8$-E$_1$, we have $P<0$. A particle has orbital energy added in the regions where $P>0$, but it decreases in the areas where $P<0$. The equilibrium points in the equatorial plane lie in the zero-gravity-power curves, which are locations where $P=0$. For example, a particle that surrounds the equilibrium point E$_5$ in a periodic orbit near-equatorial plane has half of its trajectory in region B, while another half of its path is in region C. Then, the particle increases its orbital energy in region B and decreases in region C, i.e., the orbital energy changes periodically over time.

The left side of Fig. \ref{fig:fall_4} also indicates that extreme gravity-power values are reached at locations where the terrain becomes significantly steeper. This analysis suggests more possibilities for the surface particles in regions A and C to be trapped in Bennu's Roche lobe than those of regions B and D. It also has implications for the fallen particles over the surface of asteroid Bennu. For example, regions A and C match with particle flux areas (right side of Fig. \ref{fig:to_2}), which also have higher measures. Meanwhile, regions B and D are related to the dynamics of ejected particles \cite{McMahon2020}. In addition, the right side of Fig. \ref{fig:fall_4} shows the gravity-power field across projection plane $xOz$. The black circle in the north polar region highlights the higher values of the gravity power. The following areal number density map (bottom side of Fig. \ref{fig:fall_3}), shows a greater number of impacts in the north polar region of asteroid Bennu than those found in the south polar region for prograde and retrograde orbits (Figs. \ref{fig:fall_1} and \ref{fig:fall_2}, respectively).


\section{Final Comments}
\label{sec:conc}
This study provided outcomes about the stability and evolution of simulated particles in the environment around the surface of OSIRIS-REx target asteroid (101955) Bennu. First, Bennu's geopotential was computed through the mascons approach while considering SRP perturbation. Using the most precise shape model of asteroid Bennu, we numerically examined the location and topological structure of the actual and instantaneous equilibrium points through their linear stability. We found eight equilibrium points around Bennu regarding its current shape, density, and spin period. These equilibrium points were computed for particle sizes that are not affected significantly by SRP. All eight external equilibrium points are unstable.

The gravitational force dominates particles that have a size greater than a few centimeters, while particles smaller than a centimeter are quickly removed from the orbits around Bennu by the SRP. When the density of particles is chosen as the same as meteorite analogs, then smaller particles with $\sim<0.6$\,cm radius are preferentially removed from Bennu's neighborhood. Bennu's topological stabilities show saddle-like structures around equilibrium points E$_1$, E$_3$, E$_5$, and E$_7$, and this result is compatible with the linear stability analysis. However, center-like structures appear around equilibrium points E$_2$, E$_4$, E$_6$, and E$_8$ in at least one projection plane. The inner equilibrium point E$_9$ is LS. Meanwhile, equilibrium point E$_5$ is the least unstable external equilibrium point. The linear stability is more sensitive to change for particle sizes $\sim2\times10^{-2}$\,cm. In general, particles have mostly topological structures as the sink-source-center.

Next, we performed several sets of numerical simulations of particles to visualize the dynamics near them. We have chosen three types of numerical simulations: torus cloud, local disk, and spherical cloud. For the torus cloud integrations, half the initial number of particles collided with Bennu's surface in less than $\sim 1.54$\,h. The particle flux also suggests that particles larger than a few centimeters will fall across its surface near the Kingfisher sample area or near the region diametrically opposite to it. Otherwise, almost none of them will fall near the Osprey region. The results of local disks E$_2$, E$_6$, E$_7$, and E$_8$ show that their particles will be removed from the system faster ($<300$\,h) than the group of local disks E$_1$, E$_3$, E$_4$, and E$_5$ ($2,000$\,h).

Finally, the spherical cloud was used to measure the density of reaccumulation onto Bennu's surface. Half of the number of particles in prograde orbits hits the surface of asteroid Bennu over the entire integration. Meanwhile, the particles in retrograde orbits are more stable, and they have only 30\% of their initial particles impacted with Bennu's surface during this length of time. The falls in the equatorial region are not uniformly distributed. The regions near the equilibrium point projections that are over valleys of the surface did not have a significant number of falls, while on and near those that are over peaks, the number of falls is high. In the equatorial area, the falls occur preferentially onto and near the high altitudes. However, the mid-latitudes are those more depleted of falls, as in the Nightingale and Sandpiper areas. The polar regions also show a considerable number of impacts, and they have a preferred fall location in Bennu's north pole. These results indicate the preferred locations of the falls and their connection with the geometrical altitudes of the surface of asteroid Bennu.

\pagebreak

\appendix

\section{Eigenvalues}
\label{sec:eigen}
Eigenvalues of the equilibrium points (no SRP) are presented in Table \ref{tab:equi_mas}, referring to the Bennu system.
\begin{table}
 \centering
  \caption{Eigenvalues ($\lambda_n\times 10^{-4}$) of Bennu's equilibrium points (no SRP) from Table \ref{tab:equi_mas} and their topological structures computed by the mascons technique.}
 \label{tab:app_1}
 {
 \begin{tabular}{ccccc}
  \hline
  Point & $\lambda_{1,2}$ & $\lambda_{3,4}$ & $\lambda_{5,6}$ & Topological Structure \\
  \hline
  E$_1$ & $\pm2.686$ & $\pm4.721i$ & $\pm4.235i$ & saddle--center--center \\
  E$_2$ & $-0.825 \pm2.673i$ & $0.825 \pm2.673i$ & $\pm4.481i$ & sink--source--center \\
  E$_3$ & $\pm2.537$ & $\pm4.751i$ & $\pm4.108i$ & saddle--center--center \\
  E$_4$ & $-1.123 \pm2.666i$ & $1.123 \pm2.666i$ & $\pm4.617i$ & sink--source--center \\
  E$_5$ & $\pm2.380$ & $\pm4.804i$ & $\pm3.949i$ & saddle--center--center \\
  E$_6$ & $-0.590 \pm2.640i$ & $0.590 \pm2.640i$ & $\pm4.446i$ & sink--source--center \\
  E$_7$ & $\pm2.125$ & $\pm4.567i$ & $\pm4.082i$ & saddle--center--center \\
  E$_8$ & $-0.901 \pm2.736i$ & $0.901 \pm2.736i$ & $\pm4.434i$ & sink--source--center \\
  E$_9$ & $\pm9.670i$ & $\pm6.076i$ & $\pm1.544i$ & center--center--center \\
  \hline
 \end{tabular}}
 \end{table}

\pagebreak

\section{Minor-Mercury Package}
\label{sec:minor-mercury}
We use the mascons and the polyhedra approaches to develop an integration model to deal with our numerical simulations. We built the \textit{Minor-Mercury package}: a modified version of the original Mercury package \cite{Chambers1999} to handle an irregular-shaped minor body \cite{minor-mercury}. The N-body algorithm B\"{u}lirsch--St\"{o}er \cite{Stoer1980} is used to integrate the orbits of the particles around the irregular gravity field of asteroid Bennu where a toleration accuracy parameter was set to $10^{-12}$. The particles from the Bennu centroid are integrated with gravitational and SRP equations from (\ref{eq:sim_2}) to (\ref{eq:math_3}).
\begin{figure}
  \centering
  \includegraphics[width=6.68cm]{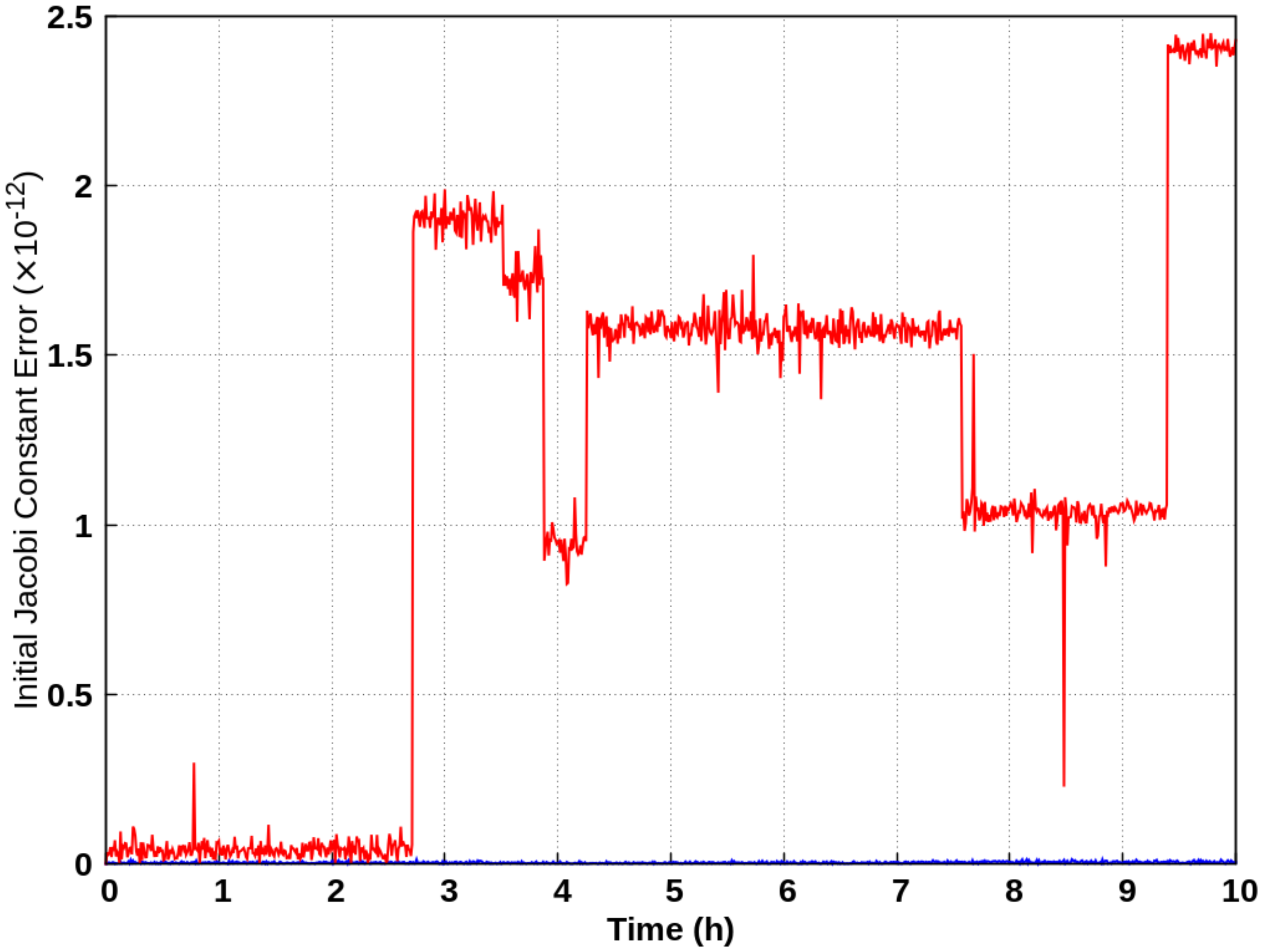}
  \includegraphics[width=6.68cm]{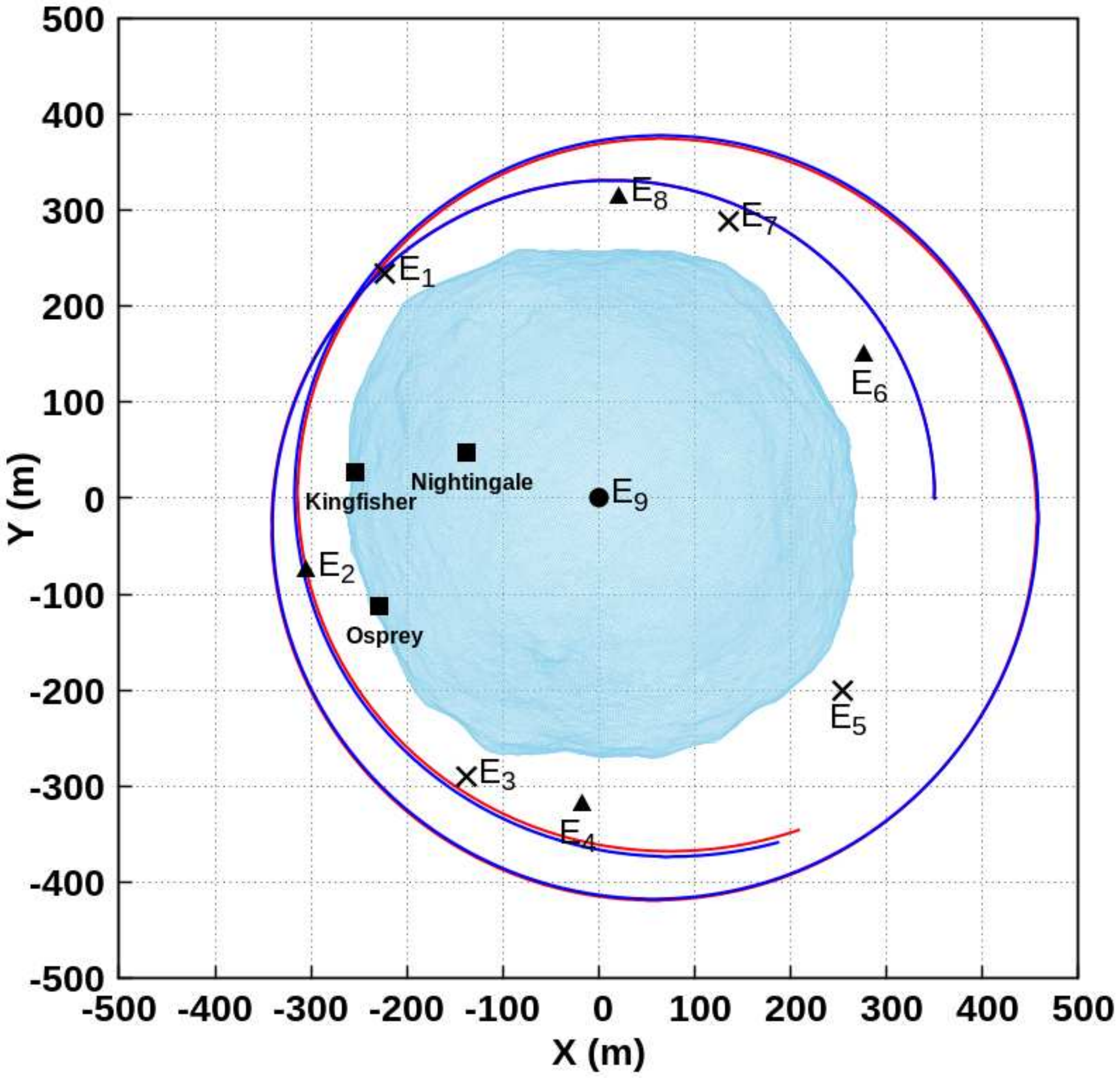}
  \caption{(left side) Relative initial Jacobi constant error ($\times 10^{-12}$) for a particle initially in a circular and planar orbit around Bennu, with a radius of 350\,m. The red line shows the time evolution of the relative initial error through the polyhedra method and the blue one the relative initial error for the mascons approach, in a time integration of 10\,h. (right side) Comparison of the particle's orbit in the inertial frame using polyhedra (red) and mascons (blue) techniques, for a 10\,h integration time.}
  \label{fig:appB_1}
\end{figure}
The left side of Figure \ref{fig:appB_1} shows the absolute relative error of the initial Jacobi constant for the polyhedra (red) \cite{Petrovic1996,Tsoulis2001} and the mascons (blue) \cite{Geissler1996,Werner1997b,Scheeres1998} techniques as a function of the time, for a particle initially placed in a circular and planar orbit around Bennu. We adopt a semimajor axis of 350\,m, and the other angles are set to be zero. The simulations are performed in the Bennu-fixed frame; however, they can be output as well in the inertial frame (right side of Fig. \ref{fig:appB_1}).

The initial Jacobi constant error (red) for the polyhedra approach is approximately $\leq 10^{-12}$ in the vicinity of Bennu's surface. The initial Jacobi constant error (blue) for the mascons technique remains smaller than $\leq 10^{-14}$. Thus, it is suitable to use this method in our numerical integrations. This procedure using the mascon model has been extensively tested in previous work to study the dynamical environment and surface characteristics of an asteroid \cite{Winter2020,Moura2020}. Let us assume that Bennu has a uniform rotation about its largest moment of inertia ($z$-axis). Consequently, the dynamical equations used in this work for motion near the uniformly rotating asteroid Bennu in its body-fixed frame are \cite{Jiang2014b}:
\begin{eqnarray}
      \ddot{x} -2\omega \dot{y} +{{\partial V}\over{\partial x}} -a_{{SRP}_{x}} &=& 0, \nonumber \\
      \ddot{y} +2\omega \dot{x} +{{\partial V}\over{\partial y}} -a_{{SRP}_{y}} &=& 0, \\
      \ddot{z} +{{\partial V}\over{\partial z}} -a_{{SRP}_{z}} &=& 0. \nonumber
\label{eq:sim_1}
\end{eqnarray}

\subsection{Collisional and Escape criteria}
\label{sec:sim:coej}
Following \citeA{Dobrovolskis1996}, we can compute an equivalent triaxial ellipsoid from the principal moments of Bennu's diagonal inertia tensor. We use the dimensions of this reference ellipsoid to implement our collisional algorithm code with the triangular mesh of asteroid Bennu. Our system configuration leads us to consider asteroid Bennu as an equivalent ellipsoid with principal semiaxes $a$, $b$, and $c$ given by:
\begin{eqnarray}
\label{eq:sim_9}
a=\sqrt{\frac{5(I_{yy}+I_{zz}-I_{xx})}{2M}}, \nonumber \\
b=\sqrt{\frac{5(I_{xx}+I_{zz}-I_{yy})}{2M}}, \nonumber \\
c=\sqrt{\frac{5(I_{xx}+I_{yy}-I_{zz})}{2M}}. \nonumber
\end{eqnarray}
From Eqs. (\ref{eq:sim_9}), the adopted shape model provides a body with principal semiaxes of $\sim 263, 255, and 222$\,m. If the trajectory of a particle goes close to the surface of the ellipsoid, then a collisional algorithm starts to check if the particle collided with Bennu's surface following the ray-casting method. This collisional algorithm also provides the approximate location of the site (face) that the particle impacted the surface of Bennu using the tetrahedrons from faces and the signs of five determinants. Let one tetrahedron face have vertices $V_0 = (x_0, y_0, z_0)$, $V_1 = (x_1, y_1, z_1)$, $V_2 = (x_2, y_2, z_2)$, $V_3 = (x_3, y_3, z_3)$ and the test point be $P = (x, y, z)$, where $V_0$ is the center of mass of the polyhedron. Then, point $P$ is in the tetrahedron and consequently within the polyhedron if all the following five determinants have the same sign:

$\left| \begin{array}{cccc}
x_0 & y_0 & z_0 & 1 \\
x_1 & y_1 & z_1 & 1 \\
x_2 & y_2 & z_2 & 1 \\
x_3 & y_3 & z_3 & 1
\end{array} \right|$
$\left| \begin{array}{cccc}
x & y & z & 1 \\
x_1 & y_1 & z_1 & 1 \\
x_2 & y_2 & z_2 & 1 \\
x_3 & y_3 & z_3 & 1
\end{array} \right|$
$\left| \begin{array}{cccc}
x_0 & y_0 & z_0 & 1 \\
x & y & z & 1 \\
x_2 & y_2 & z_2 & 1 \\
x_3 & y_3 & z_3 & 1
\end{array} \right|$
$\left| \begin{array}{cccc}
x_0 & y_0 & z_0 & 1 \\
x_1 & y_1 & z_1 & 1 \\
x & y & z & 1 \\
x_3 & y_3 & z_3 & 1
\end{array} \right|$
$\left| \begin{array}{cccc}
x_0 & y_0 & z_0 & 1 \\
x_1 & y_1 & z_1 & 1 \\
x_2 & y_2 & z_2 & 1 \\
x & y & z & 1
\end{array} \right|$

We use the escape criteria of our system as a periodic effect. An algorithm starts to check if the particle escaped from the system (from the original Mercury code). When the particle orbital radius is outside a sphere that has a 1\,km radius, then the particle is automatically removed from the simulation.

\acknowledgments

The authors thank the reviewers whose comments greatly improved the manuscript. The work was carried out with the support of the Improvement Coordination Higher Education Personnel - Brazil (CAPES) - Financing Code 001 and National Council for Scientific and Technological Development (CNPq, proc. 305210/2018-1). This research also had computational resources provided by the thematic project FAPESP (proc. 2016/24561-0) and the Center for Mathematical Sciences Applied to Industry (CeMEAI), funded by FAPESP (grant 2013/07375-0). Our generated data are publicly available through the Data Publisher for Earth $\&$ Environmental Science at \url{https://doi.org/10.1594/PANGAEA.908248} \cite{Amarante2020data1}. We are also grateful to the entire OSIRIS-REx team for making the encounter with asteroid Bennu possible.



%
%

\bibliography{biblio}

\begin{thebibliography}{}

\bibitem[\protect\citeauthoryear{%
{Amarante}%
}{%
{Amarante}%
}{%
{\protect\APACyear{2020}}%
{\protect\APACexlab{{\protect\BCnt{1}}}}}]{%
hnm-ring}%
\APACinsertmetastar{%
hnm-ring}%
{Amarante}, A.%
%
\unskip\
\newblock
\APACrefYearMonthDay{2020{\protect\BCnt{1}}}{{\APACmonth{09}}}{}.
\newblock
\APACrefbtitle{HNM-Ring.}{Hnm-ring.}
\newblock
\APACaddressPublisher{}{Zenodo}.
\newblock
 \begin{APACrefURL} \url{https://doi.org/10.5281/zenodo.4043488}
  \end{APACrefURL}
\newblock
\APACrefnote{{The source codes of this package are available on reasonable
  request.}}
\PrintBackRefs{\CurrentBib}

\bibitem[\protect\citeauthoryear{%
{Amarante}%
}{%
{Amarante}%
}{%
{\protect\APACyear{2020}}%
{\protect\APACexlab{{\protect\BCnt{2}}}}}]{%
minor-equilibria-nr}%
\APACinsertmetastar{%
minor-equilibria-nr}%
{Amarante}, A.%
%
\unskip\
\newblock
\APACrefYearMonthDay{2020{\protect\BCnt{2}}}{{\APACmonth{09}}}{}.
\newblock
\APACrefbtitle{Minor-Equilibria-NR.}{Minor-equilibria-nr.}
\newblock
\APACaddressPublisher{}{Zenodo}.
\newblock
 \begin{APACrefURL} \url{https://doi.org/10.5281/zenodo.4043461}
  \end{APACrefURL}
\newblock
\APACrefnote{{The source codes of this package are available on reasonable
  request.}}
\PrintBackRefs{\CurrentBib}

\bibitem[\protect\citeauthoryear{%
{Amarante}%
}{%
{Amarante}%
}{%
{\protect\APACyear{2020}}%
{\protect\APACexlab{{\protect\BCnt{3}}}}}]{%
minor-mercury}%
\APACinsertmetastar{%
minor-mercury}%
{Amarante}, A.%
%
\unskip\
\newblock
\APACrefYearMonthDay{2020{\protect\BCnt{3}}}{{\APACmonth{09}}}{}.
\newblock
\APACrefbtitle{Minor-Mercury.}{Minor-mercury.}
\newblock
\APACaddressPublisher{}{Zenodo}.
\newblock
 \begin{APACrefURL} \url{https://doi.org/10.5281/zenodo.4043760}
  \end{APACrefURL}
\newblock
\APACrefnote{{The source codes of this package are available on reasonable
  request.}}
\PrintBackRefs{\CurrentBib}

\bibitem[\protect\citeauthoryear{%
{Amarante}%
, {Winter}%
\BCBL{}\ \BBA{} {Sfair}%
}{%
{Amarante}%
\ \protect\BOthers{.}}{%
{\protect\APACyear{2020}}%
}]{%
Amarante2020data1}%
\APACinsertmetastar{%
Amarante2020data1}%
{Amarante}, A.%
, {Winter}, O.%
\BCBL{}\ \BBA{} {Sfair}, R.%
%
\unskip\
\newblock
\APACrefYearMonthDay{2020}{}{}.
\newblock
\APACrefbtitle{{Exploration of the Activity of Asteroid (101955)
  Bennu}}{{Exploration of the Activity of Asteroid (101955) Bennu}}\ [data
  set].
\newblock
\APACaddressPublisher{}{PANGAEA}.
\newblock
 \begin{APACrefURL} \url{https://doi.org/10.1594/PANGAEA.908248}
  \end{APACrefURL}
\newblock
\APACrefnote{Supplement to: Amarante Luiz, A et al. (submitted): Stability and
  Evolution of Fallen Particles Around the Surface of asteroid (101955) Bennu.
  Journal of Geophysical Research: Planets}
\PrintBackRefs{\CurrentBib}

\bibitem[\protect\citeauthoryear{%
{Amarante}%
\ \BBA{} {Winter}%
}{%
{Amarante}%
\ \BBA{} {Winter}%
}{%
{\protect\APACyear{2020}}%
}]{%
Amarante2020}%
\APACinsertmetastar{%
Amarante2020}%
{Amarante}, A.%
\BCBT{}\ \BBA{} {Winter}, O\BPBI C.%
%
\unskip\
\newblock
\APACrefYearMonthDay{2020}{{\APACmonth{07}}}{}.
\newblock
\BBOQ{}\APACrefatitle{{Surface dynamics, equilibrium points and individual
  lobes of the Kuiper Belt object (486958) Arrokoth}}{{Surface dynamics,
  equilibrium points and individual lobes of the Kuiper Belt object (486958)
  Arrokoth}}.\BBCQ{}
\newblock
\APACjournalVolNumPages{\mnras}{496}{4}{4154-4173}.
\PrintBackRefs{\CurrentBib}

\bibitem[\protect\citeauthoryear{%
{Barnouin}%
\ \protect\BOthers{.}}{%
{Barnouin}%
\ \protect\BOthers{.}}{%
{\protect\APACyear{2019}}%
}]{%
Barnouin2019}%
\APACinsertmetastar{%
Barnouin2019}%
{Barnouin}, O\BPBI S.%
, {Daly}, M\BPBI G.%
, {Palmer}, E\BPBI E.%
, {Gaskell}, R\BPBI W.%
, {Weirich}, J\BPBI R.%
, {Johnson}, C\BPBI L.%
\BCBL{}\ \BOthersPeriod{.}%
\unskip\
\newblock
\APACrefYearMonthDay{2019}{{\APACmonth{03}}}{}.
\newblock
\BBOQ{}\APACrefatitle{{Shape of (101955) Bennu indicative of a rubble pile with
  internal stiffness}}{{Shape of (101955) Bennu indicative of a rubble pile
  with internal stiffness}}.\BBCQ{}
\newblock
\APACjournalVolNumPages{Nature Geoscience}{12}{}{247-252}.
\PrintBackRefs{\CurrentBib}

\bibitem[\protect\citeauthoryear{%
{Burns}%
, {Lamy}%
\BCBL{}\ \BBA{} {Soter}%
}{%
{Burns}%
\ \protect\BOthers{.}}{%
{\protect\APACyear{1979}}%
}]{%
Burns1979}%
\APACinsertmetastar{%
Burns1979}%
{Burns}, J\BPBI A.%
, {Lamy}, P\BPBI L.%
\BCBL{}\ \BBA{} {Soter}, S.%
%
\unskip\
\newblock
\APACrefYearMonthDay{1979}{{\APACmonth{10}}}{}.
\newblock
\BBOQ{}\APACrefatitle{{Radiation forces on small particles in the solar
  system}}{{Radiation forces on small particles in the solar system}}.\BBCQ{}
\newblock
\APACjournalVolNumPages{\icarus}{40}{1}{1-48}.
\PrintBackRefs{\CurrentBib}

\bibitem[\protect\citeauthoryear{%
{Chambers}%
}{%
{Chambers}%
}{%
{\protect\APACyear{1999}}%
}]{%
Chambers1999}%
\APACinsertmetastar{%
Chambers1999}%
{Chambers}, J\BPBI E.%
%
\unskip\
\newblock
\APACrefYearMonthDay{1999}{{\APACmonth{04}}}{}.
\newblock
\BBOQ{}\APACrefatitle{{A hybrid symplectic integrator that permits close
  encounters between massive bodies}}{{A hybrid symplectic integrator that
  permits close encounters between massive bodies}}.\BBCQ{}
\newblock
\APACjournalVolNumPages{Monthly Notices of the Royal Astronomical
  Society}{304}{}{793-799}.
\PrintBackRefs{\CurrentBib}

\bibitem[\protect\citeauthoryear{%
{Chesley}%
\ \protect\BOthers{.}}{%
{Chesley}%
\ \protect\BOthers{.}}{%
{\protect\APACyear{2014}}%
}]{%
Chesley2014}%
\APACinsertmetastar{%
Chesley2014}%
{Chesley}, S\BPBI R.%
, {Farnocchia}, D.%
, {Nolan}, M\BPBI C.%
, {Vokrouhlick{\'y}}, D.%
, {Chodas}, P\BPBI W.%
, {Milani}, A.%
\BCBL{}\ \BOthersPeriod{.}%
\unskip\
\newblock
\APACrefYearMonthDay{2014}{{\APACmonth{06}}}{}.
\newblock
\BBOQ{}\APACrefatitle{{Orbit and bulk density of the OSIRIS-REx target Asteroid
  (101955) Bennu}}{{Orbit and bulk density of the OSIRIS-REx target Asteroid
  (101955) Bennu}}.\BBCQ{}
\newblock
\APACjournalVolNumPages{Icarus}{235}{}{5-22}.
\PrintBackRefs{\CurrentBib}

\bibitem[\protect\citeauthoryear{%
{Dobrovolskis}%
}{%
{Dobrovolskis}%
}{%
{\protect\APACyear{1996}}%
}]{%
Dobrovolskis1996}%
\APACinsertmetastar{%
Dobrovolskis1996}%
{Dobrovolskis}, A\BPBI R.%
%
\unskip\
\newblock
\APACrefYearMonthDay{1996}{{\APACmonth{12}}}{}.
\newblock
\BBOQ{}\APACrefatitle{{Inertia of Any Polyhedron}}{{Inertia of Any
  Polyhedron}}.\BBCQ{}
\newblock
\APACjournalVolNumPages{Icarus}{124}{}{698-704}.
\PrintBackRefs{\CurrentBib}

\bibitem[\protect\citeauthoryear{%
{Geissler}%
\ \protect\BOthers{.}}{%
{Geissler}%
\ \protect\BOthers{.}}{%
{\protect\APACyear{1996}}%
}]{%
Geissler1996}%
\APACinsertmetastar{%
Geissler1996}%
{Geissler}, P.%
, {Petit}, J\BHBI M.%
, {Durda}, D\BPBI D.%
, {Greenberg}, R.%
, {Bottke}, W.%
, {Nolan}, M.%
\BCBL{}\ \BOthersPeriod{.}%
\unskip\
\newblock
\APACrefYearMonthDay{1996}{{\APACmonth{03}}}{}.
\newblock
\BBOQ{}\APACrefatitle{{Erosion and Ejecta Reaccretion on 243 Ida and Its
  Moon}}{{Erosion and Ejecta Reaccretion on 243 Ida and Its Moon}}.\BBCQ{}
\newblock
\APACjournalVolNumPages{Icarus}{120}{}{140-157}.
\PrintBackRefs{\CurrentBib}

\bibitem[\protect\citeauthoryear{%
{Hamilton}%
\ \BBA{} {Burns}%
}{%
{Hamilton}%
\ \BBA{} {Burns}%
}{%
{\protect\APACyear{1991}}%
}]{%
Hamilton1991}%
\APACinsertmetastar{%
Hamilton1991}%
{Hamilton}, D\BPBI P.%
\BCBT{}\ \BBA{} {Burns}, J\BPBI A.%
%
\unskip\
\newblock
\APACrefYearMonthDay{1991}{{\APACmonth{07}}}{}.
\newblock
\BBOQ{}\APACrefatitle{{Orbital stability zones about asteroids}}{{Orbital
  stability zones about asteroids}}.\BBCQ{}
\newblock
\APACjournalVolNumPages{\icarus}{92}{1}{118-131}.
\PrintBackRefs{\CurrentBib}

\bibitem[\protect\citeauthoryear{%
Hamilton%
\ \BBA{} Krivov%
}{%
Hamilton%
\ \BBA{} Krivov%
}{%
{\protect\APACyear{1996}}%
}]{%
hamilton96}%
\APACinsertmetastar{%
hamilton96}%
Hamilton, D\BPBI P.%
\BCBT{}\ \BBA{} Krivov, A\BPBI V.%
%
\unskip\
\newblock
\APACrefYearMonthDay{1996}{{oct}}{}.
\newblock
\BBOQ{}\APACrefatitle{{Circumplanetary Dust Dynamics: Effects of Solar Gravity,
  Radiation Pressure, Planetary Oblateness, and
  Electromagnetism}}{{Circumplanetary Dust Dynamics: Effects of Solar Gravity,
  Radiation Pressure, Planetary Oblateness, and Electromagnetism}}.\BBCQ{}
\newblock
\APACjournalVolNumPages{{Icarus}}{123}{}{503-523}.
\PrintBackRefs{\CurrentBib}

\bibitem[\protect\citeauthoryear{%
{Hamilton}%
\ \protect\BOthers{.}}{%
{Hamilton}%
\ \protect\BOthers{.}}{%
{\protect\APACyear{2019}}%
}]{%
Hamilton2019}%
\APACinsertmetastar{%
Hamilton2019}%
{Hamilton}, V\BPBI E.%
, {Simon}, A\BPBI A.%
, {Christensen}, P\BPBI R.%
, {Reuter}, D\BPBI C.%
, {Clark}, B\BPBI E.%
, {Barucci}, M\BPBI A.%
\BCBL{}\ \BOthersPeriod{.}%
\unskip\
\newblock
\APACrefYearMonthDay{2019}{{\APACmonth{03}}}{}.
\newblock
\BBOQ{}\APACrefatitle{{Evidence for widespread hydrated minerals on asteroid
  (101955) Bennu}}{{Evidence for widespread hydrated minerals on asteroid
  (101955) Bennu}}.\BBCQ{}
\newblock
\APACjournalVolNumPages{Nature Astronomy}{3}{}{332-340}.
\PrintBackRefs{\CurrentBib}

\bibitem[\protect\citeauthoryear{%
{Hergenrother}%
\ \protect\BOthers{.}}{%
{Hergenrother}%
\ \protect\BOthers{.}}{%
{\protect\APACyear{2019}}%
}]{%
Hergenrother2019}%
\APACinsertmetastar{%
Hergenrother2019}%
{Hergenrother}, C\BPBI W.%
, {Maleszewski}, C\BPBI K.%
, {Nolan}, M\BPBI C.%
, {Li}, J\BPBI Y.%
, {Drouet D'Aubigny}, C\BPBI Y.%
, {Shelly}, F\BPBI C.%
\BCBL{}\ \BOthersPeriod{.}%
\unskip\
\newblock
\APACrefYearMonthDay{2019}{Mar}{}.
\newblock
\BBOQ{}\APACrefatitle{{The operational environment and rotational acceleration
  of asteroid (101955) Bennu from OSIRIS-REx observations}}{{The operational
  environment and rotational acceleration of asteroid (101955) Bennu from
  OSIRIS-REx observations}}.\BBCQ{}
\newblock
\APACjournalVolNumPages{Nature Communications}{10}{}{1291}.
\PrintBackRefs{\CurrentBib}

\bibitem[\protect\citeauthoryear{%
{Jiang}%
\ \BBA{} {Baoyin}%
}{%
{Jiang}%
\ \BBA{} {Baoyin}%
}{%
{\protect\APACyear{2014}}%
}]{%
Jiang2014b}%
\APACinsertmetastar{%
Jiang2014b}%
{Jiang}, Y.%
\BCBT{}\ \BBA{} {Baoyin}, H.%
%
\unskip\
\newblock
\APACrefYearMonthDay{2014}{Mar}{}.
\newblock
\BBOQ{}\APACrefatitle{{Orbital Mechanics near a Rotating Asteroid}}{{Orbital
  Mechanics near a Rotating Asteroid}}.\BBCQ{}
\newblock
\APACjournalVolNumPages{Journal of Astrophysics and Astronomy}{35}{1}{17-38}.
\PrintBackRefs{\CurrentBib}

\bibitem[\protect\citeauthoryear{%
{Jiang}%
, {Baoyin}%
, {Li}%
\BCBL{}\ \BBA{} {Li}%
}{%
{Jiang}%
\ \protect\BOthers{.}}{%
{\protect\APACyear{2014}}%
}]{%
Jiang2014}%
\APACinsertmetastar{%
Jiang2014}%
{Jiang}, Y.%
, {Baoyin}, H.%
, {Li}, J.%
\BCBL{}\ \BBA{} {Li}, H.%
%
\unskip\
\newblock
\APACrefYearMonthDay{2014}{Jan}{}.
\newblock
\BBOQ{}\APACrefatitle{{Orbits and manifolds near the equilibrium points around
  a rotating asteroid}}{{Orbits and manifolds near the equilibrium points
  around a rotating asteroid}}.\BBCQ{}
\newblock
\APACjournalVolNumPages{Astrophysics and Space Science}{349}{1}{83-106}.
\PrintBackRefs{\CurrentBib}

\bibitem[\protect\citeauthoryear{%
{Kaula}%
}{%
{Kaula}%
}{%
{\protect\APACyear{1966}}%
}]{%
Kaula1966}%
\APACinsertmetastar{%
Kaula1966}%
{Kaula}, W\BPBI M.%
%
\unskip\
\newblock
\APACrefYear{1966}.
\newblock
\APACrefbtitle{{Theory of satellite geodesy. Applications of satellites to
  geodesy}}{{Theory of satellite geodesy. Applications of satellites to
  geodesy}}.
\PrintBackRefs{\CurrentBib}

\bibitem[\protect\citeauthoryear{%
{Lauretta}%
\ \protect\BOthers{.}}{%
{Lauretta}%
\ \protect\BOthers{.}}{%
{\protect\APACyear{2017}}%
}]{%
Lauretta2017}%
\APACinsertmetastar{%
Lauretta2017}%
{Lauretta}, D\BPBI S.%
, {Balram-Knutson}, S\BPBI S.%
, {Beshore}, E.%
, {Boynton}, W\BPBI V.%
, {Drouet d'Aubigny}, C.%
, {DellaGiustina}, D\BPBI N.%
\BCBL{}\ \BOthersPeriod{.}%
\unskip\
\newblock
\APACrefYearMonthDay{2017}{{\APACmonth{10}}}{}.
\newblock
\BBOQ{}\APACrefatitle{{OSIRIS-REx: Sample Return from Asteroid (101955)
  Bennu}}{{OSIRIS-REx: Sample Return from Asteroid (101955) Bennu}}.\BBCQ{}
\newblock
\APACjournalVolNumPages{Space Science Reviews}{212}{}{925-984}.
\PrintBackRefs{\CurrentBib}

\bibitem[\protect\citeauthoryear{%
{Lauretta}%
, {Dellagiustina}%
\BCBL{}\ \protect\BOthers{.}}{%
{Lauretta}%
, {Dellagiustina}%
\BCBL{}\ \protect\BOthers{.}}{%
{\protect\APACyear{2019}}%
}]{%
Lauretta2019}%
\APACinsertmetastar{%
Lauretta2019}%
{Lauretta}, D\BPBI S.%
, {Dellagiustina}, D\BPBI N.%
, {Bennett}, C\BPBI A.%
, {Golish}, D\BPBI R.%
, {Becker}, K\BPBI J.%
, {Balram-Knutson}, S\BPBI S.%
\BCBL{}\ \BOthersPeriod{.}%
\unskip\
\newblock
\APACrefYearMonthDay{2019}{{\APACmonth{03}}}{}.
\newblock
\BBOQ{}\APACrefatitle{{The unexpected surface of asteroid (101955) Bennu}}{{The
  unexpected surface of asteroid (101955) Bennu}}.\BBCQ{}
\newblock
\APACjournalVolNumPages{Nature}{568}{}{55-60}.
\PrintBackRefs{\CurrentBib}

\bibitem[\protect\citeauthoryear{%
{Lauretta}%
, {Hergenrother}%
\BCBL{}\ \protect\BOthers{.}}{%
{Lauretta}%
, {Hergenrother}%
\BCBL{}\ \protect\BOthers{.}}{%
{\protect\APACyear{2019}}%
}]{%
Lauretta2019b}%
\APACinsertmetastar{%
Lauretta2019b}%
{Lauretta}, D\BPBI S.%
, {Hergenrother}, C\BPBI W.%
, {Chesley}, S\BPBI R.%
, {Leonard}, J\BPBI M.%
, {Pelgrift}, J\BPBI Y.%
, {Adam}, C\BPBI D.%
\BCBL{}\ \BOthersPeriod{.}%
\unskip\
\newblock
\APACrefYearMonthDay{2019}{{\APACmonth{12}}}{}.
\newblock
\BBOQ{}\APACrefatitle{{Episodes of particle ejection from the surface of the
  active asteroid (101955) Bennu}}{{Episodes of particle ejection from the
  surface of the active asteroid (101955) Bennu}}.\BBCQ{}
\newblock
\APACjournalVolNumPages{Science}{366}{6470}{3544}.
\PrintBackRefs{\CurrentBib}

\bibitem[\protect\citeauthoryear{%
{MacMillan}%
}{%
{MacMillan}%
}{%
{\protect\APACyear{1936}}%
}]{%
MacMillan1936}%
\APACinsertmetastar{%
MacMillan1936}%
{MacMillan}, W\BPBI D.%
%
\unskip\
\newblock
\APACrefYear{1936}.
\newblock
\APACrefbtitle{{Dynamics Of Rigid Bodies}}{{Dynamics Of Rigid Bodies}}.
\PrintBackRefs{\CurrentBib}

\bibitem[\protect\citeauthoryear{%
McMahon%
\ \protect\BOthers{.}}{%
McMahon%
\ \protect\BOthers{.}}{%
{\protect\APACyear{2020}}%
}]{%
McMahon2020}%
\APACinsertmetastar{%
McMahon2020}%
McMahon, J\BPBI W.%
, Scheeres, D\BPBI J.%
, Chesley, S\BPBI R.%
, French, A.%
, Brack, D.%
, Farnocchia, D.%
\BCBL{}\ \BOthersPeriod{.}%
\unskip\
\newblock
\APACrefYearMonthDay{2020}{}{}.
\newblock
\BBOQ{}\APACrefatitle{Dynamical Evolution of Simulated Particles Ejected from
  Asteroid Bennu}{Dynamical evolution of simulated particles ejected from
  asteroid bennu}.\BBCQ{}
\newblock
\APACjournalVolNumPages{Journal of Geophysical Research:
  Planets}{n/a}{n/a}{e2019JE006229}.
\newblock
 \begin{APACrefURL}
  \url{https://agupubs.onlinelibrary.wiley.com/doi/abs/10.1029/2019JE006229}
  \end{APACrefURL}
\newblock
\APACrefnote{e2019JE006229 2019JE006229}
\PrintBackRefs{\CurrentBib}

\bibitem[\protect\citeauthoryear{%
{Mignard}%
}{%
{Mignard}%
}{%
{\protect\APACyear{1984}}%
}]{%
Mignard1984}%
\APACinsertmetastar{%
Mignard1984}%
{Mignard}, F.%
%
\unskip\
\newblock
\APACrefYearMonthDay{1984}{{\APACmonth{01}}}{}.
\newblock
\BBOQ{}\APACrefatitle{{Effects of radiation forces on dust particles in
  planetary rings}}{{Effects of radiation forces on dust particles in planetary
  rings}}.\BBCQ{}
\newblock
\BIn{} R.~{Greenberg}\ \BBA{} A.~{Brahic}\ (\BEDS), \APACrefbtitle{IAU Colloq.
  75: Planetary Rings}{Iau colloq. 75: Planetary rings}\ (\BPG~333-366).
\PrintBackRefs{\CurrentBib}

\bibitem[\protect\citeauthoryear{%
{Moura}%
\ \protect\BOthers{.}}{%
{Moura}%
\ \protect\BOthers{.}}{%
{\protect\APACyear{2020}}%
}]{%
Moura2020}%
\APACinsertmetastar{%
Moura2020}%
{Moura}, T\BPBI S.%
, {Winter}, O\BPBI C.%
, {Amarante}, A.%
, {Sfair}, R.%
, {Borderes-Motta}, G.%
\BCBL{}\ \BBA{} {Valvano}, G.%
%
\unskip\
\newblock
\APACrefYearMonthDay{2020}{{\APACmonth{01}}}{}.
\newblock
\BBOQ{}\APACrefatitle{{Dynamical environment and surface characteristics of
  asteroid (16) Psyche}}{{Dynamical environment and surface characteristics of
  asteroid (16) Psyche}}.\BBCQ{}
\newblock
\APACjournalVolNumPages{\mnras}{491}{3}{3120-3136}.
\PrintBackRefs{\CurrentBib}

\bibitem[\protect\citeauthoryear{%
{Nolan}%
\ \protect\BOthers{.}}{%
{Nolan}%
\ \protect\BOthers{.}}{%
{\protect\APACyear{2013}}%
}]{%
Nolan2013}%
\APACinsertmetastar{%
Nolan2013}%
{Nolan}, M\BPBI C.%
, {Magri}, C.%
, {Howell}, E\BPBI S.%
, {Benner}, L\BPBI A\BPBI M.%
, {Giorgini}, J\BPBI D.%
, {Hergenrother}, C\BPBI W.%
\BCBL{}\ \BOthersPeriod{.}%
\unskip\
\newblock
\APACrefYearMonthDay{2013}{{\APACmonth{09}}}{}.
\newblock
\BBOQ{}\APACrefatitle{{Shape model and surface properties of the OSIRIS-REx
  target Asteroid (101955) Bennu from radar and lightcurve
  observations}}{{Shape model and surface properties of the OSIRIS-REx target
  Asteroid (101955) Bennu from radar and lightcurve observations}}.\BBCQ{}
\newblock
\APACjournalVolNumPages{Icarus}{226}{}{629-640}.
\PrintBackRefs{\CurrentBib}

\bibitem[\protect\citeauthoryear{%
{Park}%
, {Werner}%
\BCBL{}\ \BBA{} {Bhaskaran}%
}{%
{Park}%
\ \protect\BOthers{.}}{%
{\protect\APACyear{2010}}%
}]{%
Park2010}%
\APACinsertmetastar{%
Park2010}%
{Park}, R\BPBI S.%
, {Werner}, R\BPBI A.%
\BCBL{}\ \BBA{} {Bhaskaran}, S.%
%
\unskip\
\newblock
\APACrefYearMonthDay{2010}{{\APACmonth{01}}}{}.
\newblock
\BBOQ{}\APACrefatitle{{Estimating Small-Body Gravity Field from Shape Model and
  Navigation Data}}{{Estimating Small-Body Gravity Field from Shape Model and
  Navigation Data}}.\BBCQ{}
\newblock
\APACjournalVolNumPages{Journal of Guidance Control Dynamics}{33}{}{212-221}.
\PrintBackRefs{\CurrentBib}

\bibitem[\protect\citeauthoryear{%
{Petrovi{\'c}}%
}{%
{Petrovi{\'c}}%
}{%
{\protect\APACyear{1996}}%
}]{%
Petrovic1996}%
\APACinsertmetastar{%
Petrovic1996}%
{Petrovi{\'c}}, S.%
%
\unskip\
\newblock
\APACrefYearMonthDay{1996}{{\APACmonth{12}}}{}.
\newblock
\BBOQ{}\APACrefatitle{{Determination of the potential of homogeneous polyhedral
  bodies using line integrals}}{{Determination of the potential of homogeneous
  polyhedral bodies using line integrals}}.\BBCQ{}
\newblock
\APACjournalVolNumPages{Journal of Geodesy}{71}{}{44-52}.
\PrintBackRefs{\CurrentBib}

\bibitem[\protect\citeauthoryear{%
{Scheeres}%
}{%
{Scheeres}%
}{%
{\protect\APACyear{1994}}%
}]{%
Scheeres1994}%
\APACinsertmetastar{%
Scheeres1994}%
{Scheeres}, D\BPBI J.%
%
\unskip\
\newblock
\APACrefYearMonthDay{1994}{{\APACmonth{08}}}{}.
\newblock
\BBOQ{}\APACrefatitle{{Dynamics about Uniformly Rotating Triaxial Ellipsoids:
  Applications to Asteroids}}{{Dynamics about Uniformly Rotating Triaxial
  Ellipsoids: Applications to Asteroids}}.\BBCQ{}
\newblock
\APACjournalVolNumPages{\icarus}{110}{2}{225-238}.
\PrintBackRefs{\CurrentBib}

\bibitem[\protect\citeauthoryear{%
{Scheeres}%
, {French}%
\BCBL{}\ \protect\BOthers{.}}{%
{Scheeres}%
, {French}%
\BCBL{}\ \protect\BOthers{.}}{%
{\protect\APACyear{2019}}%
}]{%
Scheeres2019b}%
\APACinsertmetastar{%
Scheeres2019b}%
{Scheeres}, D\BPBI J.%
, {French}, A\BPBI S.%
, {McMahon}, J\BPBI W.%
, {Davis}, A\BPBI B.%
, {Brack}, D\BPBI N.%
, {Leonard}, J\BPBI M.%
\BCBL{}\ \BOthersPeriod{.}%
\unskip\
\newblock
\APACrefYearMonthDay{2019}{}{}.
\newblock
\BBOQ{}\APACrefatitle{{Implications Of The Gravity And Geophysical Environment
  Of (101955) Bennu For NEA Exploration}}{{Implications Of The Gravity And
  Geophysical Environment Of (101955) Bennu For NEA Exploration}}.\BBCQ{}
\newblock
\APACjournalVolNumPages{70th International Astronautical Congress (IAC)}{}{}{}.
\PrintBackRefs{\CurrentBib}

\bibitem[\protect\citeauthoryear{%
{Scheeres}%
\ \protect\BOthers{.}}{%
{Scheeres}%
\ \protect\BOthers{.}}{%
{\protect\APACyear{2016}}%
}]{%
Scheeres2016}%
\APACinsertmetastar{%
Scheeres2016}%
{Scheeres}, D\BPBI J.%
, {Hesar}, S\BPBI G.%
, {Tardivel}, S.%
, {Hirabayashi}, M.%
, {Farnocchia}, D.%
, {McMahon}, J\BPBI W.%
\BCBL{}\ \BOthersPeriod{.}%
\unskip\
\newblock
\APACrefYearMonthDay{2016}{{\APACmonth{09}}}{}.
\newblock
\BBOQ{}\APACrefatitle{{The geophysical environment of Bennu}}{{The geophysical
  environment of Bennu}}.\BBCQ{}
\newblock
\APACjournalVolNumPages{Icarus}{276}{}{116-140}.
\PrintBackRefs{\CurrentBib}

\bibitem[\protect\citeauthoryear{%
{Scheeres}%
, {Marzari}%
, {Tomasella}%
\BCBL{}\ \BBA{} {Vanzani}%
}{%
{Scheeres}%
\ \protect\BOthers{.}}{%
{\protect\APACyear{1998}}%
}]{%
Scheeres1998}%
\APACinsertmetastar{%
Scheeres1998}%
{Scheeres}, D\BPBI J.%
, {Marzari}, F.%
, {Tomasella}, L.%
\BCBL{}\ \BBA{} {Vanzani}, V.%
%
\unskip\
\newblock
\APACrefYearMonthDay{1998}{{\APACmonth{02}}}{}.
\newblock
\BBOQ{}\APACrefatitle{{ROSETTA mission: satellite orbits around a cometary
  nucleus}}{{ROSETTA mission: satellite orbits around a cometary
  nucleus}}.\BBCQ{}
\newblock
\APACjournalVolNumPages{Planetary Space Science}{46}{}{649-671}.
\PrintBackRefs{\CurrentBib}

\bibitem[\protect\citeauthoryear{%
{Scheeres}%
, {McMahon}%
\BCBL{}\ \protect\BOthers{.}}{%
{Scheeres}%
, {McMahon}%
\BCBL{}\ \protect\BOthers{.}}{%
{\protect\APACyear{2019}}%
}]{%
Scheeres2019}%
\APACinsertmetastar{%
Scheeres2019}%
{Scheeres}, D\BPBI J.%
, {McMahon}, J\BPBI W.%
, {French}, A\BPBI S.%
, {Brack}, D\BPBI N.%
, {Chesley}, S\BPBI R.%
, {Farnocchia}, D.%
\BCBL{}\ \BOthersPeriod{.}%
\unskip\
\newblock
\APACrefYearMonthDay{2019}{Mar}{}.
\newblock
\BBOQ{}\APACrefatitle{{The dynamic geophysical environment of (101955) Bennu
  based on OSIRIS-REx measurements}}{{The dynamic geophysical environment of
  (101955) Bennu based on OSIRIS-REx measurements}}.\BBCQ{}
\newblock
\APACjournalVolNumPages{Nature Astronomy}{3}{}{352-361}.
\PrintBackRefs{\CurrentBib}

\bibitem[\protect\citeauthoryear{%
{Stoer}%
\ \BBA{} {Bulirsch}%
}{%
{Stoer}%
\ \BBA{} {Bulirsch}%
}{%
{\protect\APACyear{1980}}%
}]{%
Stoer1980}%
\APACinsertmetastar{%
Stoer1980}%
{Stoer}, J.%
\BCBT{}\ \BBA{} {Bulirsch}, R.%
%
\unskip\
\newblock
\APACrefYear{1980}.
\newblock
\APACrefbtitle{{Introduction to Numerical Analysis}}{{Introduction to Numerical
  Analysis}}.
\PrintBackRefs{\CurrentBib}

\bibitem[\protect\citeauthoryear{%
{Tsoulis}%
\ \BBA{} {Petrovi{\'c}}%
}{%
{Tsoulis}%
\ \BBA{} {Petrovi{\'c}}%
}{%
{\protect\APACyear{2001}}%
}]{%
Tsoulis2001}%
\APACinsertmetastar{%
Tsoulis2001}%
{Tsoulis}, D.%
\BCBT{}\ \BBA{} {Petrovi{\'c}}, S.%
%
\unskip\
\newblock
\APACrefYearMonthDay{2001}{}{}.
\newblock
\BBOQ{}\APACrefatitle{{On the singularities of the gravity field of a
  homogeneous polyhedral body}}{{On the singularities of the gravity field of a
  homogeneous polyhedral body}}.\BBCQ{}
\newblock
\APACjournalVolNumPages{Geophysics}{66}{}{535}.
\PrintBackRefs{\CurrentBib}

\bibitem[\protect\citeauthoryear{%
{Walsh}%
\ \protect\BOthers{.}}{%
{Walsh}%
\ \protect\BOthers{.}}{%
{\protect\APACyear{2019}}%
}]{%
Walsh2019}%
\APACinsertmetastar{%
Walsh2019}%
{Walsh}, K\BPBI J.%
, {Jawin}, E\BPBI R.%
, {Ballouz}, R\BPBI L.%
, {Barnouin}, O\BPBI S.%
, {Bierhaus}, E\BPBI B.%
, {Connolly}, H\BPBI C.%
\BCBL{}\ \BOthersPeriod{.}%
\unskip\
\newblock
\APACrefYearMonthDay{2019}{Mar}{}.
\newblock
\BBOQ{}\APACrefatitle{{Craters, boulders and regolith of (101955) Bennu
  indicative of an old and dynamic surface}}{{Craters, boulders and regolith of
  (101955) Bennu indicative of an old and dynamic surface}}.\BBCQ{}
\newblock
\APACjournalVolNumPages{Nature Geoscience}{12}{4}{242-246}.
\PrintBackRefs{\CurrentBib}

\bibitem[\protect\citeauthoryear{%
{Werner}%
\ \BBA{} {Scheeres}%
}{%
{Werner}%
\ \BBA{} {Scheeres}%
}{%
{\protect\APACyear{1997}}%
}]{%
Werner1997b}%
\APACinsertmetastar{%
Werner1997b}%
{Werner}, R\BPBI A.%
\BCBT{}\ \BBA{} {Scheeres}, D\BPBI J.%
%
\unskip\
\newblock
\APACrefYearMonthDay{1997}{}{}.
\newblock
\BBOQ{}\APACrefatitle{{Exterior Gravitation of a Polyhedron Derived and
  Compared with Harmonic and Mascon Gravitation Representations of Asteroid
  4769 Castalia}}{{Exterior Gravitation of a Polyhedron Derived and Compared
  with Harmonic and Mascon Gravitation Representations of Asteroid 4769
  Castalia}}.\BBCQ{}
\newblock
\APACjournalVolNumPages{Celestial Mechanics and Dynamical
  Astronomy}{65}{}{313-344}.
\PrintBackRefs{\CurrentBib}

\bibitem[\protect\citeauthoryear{%
Williams%
, Kelley%
\BCBL{}\ \BBA{} {many others}%
}{%
Williams%
\ \protect\BOthers{.}}{%
{\protect\APACyear{2011}}%
}]{%
Williams2011}%
\APACinsertmetastar{%
Williams2011}%
Williams, T.%
, Kelley, C.%
\BCBL{}\ \BBA{} {many others}.%
%
\unskip\
\newblock
\APACrefYearMonthDay{2011}{March}{}.
\newblock
\APACrefbtitle{Gnuplot 4.4: an interactive plotting program.}{Gnuplot 4.4: an
  interactive plotting program.}
\newblock
\APAChowpublished{\url{http://gnuplot.sourceforge.net/}}.
\PrintBackRefs{\CurrentBib}

\bibitem[\protect\citeauthoryear{%
{Winter}%
\ \BBA{} {Amarante}%
}{%
{Winter}%
\ \BBA{} {Amarante}%
}{%
{\protect\APACyear{2017}}%
}]{%
Winter2017}%
\APACinsertmetastar{%
Winter2017}%
{Winter}, O\BPBI C.%
\BCBT{}\ \BBA{} {Amarante}, A.%
%
\unskip\
\newblock
\APACrefYearMonthDay{2017}{}{}.
\newblock
\BBOQ{}\APACrefatitle{{Mapping the Density of Particles over the Surface of
  Asteroid (101955) Bennu}}{{Mapping the Density of Particles over the Surface
  of Asteroid (101955) Bennu}}.\BBCQ{}
\newblock
\APACjournalVolNumPages{68th International Astronautical Congress (IAC)}{}{}{}.
\PrintBackRefs{\CurrentBib}

\bibitem[\protect\citeauthoryear{%
{Winter}%
\ \protect\BOthers{.}}{%
{Winter}%
\ \protect\BOthers{.}}{%
{\protect\APACyear{2020}}%
}]{%
Winter2020}%
\APACinsertmetastar{%
Winter2020}%
{Winter}, O\BPBI C.%
, {Valvano}, G.%
, {Moura}, T\BPBI S.%
, {Borderes-Motta}, G.%
, {Amarante}, A.%
\BCBL{}\ \BBA{} {Sfair}, R.%
%
\unskip\
\newblock
\APACrefYearMonthDay{2020}{{\APACmonth{03}}}{}.
\newblock
\BBOQ{}\APACrefatitle{{Asteroid triple-system 2001 SN$_{263}$: surface
  characteristics and dynamical environment}}{{Asteroid triple-system 2001
  SN$_{263}$: surface characteristics and dynamical environment}}.\BBCQ{}
\newblock
\APACjournalVolNumPages{\mnras}{492}{3}{4437-4455}.
\PrintBackRefs{\CurrentBib}

\bibitem[\protect\citeauthoryear{%
{Xin}%
, {Scheeres}%
, {Hou}%
\BCBL{}\ \BBA{} {Liu}%
}{%
{Xin}%
\ \protect\BOthers{.}}{%
{\protect\APACyear{2015}}%
}]{%
Xin2015}%
\APACinsertmetastar{%
Xin2015}%
{Xin}, X.%
, {Scheeres}, D\BPBI J.%
, {Hou}, X.%
\BCBL{}\ \BBA{} {Liu}, L.%
%
\unskip\
\newblock
\APACrefYearMonthDay{2015}{}{}.
\newblock
\BBOQ{}\APACrefatitle{{Dynamical substitutes of equilibrium points of asteroids
  under solar radiation pressure}}{{Dynamical substitutes of equilibrium points
  of asteroids under solar radiation pressure}}.\BBCQ{}
\newblock
\APACjournalVolNumPages{IAU XXIX General Assembly}{}{}{}.
\PrintBackRefs{\CurrentBib}

\end{thebibliography}

%
%
%
%
%

\end{document}